\documentclass[12pt,prd, aps, showpacs, superscriptaddress,floatfix]{revtex4}
\usepackage{graphicx}
\usepackage{bm}
\usepackage{multirow}
\usepackage{axodraw}
\usepackage{epsfig}
\usepackage{psfrag}
\usepackage{soul}
\usepackage{multirow}
\usepackage{mathptmx}
\usepackage{mathrsfs}
\usepackage{amsmath, amssymb}
\usepackage{dcolumn}
\usepackage{epstopdf}
\usepackage{color}
\usepackage{hhline}
\usepackage{changebar}
\usepackage{subfigure}
\usepackage{placeins,slashed}

\newcommand{\msbar}{\overline{\mathrm{MS}}}

\newcommand{\qslash}{\not{\! q}}
\newcommand{\qslashs}{\not{\hspace{1pt}{q}}}

\newcommand\riken{RIKEN-BNL Research Center, Brookhaven National
  Laboratory, Upton, NY 11973, USA}
\newcommand\bnl{Brookhaven National Laboratory, Upton, NY 11973, USA}
\newcommand\edinb{SUPA, School of Physics, The University of
  Edinburgh, Edinburgh EH9 3JZ, UK}
\newcommand\cu{Physics Department, Columbia University, New York,
  NY 10027, USA}
\newcommand\uconn{Physics Department, University of Connecticut,
  Storrs, CT 06269-3046, USA}
\newcommand\soton{School of Physics and Astronomy, University of
  Southampton,  Southampton SO17 1BJ, UK}
\newcommand{\maxplanck}{Max-Planck-Institut f\"ur Physik, F\"ohringer Ring 6, 80805 M\"unchen, Germany}
\newcommand{\stlouis}{Physics Department, Washington University, 
1 Brookings Drive, St. Louis, MO 63130-4899, USA}

\begin{document}
\bibliographystyle{apsrev}

\title{Lattice determination of the $K \rightarrow (\pi\pi)_{I=2}$ Decay Amplitude $A_2$}
\author{T.~Blum}\affiliation{\uconn}
\author{P.A.~Boyle}\affiliation{\edinb}
\author{N.H.~Christ}\affiliation{\cu}
\author{N.~Garron}\affiliation{\edinb}
\author{ E.~Goode}\affiliation{\soton}
\author{T.~Izubuchi}\affiliation{\bnl}\affiliation{\riken}
\author{C.~Jung}\affiliation{\bnl}
\author{C.~Kelly}\affiliation{\cu}
\author{C.~Lehner}\affiliation{\riken}
\author{M.~Lightman}\affiliation{\cu}\affiliation{\stlouis}
\author{Q.~Liu}\affiliation{\cu}
\author{A.T.~Lytle}\affiliation{\soton}
\author{R.D.~Mawhinney}\affiliation{\cu}
\author{C.T.~Sachrajda}\affiliation{\soton}
\author{A.~Soni}\affiliation{\bnl}
\author{C.~Sturm}\affiliation{\maxplanck}
\collaboration{The RBC and UKQCD Collaborations}

\mbox{}\hfill\noaffiliation{CU-TP-1202, Edinburgh 2012/10, MPP-2012-101, SHEP-1217}


\maketitle

\centerline{Abstract}
We describe the computation of the amplitude $A_2$ for a kaon to decay into two pions with isospin $I=2$. The results presented in the letter\,\cite{Blum:2011ng} from an analysis of 63 gluon configurations are updated to 146 configurations giving Re\,$A_2=1.381(46)_{\textrm{stat}}(258)_{\textrm{syst}}\,10^{-8}\,\textrm{GeV}$ and 
Im\,$A_2=-6.54(46)_{\textrm{stat}}(120)\,_{\textrm{syst}}10^{-13}\,{\rm GeV}$\,. Re\,$A_2$ is in good agreement with the experimental result, whereas the value of Im\,$A_2$ was hitherto unknown.
We are also working towards a direct computation of the $K\to(\pi\pi)_{I=0}$ amplitude $A_0$ but, within the standard model, our result for Im\,$A_2$ can be combined with the experimental results for Re\,$A_0$, Re\,$A_2$ and $\epsilon^\prime/\epsilon$ to give Im\,$A_0/$Re\,$A_0= -1.61(28)\times 10^{-4}$ . Our result for Im\,$A_2$ implies that the electroweak penguin (EWP) contribution to $\epsilon^\prime/\epsilon$ is Re$(\epsilon^\prime/\epsilon)_{\mathrm{EWP}} = -(6.25 \pm 0.44_{\textrm{stat}} \pm 1.19_{\textrm{syst}}) \times 10^{-4}$.

\section{Introduction}\label{sec:intro}

It was in $K\to\pi\pi$ decays that both indirect~\cite{Christenson:1964fg}
and direct~\cite{Burkhardt:1988yh,Gibbons:1993zq,Fanti:1999nm,AlaviHarati:2002ye} CP-violation was first discovered and a quantitative understanding of the origin of CP-violation, both within and beyond the Standard Model, remains one of the principal goals of particle physics research. Lattice QCD provides the opportunity of computing the non-perturbative QCD effects in general and in hadronic CP-violating processes in particular. The evaluation of these effects in $K\to\pi\pi$ decays is an important element in the research programme of the RBC-UKQCD collaboration and in this paper we report on the evaluation of the (complex) decay amplitude $A_2$, corresponding to the decay in which the two-pion final state has isospin 2. This is the first realistic \emph{ab initio} calculation of a weak hadronic decay.
Our final result can be found in Eq.\,(\ref{eq:results}), which we reproduce here for the reader's convenience:
\begin{equation}\label{eq:results0}
\textrm{Re}\,A_2=1.381(46)_{\textrm{stat}}(258)_{\textrm{syst}}\,10^{-8}\,\textrm{GeV},
\quad\textrm{Im}\,A_2=-6.54(46)_{\textrm{stat}}(120)\,_{\textrm{syst}}10^{-13}\,{\rm GeV}\,.
\end{equation}
This is an update of the result presented recently 
in Ref.\,\cite{Blum:2011ng} with greater statistics (146 configurations compared to 63 in \cite{Blum:2011ng}). More importantly, in this paper 
we present the details of the calculation and the analysis which could not be presented in the original letter\,\cite{Blum:2011ng}. For Re\,$A_2$ we find good agreement with the known experimental value ($1.479(4)\times10^{-8}$\,GeV obtained from $K^+$ decays), whereas the value of Im\,$A_2$ was previously unknown.

This is the first quantitative calculation of an amplitude for a realistic hadronic weak decay and hence extends the framework of lattice simulations into the important domain of non-leptonic weak decays. To reach this point has required very significant  theoretical  developments and technical progress. These are discussed in the following sections and include:
\begin{enumerate}
\item the control of $\pi\pi$ rescattering effects and  finite-volume corrections when two hadrons are present in the final state;
\item the use of carefully devised boundary conditions to tune the volume so that the decay can be simulated at physical kinematics;
\item the development of techniques for non-perturbative renormalization which has made it possible to calculate the matrix elements of the four-quark operators in the effective Hamiltonian  with good precision and without the use of lattice perturbation theory;
\item the improvement of algorithms and teraflops-scale computing which has made it possible to perform simulations at physical quark masses.
\end{enumerate}
It has therefore required a major endeavour to control all the ingredients of the calculation to arrive at the final result. The systematic errors in Eq.\,(\ref{eq:results0}) are dominated by the simple fact that the present calculation was performed at a single, rather large, value of the lattice spacing ($a\simeq .14\,$fm). With the greatly enhanced computing facilities made available to our collaboration and to others, the methods described in this paper can now be used at other lattice spacings to eliminate, or at least greatly reduce, the lattice artefacts. 

A major goal of our research programme is to calculate directly the amplitude $A_0$ for $K\to(\pi\pi)_{I=0}$ decays, in which the final-state pions have total isospin $I=0$, and $\epsilon^\prime/\epsilon$, the quantity which characterises direct CP-violation in $K\to\pi\pi$ decays, and we reviewed the status of our work in \cite{Blum:2011pu}. The evaluation of $A_0$ is considerably more difficult than the present calculation. Firstly, since the two-pion state
has vacuum quantum numbers we must evaluate disconnected diagrams with
sufficient precision. Secondly, in order to obtain physical
kinematics while avoiding the use of excited states, we must investigate
alternative methods of inducing momentum in the final state without
breaking isospin. (In the present
calculation we do break isospin symmetry through the use of different boundary conditions on the $u$ and $d$ quarks, but circumvent the issue of mixing with $I=0$ states since the final 
state has no $I=0$ component because of charge conservation; this is explained in Sec.\,\ref{sec:analysis}.)
Potential methods of improving the statistical precision in the calculation of disconnected-diagrams
include the use of advanced propagator-generation techniques
such as all-to-all propagators or low-mode/all-mode averaging. We are also investigating the use of G-parity boundary conditions\,\cite{Kim:2009fe} in order to achieve physical kinematics for decays into $I=0$ two-pion states. In the meantime, while we are developing and implementing these techniques for the direct evaluation of $A_0$, within the standard model we can combine our result for Im\,$A_2$ with the experimental values of Re\,$A_0$, Re\,$A_2$ and $\epsilon^\prime/\epsilon$ to determine the remaining unknown quantity Im\,$A_0$, so that the values of both the complex amplitudes $A_0$ and $A_2$ are now known (see Sec.\,\ref{subsec:a0}). We repeat however, that our ultimate goal is to compute $A_0$ directly, and we look forward to presenting results from a realistic computation in the future. 

This indirect determination of $A_0$ is also important in that it determines the $O(5\%)$ contribution of direct CP violation to $\epsilon$~\cite{Buras:2008nn,Buras:2010pza}. The relevance of such precision in tests of the Standard Model is due to the major recent improvement in the evaluation of the $B_K$ parameter for which recent calculations have reduced the uncertainty to less than 3\%~\cite{Colangelo:2010et}, (see Sec.\,\ref{subsec:a0}).

Since different authors use different conventions for the amplitudes we should state ours carefully. We define $A_I$ ($I=0,2$) by $\sqrt{2}A_I=\langle(\pi\pi)_I\,|\,H_W\,|K^0\rangle$ and the corresponding experimental results are Re\,$A_2\simeq |A_2|=1.479(4)\times 10^{-8}$\,GeV and Re\,$A_0\simeq |A_0|=3.320(2)\times 10^{-7}$\,GeV. Expressions for the widths for $K^+\to\pi^+\pi^0$, $K_S\to\pi^+\pi^-$ and $K_S\to\pi^0\pi^0$ decays in terms of the amplitudes are given in Eqs.\,(\ref{eq:kpippi0width}), (\ref{eq:kpippimwidth}) and (\ref{eq:kpi0pi0width}) and the surrounding discussion.

The structure of the remainder of the paper is as follows. In the next section we present the details of the simulation and explain the properties of the ensembles which were used. This is followed in Sec.\,\ref{sec:analysis} by a description of our analysis together with the final results. A presentation of the technical details of some of the components of the analysis, including the determination of the systematic errors are postponed to later sections. The renormalization of the operators present in the effective weak Hamiltonian is described in Sec.\,\ref{sec:npr} and the remaining sections are devoted to a detailed discussion of the systematic errors. Since the matrix elements  were calculated on a single coarse lattice, the corresponding artefacts are the largest component of the systematic error and we explain how we estimate them in Sec.\,\ref{sec:artefacts}. In Sec.\,\ref{sec:pq} we discuss the errors due to partial quenching and in Sec.\,\ref{sec:errors} we present the remainder of the error budget. Finally in Sec.\,\ref{sec:concs} we summarise and discuss the prospects for further work.

\section{Details of the Simulation}\label{sec:simulation} 
In this section we start with an explanation of the discrete QCD action used in our simulations (Subsec.\,\ref{subsec:action}). We then present the quark masses which we use and discuss the determination of the lattice spacing~(Subsec.\,\ref{subsec:parameters}) and finally in Subsec.\,\ref{subsec:sources} we discuss some technical issues concerning the calculation of the correlation functions from which the required matrix elements are determined.

\subsection{Lattice Action}\label{subsec:action}
For the quarks, we choose to use the domain wall fermion formulation~\cite{Kaplan:1992bt,Shamir:1993zy,Furman:1994ky}. This is a five dimensional description of QCD on a hypercubic grid, in which the fifth dimension of length $L_s$ serves to separate the left- and right-handed fermion chiralities which appear as surface states bound to opposite four-dimensional faces of the fifth dimension. The elusive chiral symmetry is restored in the limit $L_s\rightarrow \infty$. 
At finite $L_s$ the chiral symmetry is explicitly broken as the chiral modes can propagate across the fifth dimension. The symmetry breaking can be parametrised by the quantity $m_\mathrm{res}$, the residual mass, which additively renormalizes the bare quark masses. Its magnitude is governed by the density of eigenmodes of the 4D Hamiltonian obtained from the transfer matrix in the fifth dimension~\cite{Antonio:2008zz}. 
The contributions of the extended eigenmodes with eigenvalues above the mobility edge (which separates the localized low-modes from the extended high-modes) are dominant at small $L_s$ but fall exponentially as $L_s$ is increased. In modern simulations with large $L_s$, $m_\mathrm{res}$ is dominated by the density of near-zero eigenmodes; these are associated with localized and short-lived dislocations, or \textit{tears}� in the gauge fields which cause a change in the gauge field topology. 

We now discuss the choice of the gauge action. Until recently, our simulations~\cite{Allton:2008pn,Aoki:2010dy} have been performed using the Iwasaki renormalization-group improved gauge action, which has been shown to allow adequate gauge-field topology change while retaining good chiral properties in Monte Carlo simulations when used in conjunction with domain wall fermions. The lightest (unitary) pions in these simulations had masses of about 290\,MeV and the results were extrapolated to the physical value, $m_\pi\simeq 140$\,MeV. In the present computation of $K\to\pi\pi$ decay amplitudes, we perform the simulations with sufficiently light quark masses that the pions have (almost) their physical masses. However as the quark masses are decreased, the pions propagate over larger distances and they are more strongly affected by finite-volume effects; this necessitates the use of physically larger lattices. In order to make the simulation affordable, the large lattice is achieved by increasing the lattice spacing $a$ (decreasing the inverse gauge coupling $\beta$ used in the simulation). Unfortunately, as $\beta$ is lowered, the dislocations appear more frequently and thus $m_\mathrm{res}$ becomes large. To counter this effect we modify the Iwasaki gauge action with a weighting factor known as the Dislocation Suppressing Determinant Ratio (DSDR)~\cite{Vranas:1999rz,Vranas:2006zk,Fukaya:2006vs,Renfrew:2009wu}, allowing us to tune the molecular dynamics force in the gauge evolution to suppress configurations with large numbers of near-zero modes while retaining adequate topological change. This is discussed in more detail in ref.\,\cite{dsdrpaper}. For the remainder of this paper we label this action and the corresponding ensembles by IDSDR (representing Iwasaki + DSDR). The gauge action and ensembles without the DSDR correction are referred to simply by the label ``Iwasaki".

\subsection{Parameters of the Simulation}\label{subsec:parameters}

We have generated two ensembles of $2+1$ flavor domain wall fermions with the IDSDR gauge action at $\beta=1.75$ (corresponding to $a^{-1}=1.364\,$GeV, see below) and a lattice size of $32^3\times 64\times 32$, 
where the final number is $L_s$, the length of the fifth dimension. We determine the residual mass to be $m_\mathrm{res}=0.001843(8)$, approximately equal in size to the
 $3.6$\,MeV average of the up and down quark masses~\cite{dsdrpaper}. 
(Masses written without explicit units are to be understood as being in lattice units, so that for example, $m_{\mathrm{res}}=0.001843(8)$ should be read as 
$am_{\mathrm{res}}=0.001843(8)$.) The ensembles are generated with a simulated strange-quark mass of $m_h = 0.045$ and have light-quark masses of 
$m_l=0.001$ and $m_l=0.0042$, with corresponding unitary pion masses of approximately $170$~MeV and $250$~MeV respectively. 
For the determination of the lattice spacing $a$ and the physical bare quark masses used in the current project, as well as for the computation of the particle spectrum, 
decay constants and the kaon bag parameter $B_K$, we generate quark propagators with three heavy valence masses, $0.055$, $0.045$ and $0.035$, and four 
light valence quark masses, $0.008$, $0.0042$, $0.001$ and $0.0001$. The lightest partially-quenched pion has a near-physical mass of approximately $140$\,MeV. 
The analysis presented in this paper is performed using 146 configurations from the $0.001$ ensemble, each separated by  8 molecular dynamics time units, 
with additional strange quark propagators with $m_h=0.049$ corresponding to our original estimate of the physical value of the (bare) strange quark mass, and light-quark propagators with a valence mass of $0.0001$ . The subsequent detailed 
analysis with greater statistics and improved procedures have yielded the value 0.0472(6)
for the bare physical strange quark mass.

We obtain the lattice spacing and the two physical quark masses $m_{ud}$ and $m_s$ using a combined analysis of these IDSDR ensembles and our $32^3\times 64\times 16$ and $24^3\times 64\times 16$ domain wall fermion configurations with the Iwasaki gauge action at $\beta=2.25$ and $\beta = 2.13$ respectively~\cite{Allton:2008pn,Aoki:2010dy}. This involves a combined fit of the pion and kaon masses and decay constants and the mass of the $\Omega$-baryon as functions of the quark masses and lattice spacing. We use three different ans\"{a}tze for the quark-mass dependence in order to estimate the systematic error on the chiral extrapolations. Two of these are obtained from next-to-leading order (NLO) partially-quenched chiral perturbation theory with and without finite-volume corrections, and the third assumes a simple linear mass dependence (labelled \emph{analytic} in the following). Following our 2010 analysis~\cite{Aoki:2010dy} of the two Iwasaki lattices, we extrapolate to the continuum limit along a family of scaling trajectories (lines of constant physics) that are defined by constant values of $m_\pi$, $m_K$ and $m_{\Omega}$;  
i.e. by imposing the condition that these masses have no lattice cutoff dependence on the scaling trajectory. The leading dependence on $a$ of the remaining quantities is expected to be $O(a^2)$ and in our fits we assume such a quadratic dependence. Note that the coefficients of the $a^2$ terms are not constrained to be equal for the two different lattice actions. From the combined chiral and continuum fits
we determine the lattice spacings and physical quark masses required for the pion, kaon and $\Omega$ masses to match their physical values, obtaining for the IDSDR ensembles an inverse-lattice spacing of $a^{-1}=1.364(9)$\,GeV and dimensionless physical quark masses of $\tilde m_l  = 0.00178(3)$ and 
$\tilde m_s = 0.0490(6)$, which correspond to $3.09\pm 0.11$ and $84.1\pm 2.0$\,MeV respectively when expressed in physical units in the $\overline{\rm MS}$ scheme
at 3\,GeV. Here $\tilde m = m+m_\mathrm{res}$ and the quoted errors contain both statistical and systematic contributions estimated using the procedures developed in ref.~\cite{Aoki:2010dy}.

The numbers presented above were all obtained from an analysis of the 146~configurations used below in the evaluation of the $K\to\pi\pi$ matrix elements. Ref.\,\cite{dsdrpaper} contains a detailed analysis on an extended set of ensembles (including  180~configurations for $m_l=0.001$). The corresponding values in Ref.\,\cite{dsdrpaper} include $a^{-1}=1.371(8)$\,GeV for the inverse lattice spacing, $\tilde m_l  = 0.00176(2)$ and 
$\tilde m_s = 0.0486(6)$ for the dimensionless physical  quark masses and  $3.05\pm 0.11$ and $83.6\pm 2.1$\,MeV respectively for the quark masses in physical units in the $\overline{\rm MS}$ scheme
at 3\,GeV. 

In order to correctly propagate the correlations between the data used
in the determination of the lattice spacings and physical quark masses
with that of the present calculation of the $K\to\pi\pi$ matrix elements we make use of the super-jackknife
method, in which the statistical fluctuations associated with each
ensemble are maintained separately, and the total error is determined by
combining these contributions in quadrature. This prevents accidental
correlations between the statistically independent data on each of the
ensembles, and therefore improves on the bootstrap and standard
jackknife methods for combining independent data. (The super-jackknife
also does not require the number of samples on each ensemble to be the
same, a limitation of the traditional jackknife.) A clear description of the super-jacknife technique can be found in~\cite{Bratt:2010jn}.

\subsection{Evaluation of the Correlation Functions}\label{subsec:sources}

\begin{figure}[t]
\begin{center}
\includegraphics[width = 0.6 \textwidth]{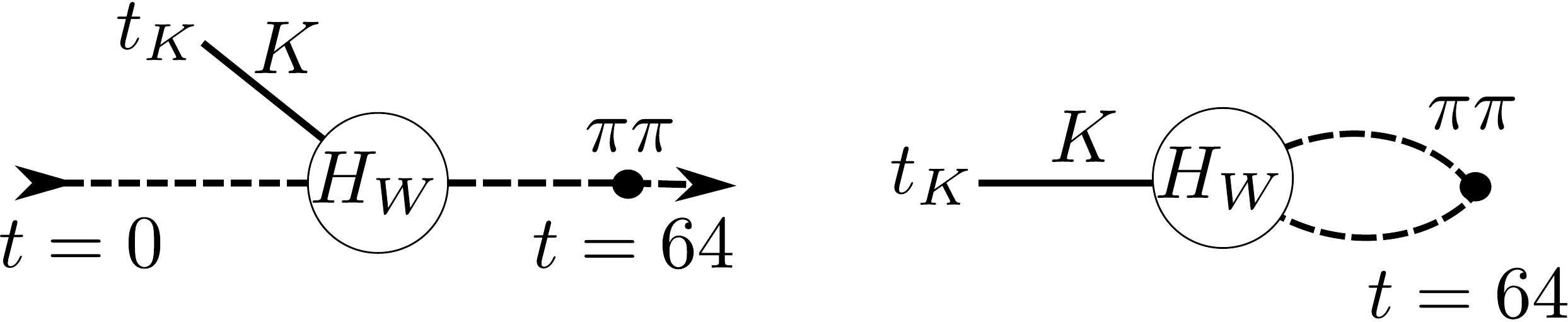}
\caption{\label{fig:roundtheworld} An illustration of around-the-world effects in the $K\to\pi\pi$ correlation function. In the left-hand figure the two-pion operator at $t=64$ annihilates one pion and creates another, while the weak Hamiltonian annihilates the kaon and a pion and creates a pion. The right-hand diagram illustrates the $K\to\pi\pi$ transition whose matrix element we evaluate.}\end{center}
\end{figure}

We now explain some technical details concerning the evaluation of the correlation functions from which the matrix elements for $K\to\pi\pi$ decays are evaluated. Quark propagators with periodic and antiperiodic boundary conditions in the time direction were computed on each configuration with a source at $t=0$. They were then combined so as to effectively double the time extent of the lattice. Meson correlation functions formed using the sum of the propagators
with periodic and antiperiodic boundary conditions can be interpreted as
containing forward propagating mesons originating at time $t=0$,  whereas
those calculated with the difference can be interpreted
as containing backward propagating mesons originating from a source at $t
= 64$. The purpose of this procedure is to suppress the so called ``around the world'' effects. An example of such effects
can be seen in the two-pion correlation function, $C_{\pi\pi}(t)$:
\begin{equation}
C_{\pi\pi}(t)=\langle\,0\,|\,J_{\pi\pi}(t)\,J_{\pi\pi}^\dagger(0)\,|\,0\,\rangle= |\langle\,0|\,J_{\pi\pi}(0)\,|\,\pi\pi\,\rangle|^2e^{-E_{\pi\pi} t}+\cdots\,.\label{eq:cpipidef}\end{equation}
The term on the right-hand side of (\ref{eq:cpipidef}) corresponds to the creation of two pions at time zero by $J_{\pi\pi}^\dagger$ and their annihilation by $J_{\pi\pi}$ at $t$. The corresponding functional integral however, also has a contribution where each of $J_{\pi\pi}^\dagger(0)$ and $J_{\pi\pi}(t)$ annihilate one pion and create another, so that a single pion propagates across the entire lattice. This contribution to the correlation function is independent of $t$, and although it contains the small factor $e^{-E_\pi\,T}$, where $T$ is the temporal size of the lattice, it may nevertheless lead to a loss of precision. Combining the propagators obtained with periodic and antiperiodic boundary conditions effectively replaces $T$ by $2T$ thus suppressing this unwanted contribution.
A similar effect can occur in the $K \rightarrow \pi\pi$ correlator
if the weak operator in the effective Hamiltonian annihilates the kaon and one pion and creates a new pion, before the two-pion interpolating operator annihilates
this pion and creates another (see Fig.\,\ref{fig:roundtheworld}).
Strange-quark propagators, with periodic + antiperiodic combinations, were generated with sources at $t_K = $ 20, 24, 28, 32, 36, 40 and 44
in order to calculate $K\rightarrow \pi\pi$ correlation functions with kaon sources at these times, while the two-pion sources remained at either $t=0$ or $t=64$. Thus we could achieve time separations between the kaon and two pions of 20, 24, 28 and 32 lattice time units in two different ways which increased the statistics. These separations were chosen so that the signals from the kaon and two pions did not decay into noise before reaching the four-quark weak operator.

We end this section with an explanation of the sources which were used for the quark propagators and hence of the operators which create and annihilate the mesons. For the propagators of the $u$ and $s$ quarks, which have periodic spatial boundary conditions, we use Coulomb gauge-fixed wall sources. For the $d$-quark on the other hand we impose antiperiodic boundary conditions in some spatial directions and use Coulomb gauge-fixed momentum wall sources of the ``cosine'' type
\begin{equation}\label{eq:cossources}
s_{\mathbf{p},\cos}(\mathbf{x}) = \cos\left(p_x x\right) \cos\left(p_y y\right) \cos\left(p_z z\right).
\end{equation}
Here the components of momentum are given by $p_i=n_i(\pi/L)$ where $n_i$ is an even or odd integer depending on whether periodic or antiperiodic boundary conditions were imposed on the quark field in direction $i$. For our lattice, the choice $n_1=n_2=1$ and $n_3=0$ (or permutations) corresponds approximately to the kinematics of a physical $K\to\pi\pi$ decay. 

As explained at the beginning of Sec\,\ref{sec:analysis}, we use the Wigner-Eckart theorem to relate the physical amplitude $A_2$ which we wish to determine to unphysical $K^+\to\pi^+\pi^+$ matrix elements which we compute directly.
When studying the propagation of two $\pi^+$ mesons we use the same cosine source for each $d$-quark, which introduces cross terms in correlation functions that couple to two-pion states with non-zero total momentum. For illustration, consider the case $\mathbf{p}=(\pi/L,0,0)$ so that the  product of the sources of the two $d$-quarks is
\begin{equation}
\begin{split}
 s_{\mathbf{p},\text{cos}}(\mathbf{x}_1)&s_{\mathbf{p},\text{cos}}(\mathbf{x}_2)  = \cos\left(\frac{\pi}{L} x_1\right) \cos\left(\frac{\pi}{L}x_2\right) \\
& = \frac{1}{4}\left( e^{i\frac{\pi}{L}x_1}e^{i\frac{\pi}{L}x_2} + e^{i\frac{\pi}{L}x_1}e^{-i\frac{\pi}{L}x_2}
+ e^{-i\frac{\pi}{L}x_1}e^{i\frac{\pi}{L}x_2} +e^{-i\frac{\pi}{L}x_1}e^{-i\frac{\pi}{L}x_2} \right).
\end{split}
\label{eq:multcos}
\end{equation}
We require the two pions to have individual momenta $\mathbf{p}_1 = \frac{\pi}{L}\mathbf{\hat x}$ and
$\mathbf{p}_2 = -\frac{\pi}{L}\mathbf{\hat x}$ (or vice-versa), but the first and last terms on the right hand side of Eq.\,\eqref{eq:multcos} couple to two-pion states with total momentum $2\frac{\pi}{L}$ and $-2\frac{\pi}{L}$ respectively. We eliminate the unwanted terms in the two-pion correlation functions by using different sinks, $\exp(\pm i \pi x_i/L)$, for the
    two $d$ quarks ensuring that they carry equal and opposite
    momenta which constrains the final state to have zero total momentum. In the $K\rightarrow \pi\pi$ correlation functions, the kaon has zero 
momentum and the sum over the spatial position of the weak operator then ensures that the two-pion final state also has zero total momentum.

The advantage of using the cosine sources is that it halves the number of inversions which have to be performed for the $d$-quark. Had we used the more conventional momentum source,
\begin{equation}
 s_{\mathbf{p}}(\mathbf{x}) = e^{i\mathbf{p}\cdot\mathbf{x}}\,,
\end{equation}
we would have needed to perform two separate $d$-quark inversions with momentum $+\mathbf{p}$ for one and $-\mathbf{p}$ for the other. The cosine source eliminates one of these inversions. In practice we only compute $d$-quark propagators with antiperiodic boundary conditions in $0$ or $2$ spatial directions, corresponding to pions with ground-state momenta $|\mathbf{p}|=0$ and $|\mathbf{p}| = \sqrt{2}\pi/L$. As explained above, this choice is motivated by the expectation that, with our choice of quark masses, $|\mathbf{p}|=\sqrt{2}\pi/L$
corresponds to on-shell kinematics, i.e. that the energy of the two-pion state is (almost) equal to $m_K$.

We mention one further subtlety. As explained above, the use of antiperiodic boundary conditions in two spatial directions for the $\bar{d}$ quark enabled us to match the two-pion energy with $m_K$. 
It was shown in \cite{Sachrajda:2004mi} that it is sufficient to use the antiperiodic boundary conditions only on the valence down anti-quarks in the $\pi^+$ mesons, and to use periodic boundary conditions for the sea quarks used in the simulations. Thus we can use the gluon configurations already generated in which periodic boundary conditions were imposed on all the sea quarks.

\section{The Analysis}\label{sec:analysis}

In this section we describe the evaluation of $A_2$. While the results presented in Eq.\,(\ref{eq:results}) towards the end of this section contain our estimates of the uncertainties, we postpone the detailed discussion of the determination of the systematic errors to the subsequent sections. 

The generic form of the effective Hamiltonian for $K\to(\pi\pi)_{I=2}$ decays is
\begin{equation}\label{eq:ope}
H_{\rm eff}=\frac{G_F}{\sqrt{2}}\,V_{us}^\ast V_{ud}\sum_{i}\,C_i\,Q^{3/2}_i\,,
\end{equation}
where $G_F$ is the Fermi constant, $V_{ud}$ and $V_{us}$ are CKM-matrix elements, 
$V_{ud} = 0.97429$, $V_{us}=0.2253$ and the $C_i$ are Wilson coefficient functions. The $C_i$ contain a dependence on $\tau = -V^*_{ts}V_{td}/V^*_{us}V_{ud} = 0.0014606 - 0.00060408 i$, as explained below.

The three four-quark operators contributing to the effective Hamiltonian for $\Delta I=3/2$ decays are
\begin{eqnarray}\label{eq:q271}
Q^{3/2}_{(27,1)}&=&(\bar{s}^id^i)_L\,\big\{(\bar{u}^ju^j)_L-(\bar{d}^jd^j)_L\big\}+(\bar{s}^iu^i)_L\,(\bar{u}^jd^j)_L \label{eq:o271def}\\ \label{eq:q88}
O^{3/2}_{(8,8)}&=&(\bar{s}^id^i)_L\,\big\{(\bar{u}^ju^j)_R-(\bar{d}^jd^j)_R\big\}+(\bar{s}^iu^i)_L\,(\bar{u}^jd^j)_R \label{eq:o732def}\\
O^{3/2}_{(8,8)\mathrm{mix}}&=&(\bar{s}^id^j)_L\,\big\{(\bar{u}^ju^i)_R-(\bar{d}^jd^i)_R\big\}+(\bar{s}^iu^j)_L\,(\bar{u}^jd^i)_R\,,\label{eq:q88mix}
\end{eqnarray}
where the superscript $3/2$ denotes $\Delta I=3/2$ transitions and the subscripts denote how the operators transform under the
$\text{SU}(3)_L \times \text{SU}(3)_R$ chiral symmetry. $i,j$ are color labels which run from 1 to 3 and $L,R$ denote left and right (e.g. $(\bar s d)_L (\bar u u)_L=(\bar s\gamma^\mu(1-\gamma^5)d)\,(\bar u\gamma^\mu(1-\gamma^5)u)$ and $(\bar s d)_L (\bar u u)_R=(\bar s\gamma^\mu(1-\gamma^5)d)\,(\bar u\gamma^\mu(1+\gamma^5)u)$ with the spinor labels contracted within each pair of parentheses)\,. 

In physical $K^+\to\pi^+\pi^0$ decays the third component of isospin, $I_z$, changes by 1/2, $\Delta I_z=1/2$.
As proposed and first explored in \cite{Kim:2003xt,Kim:2005gka}, it is particularly convenient to use the Wigner-Eckart theorem to relate the matrix elements of the operators in (\ref{eq:q271})\,-\,(\ref{eq:q88mix}) between $|K^+\rangle$ and $|\pi^+\pi^0\rangle$ states to those of the corresponding operators with $\Delta I_z=3/2$ for the unphysical process $K^+\to\pi^+\pi^+$:
\begin{equation}\label{eq:wigner_eckart}
\langle\pi^+\pi^0\,|Q^{\Delta I=3/2}_{\Delta I_z=1/2}|\,K^+\rangle=\frac{\sqrt3}2\langle \pi^+\pi^+\,|Q^{\Delta I=3/2}_{\Delta I_z=3/2}|\,K^+\rangle\,.
\end{equation}
On the left-hand side of Eq.\,(\ref{eq:wigner_eckart}) the operator $Q^{\Delta I=3/2}_{\Delta I_z=1/2}$ is one of the three operators in Eqs.\,(\ref{eq:q271})-(\ref{eq:q88mix}), whereas on the right-hand side the operators $Q^{\Delta I=3/2}_{\Delta I_z=3/2}$ operators are $\sqrt{3}\,Q_i$, where $i$ runs over the labels $(27,1),\,(8,8)$ and $(8,8)_{\mathrm{mix}}$ and
\begin{equation}\label{eq:Qdef}
Q_{(27,1)}=(\bar{s}^id^i)_L\,(\bar{u}^jd^j)_L,\quad
Q_{(8,8)}=(\bar{s}^id^i)_L\,(\bar{u}^jd^j)_R, \quad
Q_{(8,8){\rm mix}}=(\bar{s}^id^j)_L\,(\bar{u}^jd^i)_R \,.
\end{equation}
$\sqrt{3}/2$ in Eq.\,(\ref{eq:wigner_eckart}) is the Clebsch-Gordan factor and, neglecting violations of isospin, Eq.\,(\ref{eq:wigner_eckart}) is exact. $\text{A}_2$ can therefore be determined by computing the matrix elements of the three operators in Eq.\,(\ref{eq:Qdef}) and indeed it is these three matrix elements which we compute directly. For compactness of notation we suppress the labels $\Delta I$ and $\Delta I_z$ on the operators both in Eq.\,(\ref{eq:Qdef}) and in the following. 

The use of the Wigner-Eckart theorem to replace the operators in Eqs.\,(\ref{eq:q271})\,-\,(\ref{eq:q88mix}) by those in Eq.\,(\ref{eq:Qdef}) leads to very significant practical simplifications. All the quarks participating directly in $\Delta I=3/2$ decays are valence quarks and in such cases the effects of introducing partially-twisted boundary conditions (for which the valence and sea quarks satisfy different boundary conditions) 
are exponentially small~\cite{Sachrajda:2004mi}.
In particular, we assign anti-periodic boundary conditions in some directions to the valence $d$ quarks, so that the corresponding components of the momenta of the final-state $\pi^+$ mesons are $(2n+1)\,\pi/L$ , where $n$ is an integer and $L$ is the spatial extent of the lattice. The volume of the lattice has been chosen so that for pions with momenta $\sqrt{2}\pi/L$, the energy of the two-pion state $E_{\pi\pi}$ is very close to $m_K$, the mass of the kaon, $m_K\simeq E_{\pi\pi}$, corresponding to a physical decay. (The total three-momentum of the kaon and of the two-pion state are zero.) The most significant simplification is that the two-pion state is the lightest one with these boundary conditions. With periodic boundary conditions, the lightest two-pion state is one with each of the two pions at rest and so when computing physical $K\to\pi\pi$ amplitudes one is obliged to consider excited two-pion states~\cite{Lellouch:2000pv}. Moreover, for $K^+\to\pi^+\pi^0$ matrix elements even with anti-periodic boundary conditions on one or more of the quark fields, the momentum of the $\pi^0$ is $2n\pi/L$ with integer $n$, negating the advantages described above.
Finally we note that by using anti-periodic boundary conditions one can achieve the kinematics of a physical decay on a smaller lattice than with periodic boundary conditions. 

\begin{figure}[t]
 \centering
\subfigure[\label{sfig:pion_eff_mass} Effective mass plot for the pion]{\includegraphics*[width=0.45\textwidth]{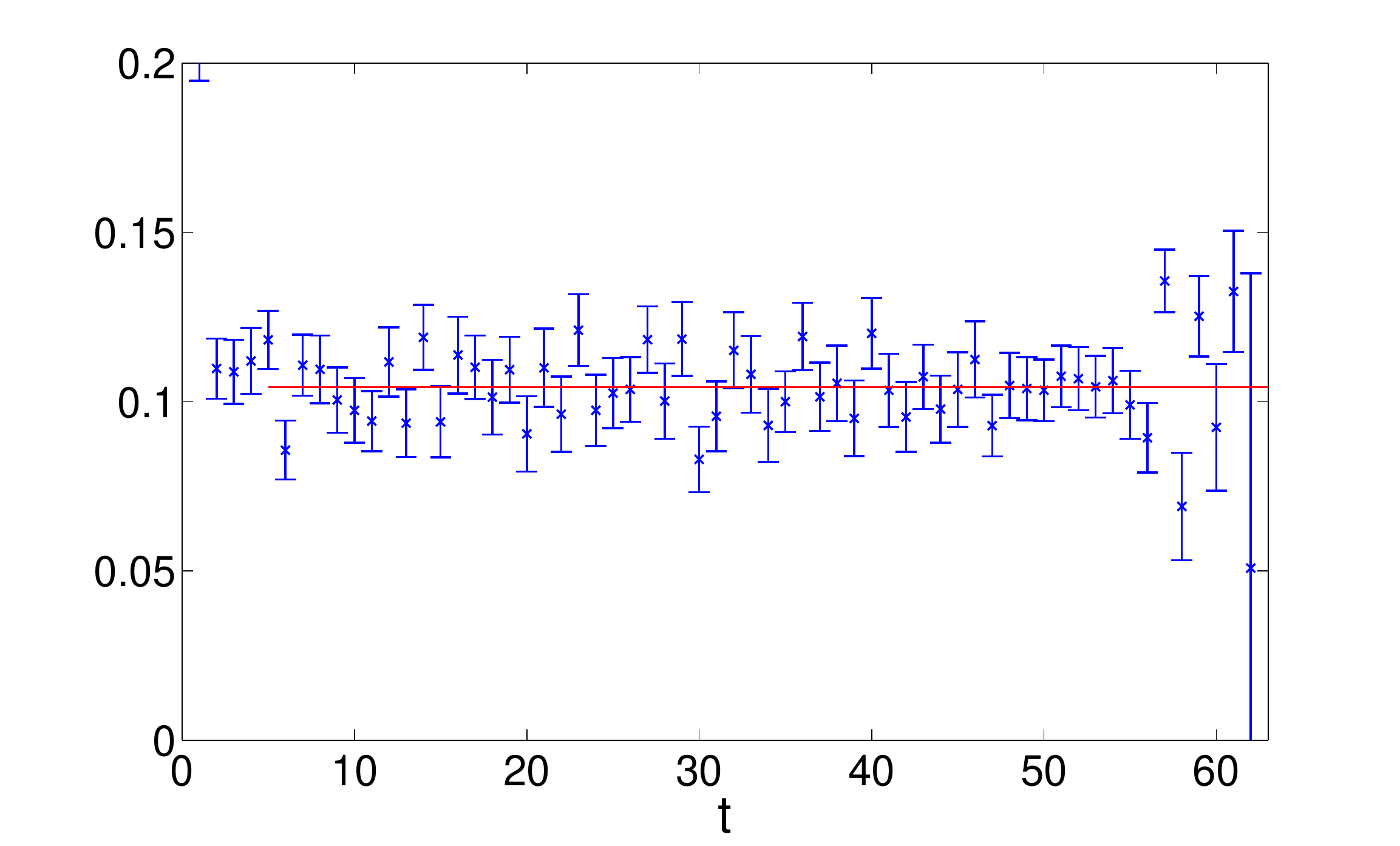}}\hspace{0.25in}
\subfigure[\label{sfig:kaon_eff_mass} Effective mass plot for the kaon]{\includegraphics*[width=0.45\textwidth]{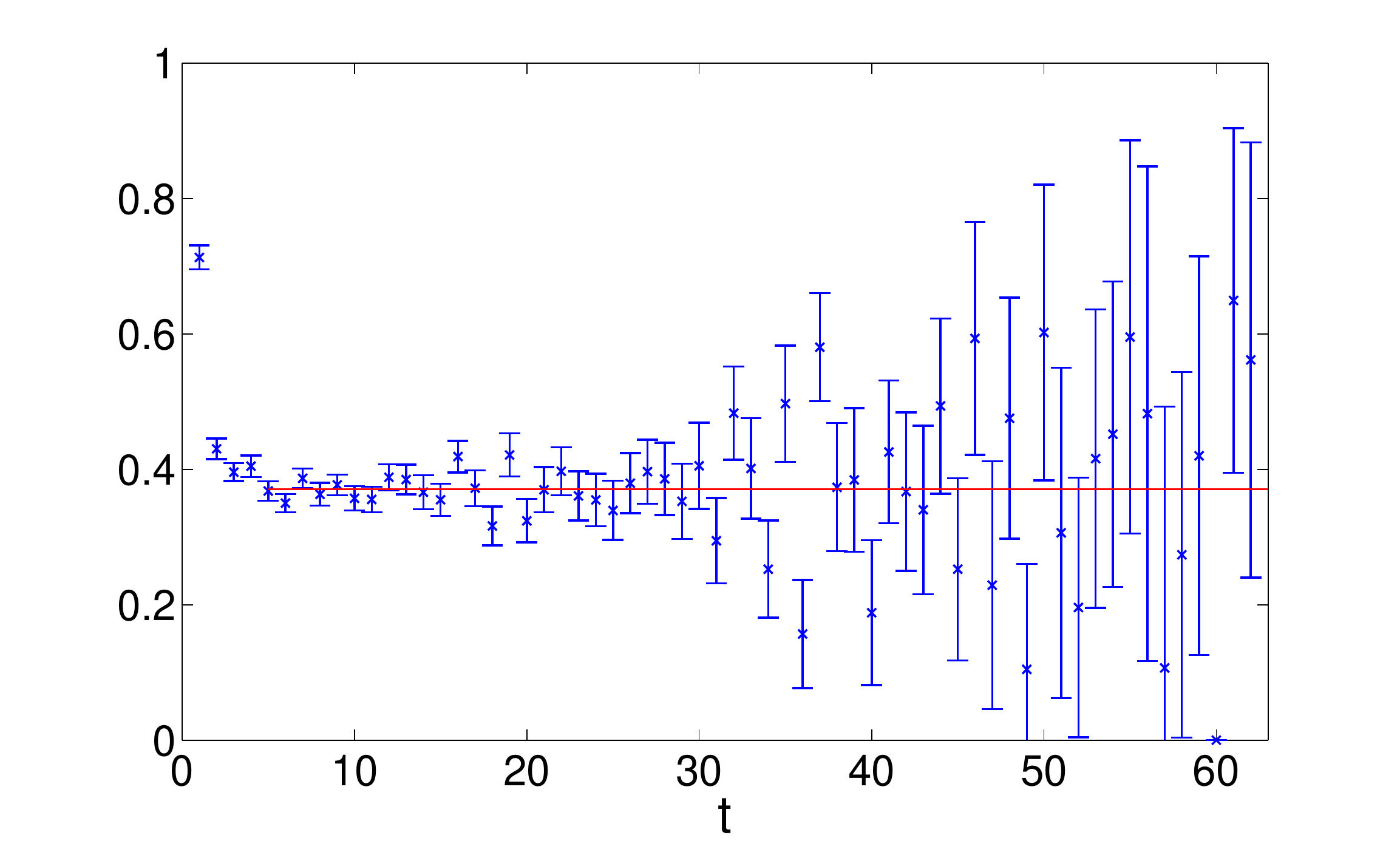}}
\caption{\label{fig:meson_effmass} Effective mass plots for the pion and kaon. Results for $m_\pi$ and $m_K$ obtained from the fits of the correlation functions to Eq.\,(\ref{eq:mesoncorr}) are shown as the horizontal lines in each plot.}
\end{figure}

We now turn to the determination of the matrix elements. The pion and kaon two-point correlation functions at zero momentum are fit to the form
\begin{equation}
\label{eq:mesoncorr}
C_{P}(t) =  \langle 0\,|\,J_P(t)\,J_P^\dagger(0)\,|0\rangle =
|Z_{P}|^2 \left( e^{-m_{P}t} + e^{-m_{P}(T - t)} \right),
\end{equation}
where $T = 128$ is the total effective time extent of the lattice and $m_P$ is the mass of 
pseudoscalar meson $P$. $J_P$ and $J_P^\dagger$ are interpolating annihilation and creation operators for the meson $P$ and Equation (\ref{eq:mesoncorr})
defines $Z_{\pi}$ and $Z_{K}$ for $P=\pi$ and $P=K$ respectively.
For both the pion and kaon the final results are obtained by fitting between $t=5$ and $t=63$. 
The masses extracted from these fits are superimposed on the effective mass plots in Figs.\,\ref{sfig:pion_eff_mass} and
 \ref{sfig:kaon_eff_mass}, and the numerical results are given in Tab.\,\ref{tab:masses}. The effective mass in these plots, $m_{P,\,\mathrm{eff}}$ is defined by $C_P(t)/C_P(t+1)=\cosh(m_{P,\,\mathrm{eff}}(t-T/2))/\cosh(m_{P,\,\mathrm{eff}}(t+1-T/2))$.

\begin{table}[t]
\begin{center}\vspace{0.15in}
\begin{tabular}{|c|c|c|c|c|c|c|}
\hline
units&$m_{\pi}$ & $m_K$ & $E_{\pi,2}$ & $E_{\pi\pi,0}$ & $E_{\pi\pi,2}$ & $m_K - E_{\pi\pi,2}$ \\
\hline
lattice &0.10421(22) & 0.37066(68) & 0.17386(91) & 0.21002(43)& 0.3560(23) & 0.0146(23)\\
MeV &142.11(94) & 505.5(3.4) & 237.1(1.8)& 286.4(1.9)& 485.5(4.2) &20.0(3.1)\\
\hline
\end{tabular}
\caption{Results for meson masses and energies. The subscripts $0$, $2$ denote
$p= 0$ and $p = \sqrt{2}\pi/L$ respectively, where $p=|\mathbf{p}|$. \label{tab:masses}}\end{center}
\end{table}

\begin{figure}[t]
\begin{center}
\includegraphics[width = 0.45 \textwidth]{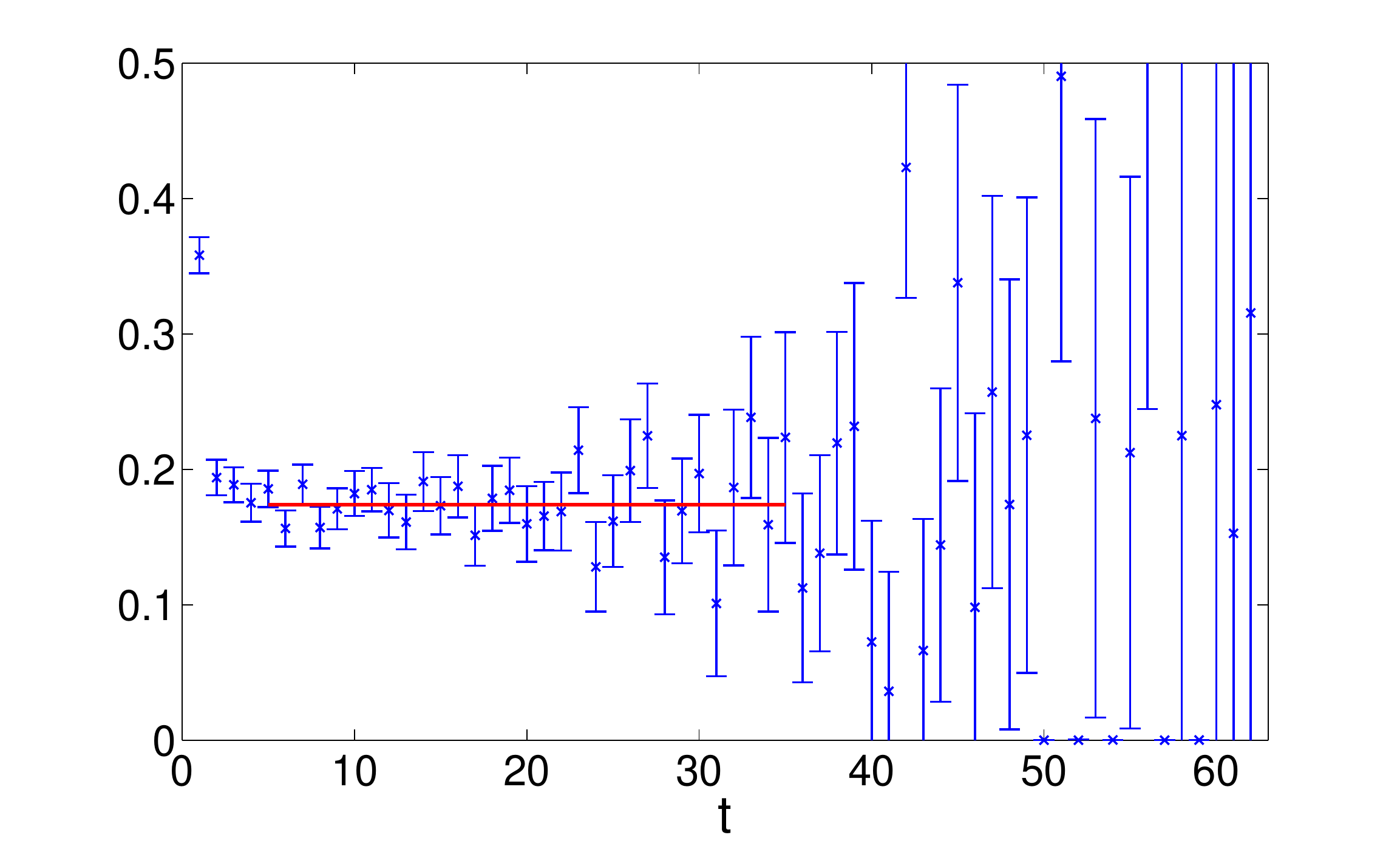}
\caption{\label{fig:pionP2_eff_mass} Effective energy plot for a pion with momentum
$p=\sqrt{2} \pi/L$. The horizontal line corresponds to value of $E_\pi$ obtained from a fit to Eq.\,(\ref{eq:pion_p2}).}\end{center}
\end{figure}

The pions in the final state for $K\to\pi\pi$ decays have momentum $|\mathbf{p}|=\sqrt{2}\,\pi/L$ and in 
Fig.\,\ref{fig:pionP2_eff_mass}  we plot the effective energy for a pion with this momentum. Since the correlation functions become noisier when the pion has a non-zero momentum, we now fit over the time interval $t=[5,35]$ where we can neglect the contribution from the backward propagating pion and use the form,
\begin{equation}
\label{eq:pion_p2}
C_{\pi}(t,p=\sqrt{2}\pi/L) = |Z_{\pi}(p=\sqrt{2}\pi/L)|^2 e^{-E_{\pi} t}\,,
\end{equation}
where $p=|\mathbf{p}|$ and $E_\pi$ is the corresponding energy. The value $E_{\pi,2}=0.17386(91)$ obtained from the fit (see Tab.\,\ref{tab:masses}) is nicely consistent with the (continuum) dispersion relation for a pion with mass 0.10421(22). The subscript 2 in $E_{\pi,2}$ indicates that the momentum of the pion is $\sqrt{2} \pi/L$, i.e. that anti-periodic boundary conditions have been imposed on the $d$ quark in two directions.

Next we consider the two-pion correlation function which has a larger statistical error. Having suppressed the \textit{around-the-world} contributions by combining propagators with periodic and antiperiodic boundary conditions in time and neglecting the contributions from excited states, the expected behavior of the two-pion correlation function is  
\begin{equation}
\label{eq:pipi_corr}
  C_{\pi\pi}(t) \equiv \langle 0\,|\,J_{\pi\pi,\,\mathrm{e}}(t)\,J^\dagger_{\pi\pi,\mathrm{c}}(0)\,|\,0\rangle
 = \frac{1}{2^{n_\mathrm{tw}}}\mathrm{}\, |Z_{\pi\pi,\,\mathrm{e}}|^2\,\left( e^{-E_{\pi\pi}t} + e^{ -E_{\pi\pi}(T - t)} \right)\,,
\end{equation}
where the labels $\mathrm{c}$ and $\mathrm{e}$ refer to the cosine and exponential sources discussed in Sec.\,\ref{subsec:sources} and $n_\mathrm{tw}$ is the number of directions with anti-periodic boundary conditions on the $d$ quark.
The leading around-the-world effects would manifest themselves as a time-independent constant on the right-hand side of Eq.\,(\ref{eq:pipi_corr}).

\begin{figure}[t]
 \centering
\subfigure[\label{sfig:Epipi_p0} $C_{\pi\pi}(t)/C_{\pi}^2(t), p=0$ ]{\includegraphics*[width=0.45\textwidth]{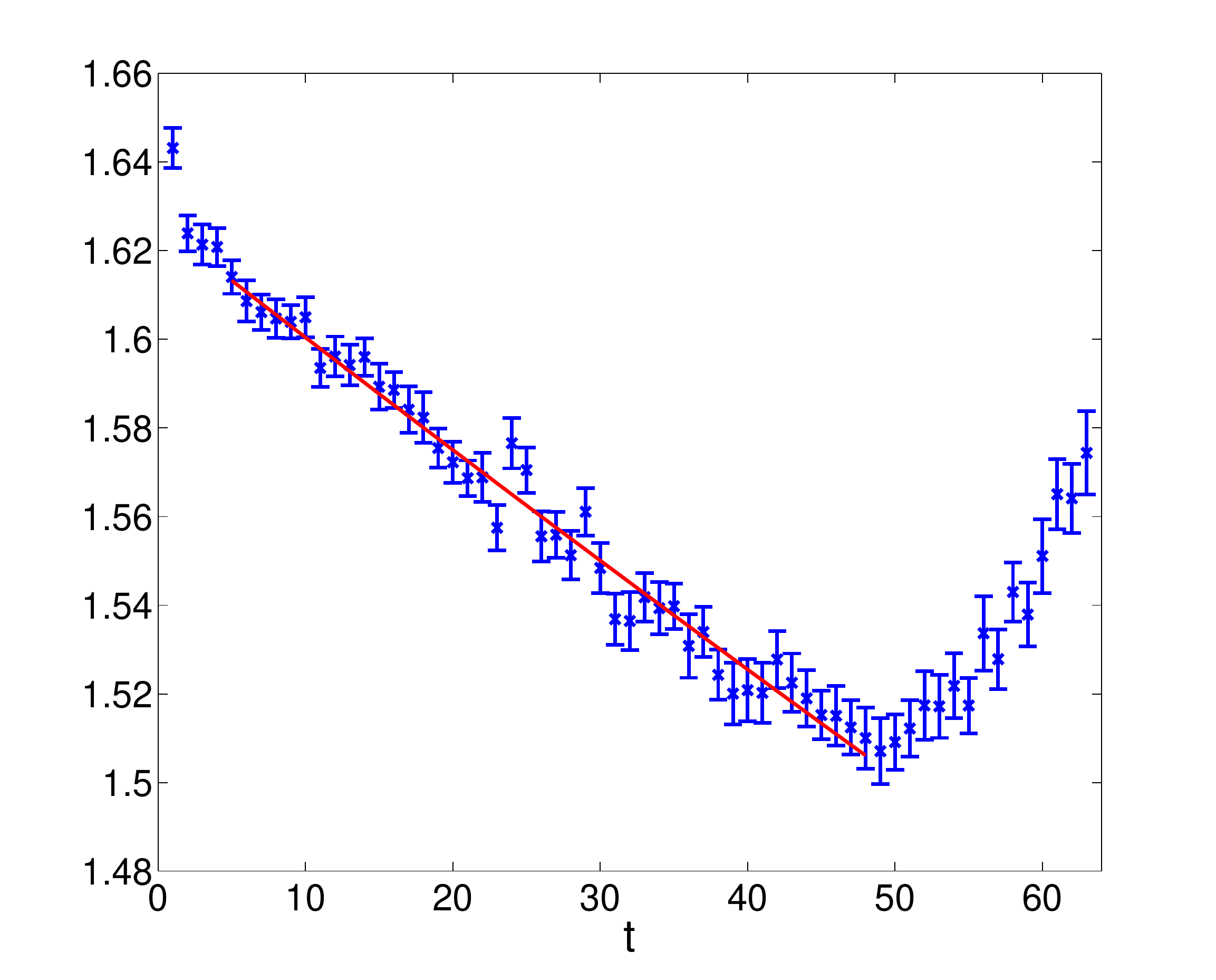}}\qquad
\subfigure[\label{sfig:Epipi_p2} $C_{\pi\pi}(t)/C_{\pi}^2(t), p=\sqrt{2}\pi/L$]{\includegraphics*[width=0.45\textwidth]{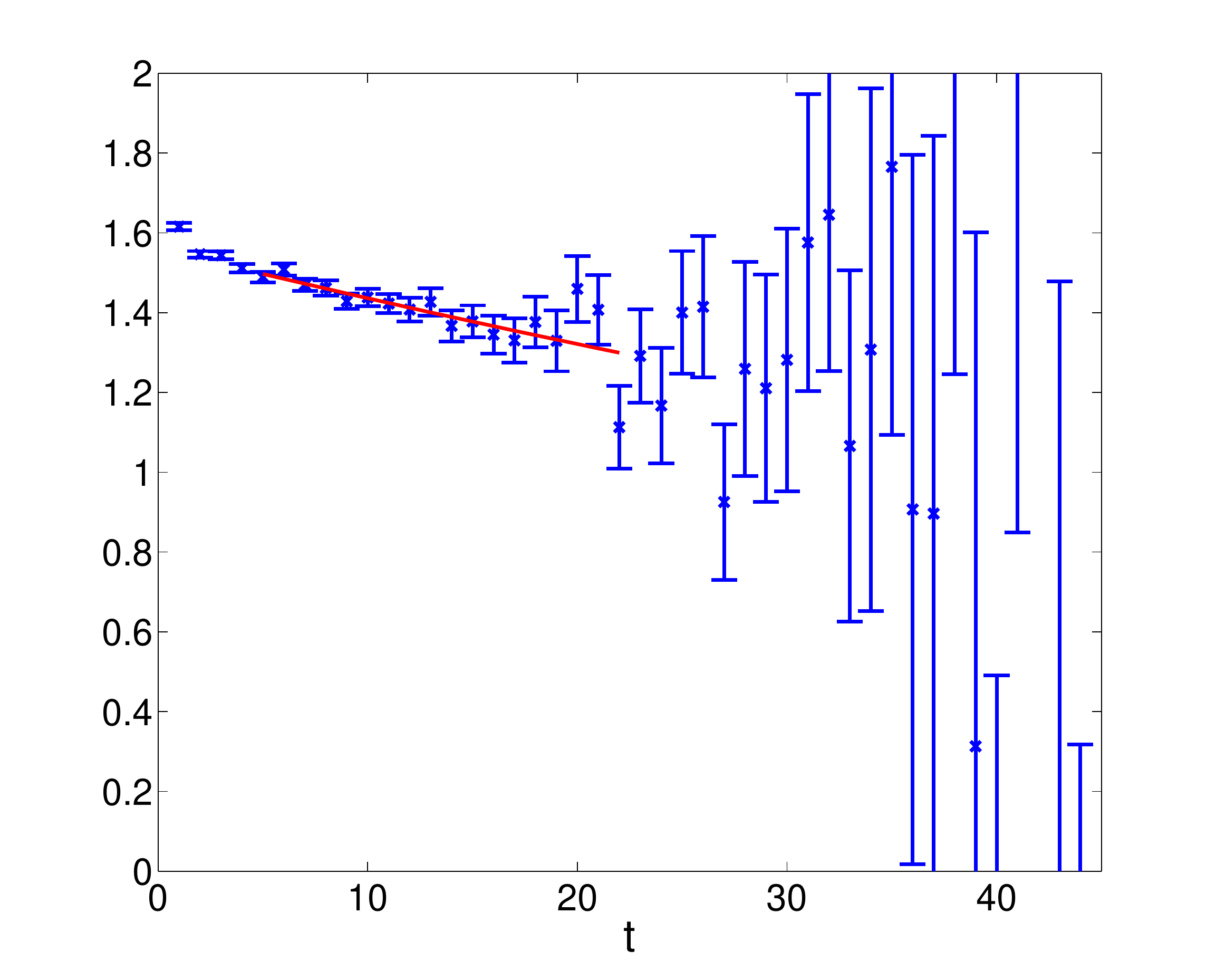}}
\caption{\label{fig:quot_E2pion} The ratios $C_{\pi\pi}(t)/(C_\pi(t))^2$ defined in Eq.\,(\ref{eq:pion_quot}) at $p=0$ (left-hand plot) and at $p=\sqrt{2}\pi/L$ (right-hand plot). The minimum seen in the left-hand panel around $t=52$ results
from the different large-time behavior of the numerator and
denominator.  While the denominator decreases exponentially
as $t$ increases from 0 to 64, the numerator contains a small
$t$-independent constant (caused by one backward propagating
pion) which lessens its decrease at large time.  If examined
for $0 \le t \le 128$ the ratio shown in the left-hand panel
is symmetrical about the point $t=64$.}
\end{figure}

We find it effective in reducing the statistical errors to calculate the quotient of two-pion and single-pion correlators and fit the ratio to the form
\begin{equation}
\label{eq:pion_quot}
 \frac{C_{\pi\pi}(t)}{(C_{\pi}(t))^2} \simeq R^2 e^{- \Delta E\, t}\,,
\end{equation}
where $ \Delta E = (E_{\pi\pi} - 2E_{\pi})$ and $R^2=\frac{|Z_{\pi\pi,\,\mathrm{e}}|^2}{2^{n_{\mathrm{tw}}}\,|Z_{\pi}|^4}$. The energy difference $ \Delta E$ is not equal to zero because of the repulsive interaction between the two pions with isospin 2 in a finite volume. 
The two-pion energy $E_{\pi\pi}$ is then given by $E_{\pi\pi} = \Delta E + 2 E_{\pi}$, and
$Z_{\pi\pi,\,\mathrm{e}}$ is found from
\begin{equation}
 Z_{\pi\pi,\,\mathrm{e}} = (2^{\frac{n_\mathrm{tw}}{2}})\,Z^2_{\pi}\, R\,.
\end{equation}
We can use Eq.\,(\ref{eq:pion_quot}) for values of $t$ which are sufficiently large to neglect excited states and sufficiently smaller than $T/2$ so that the backward propagating states (and the around-the-world effects) can also be neglected. In practice, in order to improve the statistical precision, we \emph{fold} the correlation functions, averaging the equivalent results at $t$ and $T-t$. We calculate the ratio in Eq.\,(\ref{eq:pion_quot}) for $p = 0$, in which case
$Z_\pi$ and $E_\pi$ are just the normalization factor and pion
mass found from the fit to Eq.\,(\ref{eq:mesoncorr}) and for $p = \sqrt{2}\pi/L$
in which case $Z_\pi$ and $E_\pi$ are taken from the fit to Eq.\,(\ref{eq:pion_p2}). The fit regions for the quotients are $t=[5,48]$ for $p=0$ and $[5,22]$ for $p= \sqrt{2}\pi/L$. Plots of the quotients at the two values of $p$ are shown in Figure \ref{fig:quot_E2pion}. The results for all the meson masses and energies are presented in Tab.\,\ref{tab:masses}. We also present the results for  $m_K - E_{\pi\pi}$ to demonstrate that our kinematics are close to being energy conserving.

The momentum $k_{\pi}$ of each pion in the two-pion state is defined from the two-pion energy using the
dispersion relation $E_{\pi\pi} = 2 \sqrt{m^2_{\pi} + k^2_{\pi}}$. The interactions between the two pions lead to $k_\pi$ being different from $0$ or $\sqrt{2}\pi/L$.

Next we  come to the evaluation of the $K\to\pi\pi$ matrix elements. In the calculation as described below, we place the two-pion source at time $t_{\pi\pi}=0$ (or equivalently at $64$) and vary the position of the kaon source $t_K$. The operators of the weak Hamiltonian are inserted between $t_{\pi\pi}$ and $t_K$. 
The symmetries of lattice QCD (including translation invariance and time-reversal) allow us to translate the results into $K\to\pi\pi$ matrix elements. 

\begin{figure}[t]
\centering
\subfigure[$(27,1)$ operator]{\includegraphics*[width=0.31\textwidth]{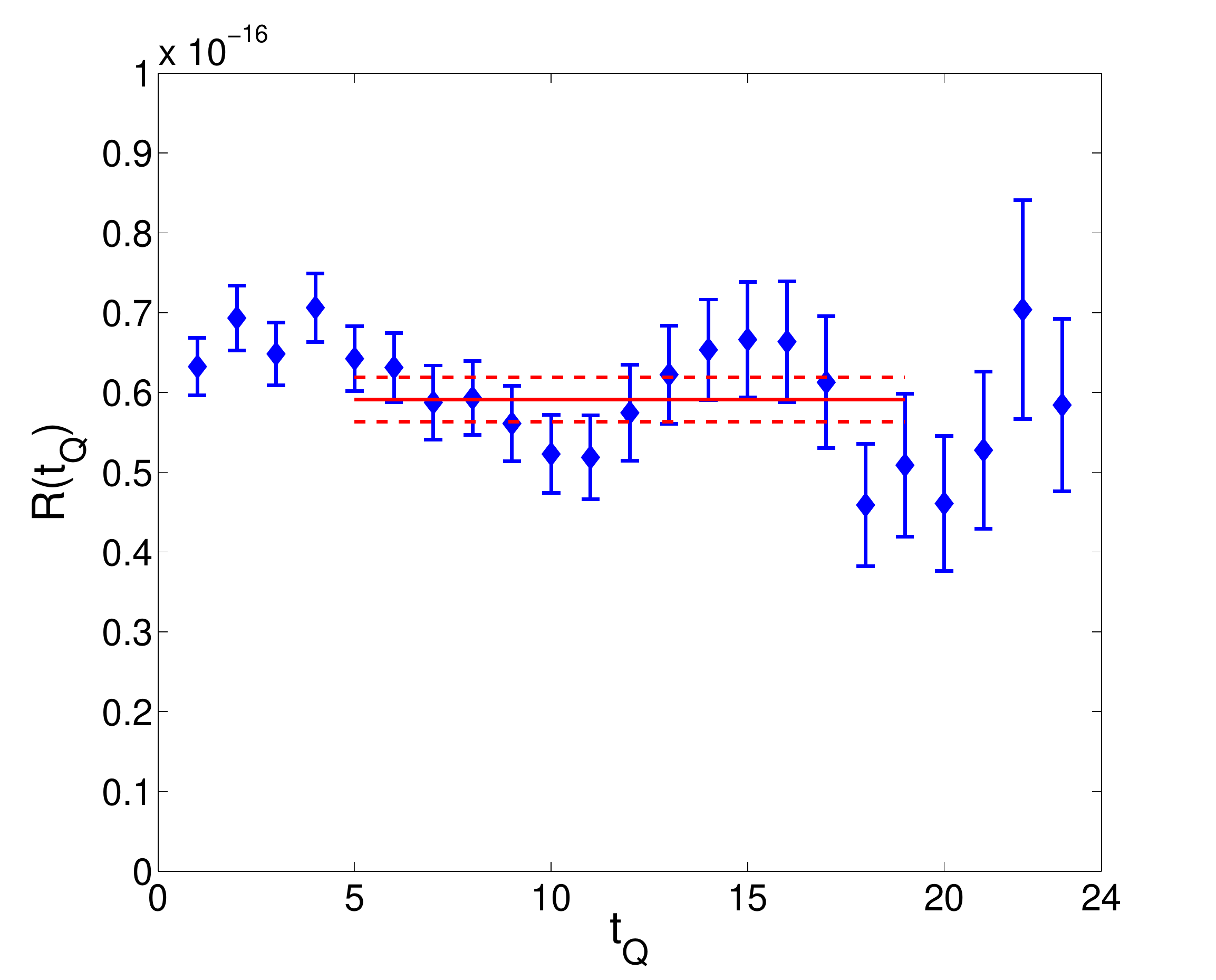}}
\subfigure[$(8,8)$ operator]{\includegraphics*[width=0.31\textwidth]{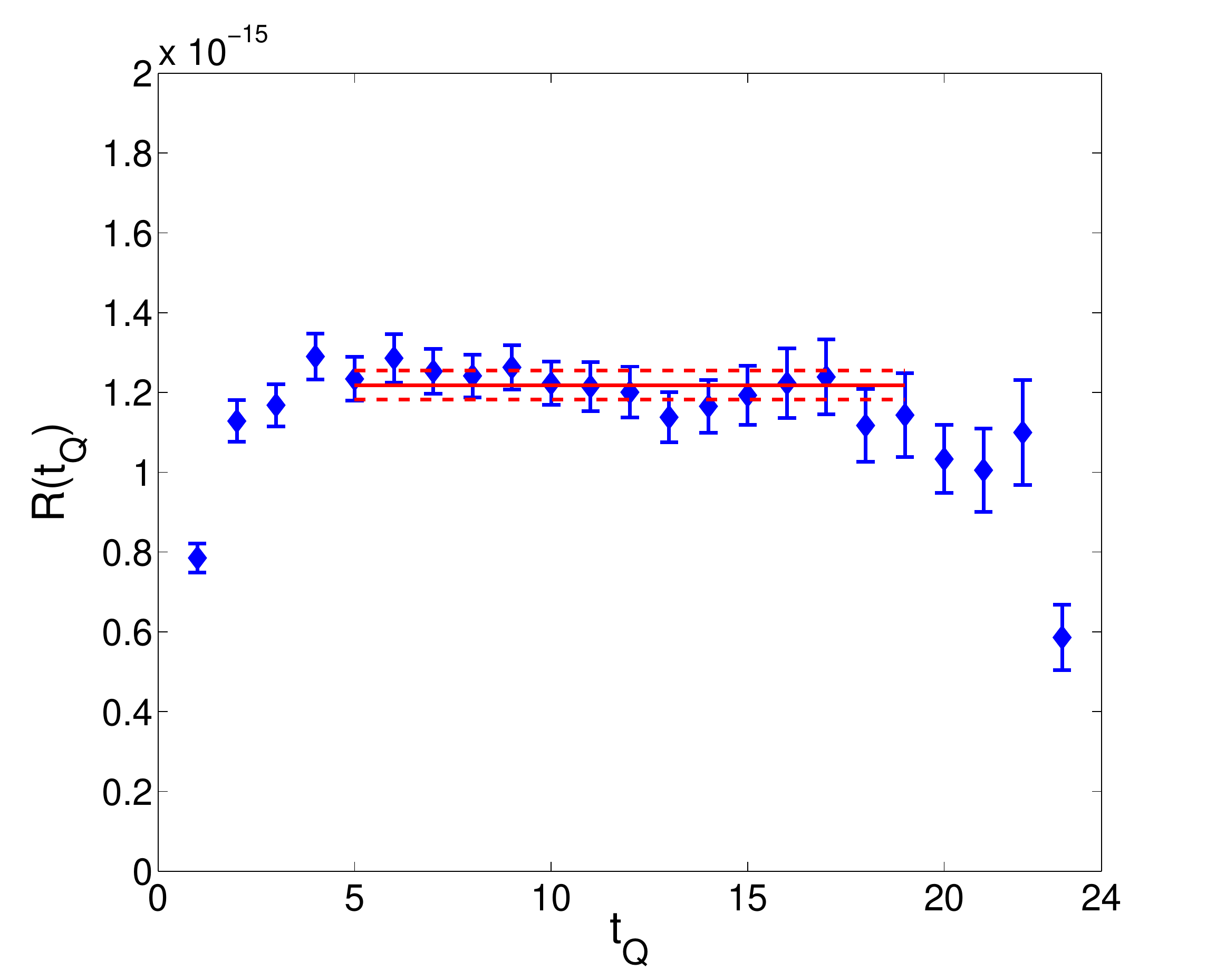}}
\subfigure[$(8,8){\text{mix}}$ operator]{\includegraphics*[width=0.31\textwidth]{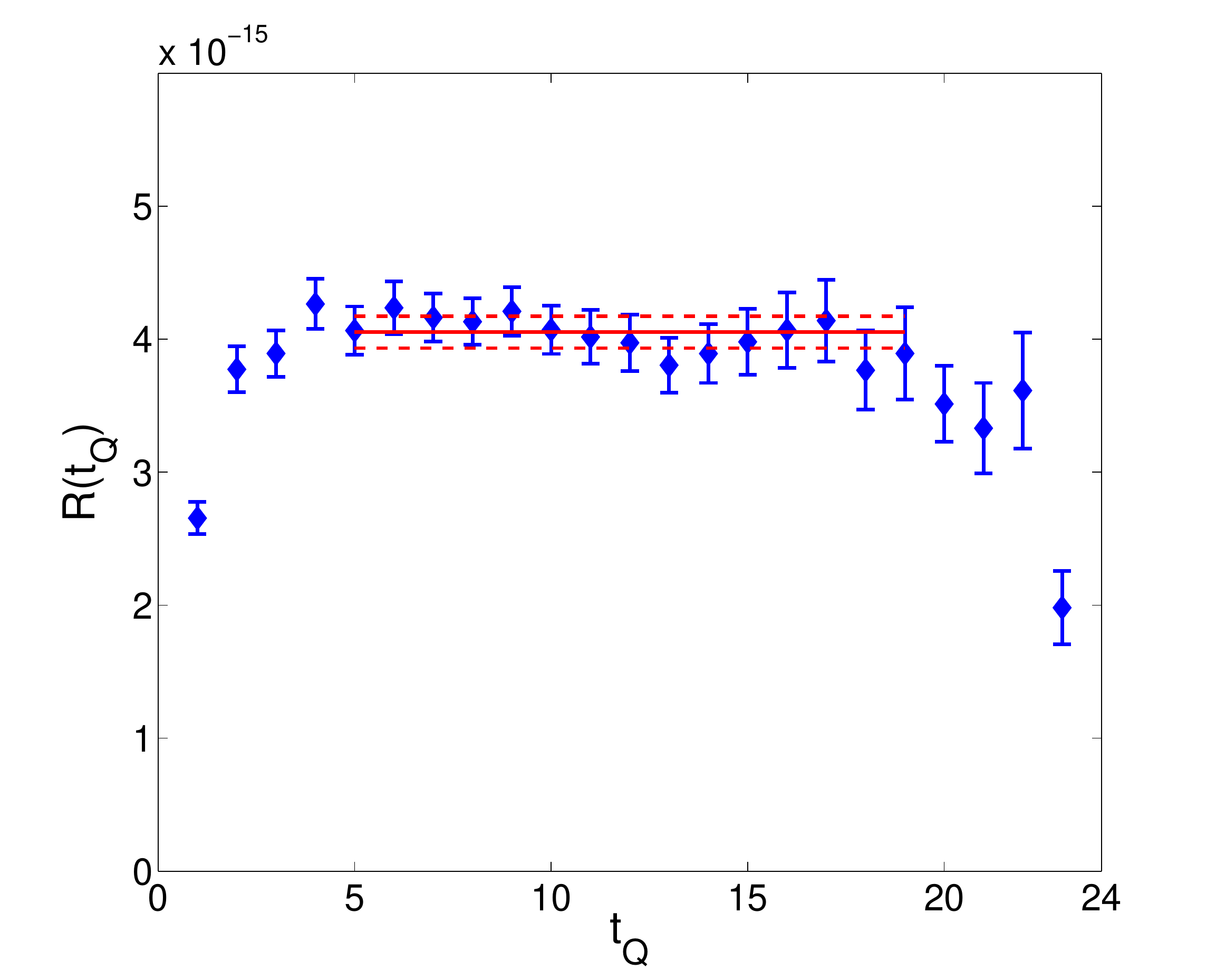}}
\caption{\label{fig:q_plot} The ratios defined in Eq.\,(\ref{eq:quot1}) for $p= \sqrt{2}\pi/L$.
The two-pion source is at $t=0$ while the kaon source is at $t_K=24$. The dashed line shows the error on the fit}
\end{figure}

For each of the three operators $Q_i$ in Eq.\,(\ref{eq:Qdef}), where $i$ labels the operator, the corresponding
$K \rightarrow \pi\pi$ matrix element $\mathcal{M}_i\equiv\langle\pi^+\pi^+\,|\,Q_i\,|K^+\rangle$ is extracted by calculating the ratios
\begin{equation}
 \frac{C^i_{K\pi\pi}(t)}{C_K(t_K-t)C_{\pi\pi}(t)} = \frac{\mathcal{M}_i}{Z_K Z_{\pi\pi,\,\mathrm{e}}}
\label{eq:quot1}
\end{equation}
and fitting to a constant in time $t$. The quantity $C^i_{K\pi\pi}$ is the $K \rightarrow \pi\pi$ correlator with the operator $Q_i$ inserted at $t$ and the kaon and two-pion interpolating operators placed at fixed times $t_K$ and $0$ respectively. 
$Z_K$ and $Z_{\pi\pi,\,\mathrm{e}}$ are determined from the kaon and two-pion correlation functions using eqs.\,(\ref{eq:mesoncorr})  and (\ref{eq:pipi_corr}). For illustration, the left-hand side of equation \eqref{eq:quot1} is plotted in Fig.\,\ref{fig:q_plot} for each of the three operators for the choice $t_K=24$. The figure demonstrates that sufficiently far from
the kaon and two-pion sources the data is indeed consistent with the expected constant behavior. 
We determine the matrix elements by fitting the data between $t=5$ and $t=t_K-5$, where $t$ denotes the time distance from the two-pion
source. The results for $\mathcal{M}_i/(Z_K Z_{\pi\pi,\,\mathrm{e}})$ obtained from the fits are indicated on the plot together with their errors. 

The finite-volume matrix elements computed in the lattice simulations ${\cal M}_i$ are related to the corresponding infinite-volume ones ${\cal A}_i$ by the Lellouch-L\"uscher factor~\cite{Lellouch:2000pv,Lin:2001ek}:
\begin{equation}
{\cal A}_i = \left[ \frac{\sqrt{2^{n_\mathrm{tw}}}}{2\pi q_{\pi}} \sqrt{\frac{\partial \phi}{\partial q_{\pi}} + \frac{\partial \delta}{\partial q_{\pi}} } \,\right]
\frac{2}{\sqrt{2^{n_\mathrm{tw}}}}\,L^{3/2}\sqrt{m_K} E_{\pi\pi} \mathcal{M}_i\,,
\label{eq:amp}
\end{equation}
where the quantity in square brackets (denoted by LL in Tab.\,\ref{tab:derivs})
contains the effects of the Lellouch-L\"uscher factor beyond the free-field normalization. $\delta$ is the $s$-wave phase shift, 
$q_{\pi}$ is a dimensionless quantity related to the pion momentum $k_{\pi}$ by
$q_{\pi} = k_{\pi}L/2\pi$ and $\phi$ is a kinematic function defined in \cite{Lellouch:2000pv}. Once $E_{\pi\pi}$ has been measured and $q_{\pi}$ determined, $\delta$ can be calculated using the
L\"uscher quantisation condition \cite{Luscher:1990ux}:
\begin{equation}
 n\pi = \delta(k_{\pi}) + \phi(q_{\pi}).
\end{equation}
Results for $E_{\pi\pi}$, $k_{\pi}$, $q_{\pi}$ and $\delta$ are presented in Tab.\,\ref{tab:p_delta}.

\begin{table}[t]
\centering
\begin{tabular}{|c|c|c|c|c|}
\hline
$p$ & $E_{\pi\pi}$ (MeV) & $k_{\pi}$ (MeV)& $q_{\pi}$ & $\delta$ (degrees) \\
\hline
0 & 286.4(1.9)& 17.63(36) & 0.0659(13) & -0.311(18)\\
$\sqrt{2}\pi/L$ & 485.5(4.2)& 196.8(2.2) & 0.7350(72) & -7.96(2.07)\\
\hline
\end{tabular}
\caption{The two-pion energy $E_{\pi\pi}$, $k_\pi$, $q_\pi$ and $s$-wave phase shift \label{tab:p_delta}}
\end{table}

\begin{figure}[t]
 \begin{center}
\includegraphics*[width=0.4\textwidth,angle=0]{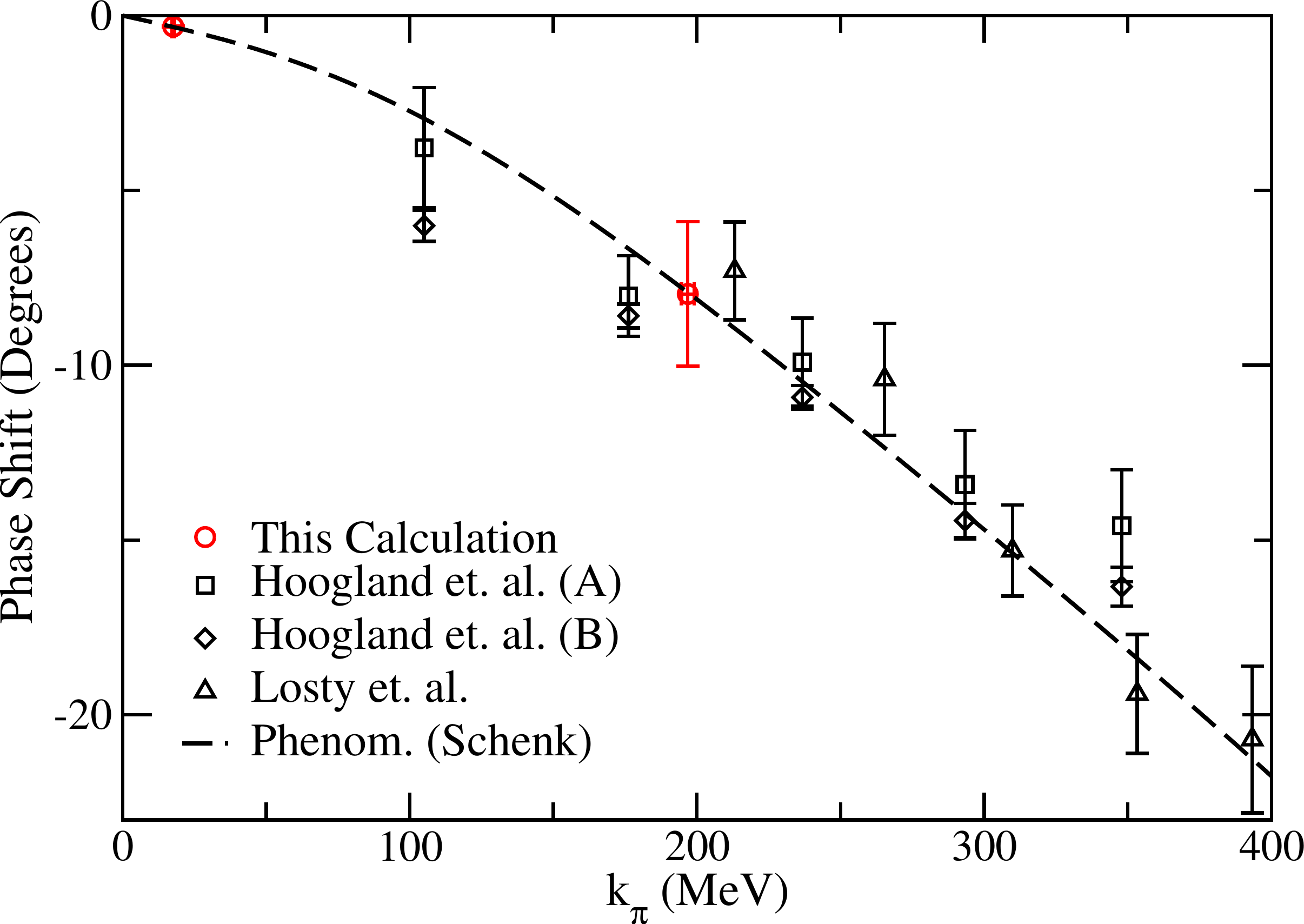} 
\qquad\qquad
\includegraphics*
[width=0.385\textwidth,angle=0]{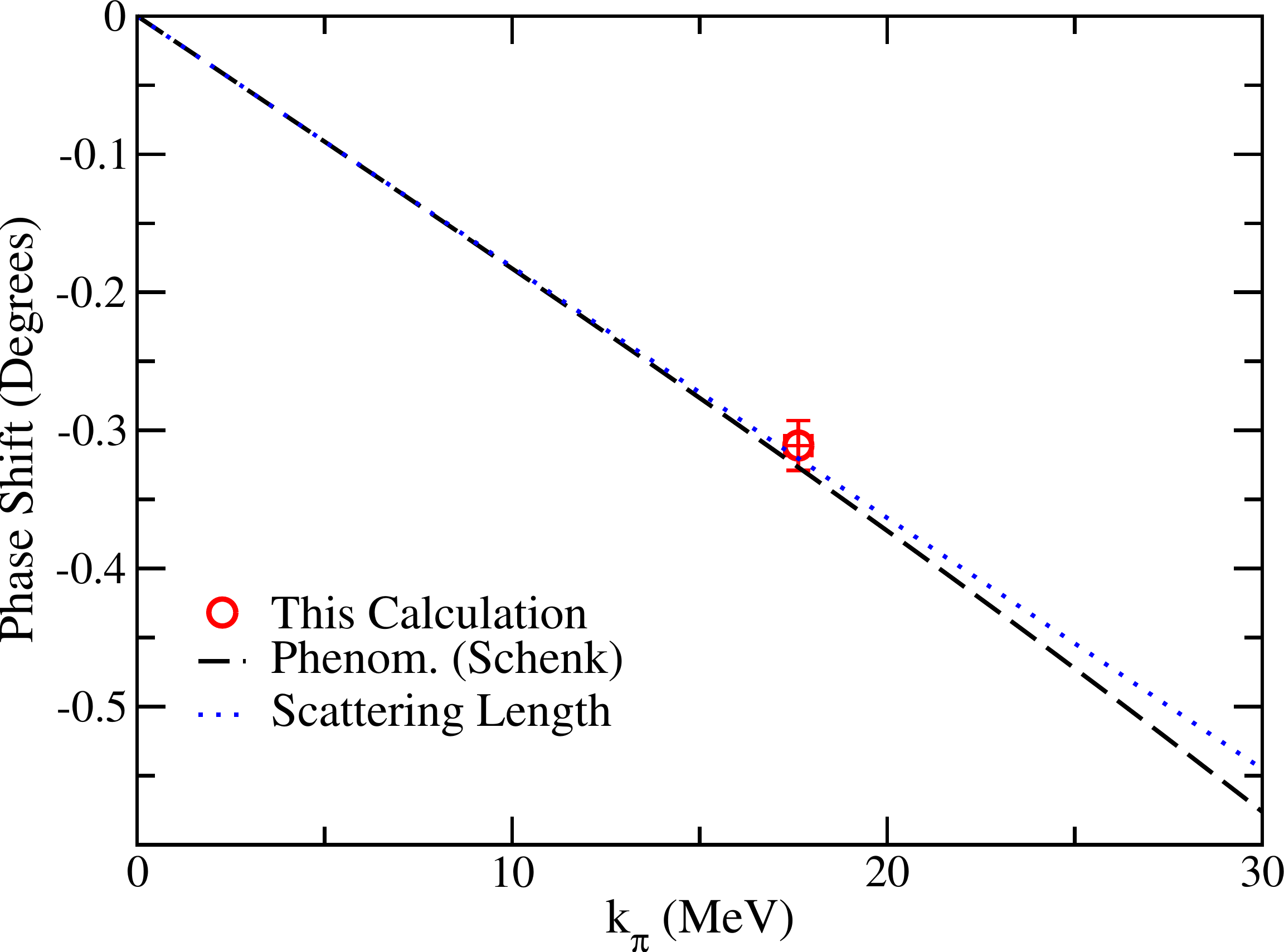}
\caption{\label{fig:phase_shift}Plots of the $I=2$ two-pion s-wave phase shift against momentum $k_{\pi}$. Our results at $p=0$ and $p=\sqrt{2}\pi/L$ are denoted by the red circles and the dashed curve is the phenomenological representation from ref.\,\cite{schenk_curve}. The left-hand plot is a comparison of the calculated phase shift with experimental results~\cite{Hoogland:1977kt, Losty:1973et, schenk_curve}. The right-hand plot 
is a zoom into the small $k_\pi$ region, demonstrating the approximate linear behavior of the phenomenological curve in the region of $p=0$. The scattering length used in the straight (dotted) line is calculated using chiral perturbation theory~\cite{Colangelo:2001df}.}
\end{center}\end{figure}

Since $\partial \phi/\partial q_{\pi}$ can be calculated
analytically the only unknown in equation \eqref{eq:amp} is $\partial \delta/\partial q_{\pi}$.
The results for the phase shift are plotted against $k_{\pi}$ and compared with experimental results  \cite{Hoogland:1977kt, Losty:1973et} in the left-hand plot of Fig.\,\ref{fig:phase_shift}; we see good agreement. Near $p=0$ we assume that $\delta$ is linear in $k_{\pi}$ in order to
calculate $\partial \delta/ \partial q_{\pi}$ (see the right-hand plot of Fig.\,\ref{fig:phase_shift}). For $p=\sqrt{2}\pi/L$ we use the phenomenological curve \cite{schenk_curve} shown in Fig.\,\ref{fig:phase_shift} to calculate the derivative of the phase shift at the corresponding value of $q_{\pi}$. The derivative of the phase shift is
found to be a small term in comparison with $\partial \phi/ \partial q_{\pi}$.
Results for $\partial \phi/\partial q_{\pi}$
and $\partial \delta / \partial q_{\pi}$ are presented in Tab.\,\ref{tab:derivs}.

\begin{table}[t]
\centering
\begin{tabular}{|c|c|c|c|}
\hline
$p$ & $\partial\phi/\partial q_{\pi}$ & $ \partial \delta / \partial q_{\pi} $& LL \\
\hline
0 & 0.2413(90)& -0.0824(32) &0.9632(14) \\
$\sqrt{2}\pi/L$ & 5.014(21)& -0.2911(23)& 0.9411(71) \\
\hline
\end{tabular}
\caption{Contributions to Lellouch-L\"uscher factor.
The second and third columns provide numerical values for
two of the quantities entering the Lellouch-L\"uscher factor
given within the square brackets in Eq.\,(\ref{eq:amp}), while the fourth
column gives the value of the complete factor.
 \label{tab:derivs}}
\end{table}

We perform the analysis for four separations $\delta t$ between the kaon and two-pion sources, $\delta t =20, 24, 28$ and 32. The physical decay amplitude $A_2$ is given in terms of the matrix elements ${\cal A}_i$ by
\begin{equation}\label{eq:A2deltat}
A_2^{\delta t} = a^{-3} \frac{\sqrt{3}}{2} \frac{G_F}{\sqrt{2}} V_{ud} V^*_{us} \sum_{i,j} C_i(\mu) Z_{ij}(\mu a) {\cal A}_j^{\delta t}\,,
\end{equation}
where we have added the label $\delta t$ to indicate the separation being used and the labels $i$ and $j$ run 
over the three operators in Eq.\,(\ref{eq:Qdef}). $C_i$ are the Wilson coefficients, which are generally calculated in schemes based on dimensional regularization;
 we take them to be in the $\overline{\mathrm{MS}}$-NDR scheme. The $Z_{ij}$ are the renormalization constants which relate the bare weak operators defined in the lattice theory (where the lattice spacing $a$ acts
as a cut-off) to those in the $\overline{\mathrm{MS}}$-NDR scheme at scale $\mu$. 
The $(27,1)$ operator renormalizes multiplicatively, whereas the $(8,8)$ and $(8,8)_{\mathrm{mix}}$ operators mix under renormalization. 
The calculation of the $Z_{ij}$ is described in detail in Sec.\,\ref{sec:npr} and involves a non-perturbative calculation of the renormalization constants in RI-SMOM 
schemes, step-scaling to run the results to  $\mu=3\,$GeV and matching perturbatively to the $\overline{\mathrm{MS}}$-NDR scheme at 3\,GeV. As explained in Sec.\,\ref{sec:npr}, four possible choices for the intermediate RI-SMOM schemes are considered. The results presented in Tab.\,\ref{tab:A2} are calculated using the renormalization constants with the intermediate scheme $(\mathrm{I_v},\mathrm{I_q})=(\slashed{q},\slashed{q})$ (see Sec.\,\ref{sec:npr}).
The factor of $\sqrt{3}/2$ on the right-hand side of Eq.\,(\ref{eq:A2deltat}) is needed to convert from the unphysical $K^+ \rightarrow \pi^+ \pi^+$ amplitudes back to the physical $K^+ \rightarrow \pi^+ \pi^0$ amplitudes.
 
Results for Re\,$A_2$ and Im\,$A_2$ for the four different separations $\delta t$ are shown in Tab.\,\ref{tab:A2} for the (almost) physical choice $p=\sqrt{2}\pi/L$.  
Our final result for $A_2$ is an error weighted average (EWA) over the four separations, defined by
\begin{equation}
  \label{eq:ewa}
A_2^{\text{EWA}} = \frac{\sum_{\delta t} A_2^{\delta t} / (e_{\delta t})^2}
{\sum_{\delta t} 1/(e_{\delta t})^2}\,,
\end{equation}
where $e_{\delta t}$ is the statistical error in the evaluation of $A_2^{\delta t}$.

\begin{table}[h]
\centering
\begin{tabular}{|c|c|c|}
\hline
$\delta t$& Re\,$A_2$(units of $10^{-8}$ GeV) & Im\,$A_2$(units of $10^{-13}$ GeV) \\
\hline
20 &1.411(56)& -6.59(19)\\
24 &1.346(64)& -6.67(22)\\
28 &1.427(73)& -6.28(25)\\
32 &1.295(94)& -6.56(33)\\
\hline
EWA(a) &1.381(38)& -6.54(15)\\
\hline
EWA(b) & 1.381(44)(12)&  -6.54(19)(42)\\
\hline
\end{tabular}
\caption{\label{tab:A2}Final results for $A_2$. The errors on each $A_2^{\delta t}$, on EWA(a) and the first error in EWA(b)
(EWA = error weighted average) are the statistical errors only. In the EWA(b) result the second error is that due from the uncertainty in the
evaluation of the renormalization constants as explained in Sec.\,\ref{sec:npr} below. }
\end{table}

The errors in the results labelled by EWA(a) in Tab.\,\ref{tab:A2} are due to the statistical fluctuations on the ${\cal A}_i$ calculated using Eq.\,(\ref{eq:amp}). In the row marked EWA(b) the first error combines the uncertainty due to these fluctuations with the statistical uncertainty in the value of the lattice spacing and the 
second error is $\Delta_Z$, which arises from the statistical uncertainty in the evaluation of the renormalization constants $Z_{ij}$. This is calculated using:
\begin{equation}\label{eq:DeltaZdef}
\Delta_Z^2=\left[C_{(27,1)}\,\delta Z_{(27,1)}\,{\cal A}_{(27,1)}\right]^2 + \sum_{i,j}\Big[\, C_i\,\delta Z_{ij}\,{\cal A}_j\,\Big]^2\,,
\end{equation}
where $i,j$ run over $(8,8)$ and $(8,8)_\mathrm{mix}$ and the $\delta Z$ are the statistical uncertainties in the corresponding renormalization constants as explained in Sec.\,\ref{sec:npr}. The presence of the four terms in the sum over $i$ and $j$ reflects the mixing of $Q_{(8,8)}$ and $Q_{(8,8)_{\mathrm{mix}}}$ under renormalization. ${\cal A}_{(27,1)}$, ${\cal A}_{(8,8)}$ and ${\cal A}_{(8,8)_\mathrm{mix}}$ on the right-hand side of  Eq.(\ref{eq:DeltaZdef}) are obtained from the corresponding bare matrix elements using Eq.\,(\ref{eq:amp}). The numerical results presented here were obtained by using the statistical errors $e_{\delta t}$ in the evaluation of $A_2$ so that for example:
\begin{equation}
{\cal A}_{(27,1)}=\frac{\sum_{\delta t} {\cal A}_{(27,1)}^{\delta t} / (e_{\delta t})^2}
{\sum_{\delta t} 1/(e_{\delta t})^2}\,,
\end{equation}and similarly for the remaining operators. We have checked that performing the error weighted average on each operator using the statistical error corresponding to the operator makes only a negligible difference to the estimate of the final errors. 

For the Wilson coefficients we use the standard notation $C_i = z_i(\mu) + \tau y_i(\mu)$ where, as explained above, $\tau = -V^\ast_{ts}V_{td}/V^\ast_{us}V_{ud}$. The Wilson coefficients are calculated using the equations in \cite{Buchalla:1995vs}, which uses a 10-operator basis for the effective Hamiltonian. The equations in \cite{Buchalla:1995vs} are based on the pioneering Next-to-Leading Order QCD and QED calculations from the Munich and Rome groups~\cite{Buras:1993dy,Ciuchini:1992tj,Ciuchini:1995cd}.
The Wilson coefficients in the 10-operator basis are related to 
the three $\Delta I = 3/2$ Wilson coefficients by
\begin{equation}
C_{(27,1)}(\mu) = \frac{C_1(\mu) + C_2(\mu)}{3} + \frac{C_9(\mu) + C_{10}(\mu)}{2},~~
 C_{(8,8)}(\mu) = \frac{C_7(\mu)}{2}\,,~
C_{(8,8)_\text{mix}}(\mu) = \frac{C_8(\mu)}{2}\,.
 \end{equation}
Results for $z_i$ and $y_i$ at $ \mu= 3$\,GeV in the $\overline{\text{MS}}$-$\text{NDR}$ scheme
are presented in Tab.\,\ref{tab:wilson}. We observe that the Wilson coefficients are sensitive to the
value of $\alpha_s$. This calculation is based on $\alpha^{(3)}_s(3\, \text{GeV}) = 0.24544$ which is found
by solving the 4-loop running formula for $\alpha_s$ \cite{vanRitbergen:1997va} with initial condition
$\alpha_s^{(5)}(M_Z)=0.1184$ for $M_Z = 91.1876$\,MeV\,\cite{Nakamura:2010zzi}. The superscript $(n)$ indicates
the number of flavors.

\begin{table}[t]
\centering
\begin{tabular}{|c|c|c|}
\hline
   weak operator   & $z_i$ & $y_i$\\
\hline
$Q_1$ &  -0.241415   &0 \\
$Q_2$ &  1.11228     &0\\
$Q_3$ &  -0.00392423 &0.0211096 \\
$Q_4$ &   0.0169695    &-0.0558734 \\
$Q_5$ &  -0.00349963 & 0.0117843 \\
$Q_6$ &   0.0120747  &-0.0610235 \\
$Q_7$ &  0.0000940198&-0.000161911 \\
$Q_8$ &  -0.000104478 &0.000652032 \\
$Q_9$ &  0.0000275290&-0.0103828 \\
$Q_{10}$&  0.0000798557&0.00243775 \\
\hline
$Q_{(27,1)}$ & 0.290342 & -0.00397252\\
$Q_{(8,8)}$ & 4.70099 $\times 10^{-5}$ & -8.09555$\times 10^{-5}$\\
$Q_{(8,8)_\text{mix}}$ & -5.22390$\times 10^{-5}$ & 3.26016 $\times 10^{-4}$ \\
\hline
\end{tabular}
\caption{\label{tab:wilson} Wilson coefficients at 3\,GeV in the $\overline{\text{MS}}$-NDR scheme.}
\end{table}

Using the procedures described above, we obtain our final results for the complex amplitude $A_2$:
\begin{equation}\label{eq:results}
\textrm{Re}\,A_2=1.381(46)_{\textrm{stat}}(258)_{\textrm{syst}}\,10^{-8}\,\textrm{GeV},
\quad\textrm{Im}\,A_2=-6.54(46)_{\textrm{stat}}(120)\,_{\textrm{syst}}10^{-13}\,{\rm GeV}\,.
\end{equation}
The result for Re\,$A_2$ agrees well with the experimental value of $1.479(4)\times10^{-8}$\,GeV obtained from $K^+$ decays and 
$1.573(57)\times10^{-8 }$\,GeV obtained from $K_S$ decays
(the difference
arises from the unequal $u$ and $d$ quark masses and from
electromagnetism, two small effects not included
in our calculation). Im\,$A_2$ is unknown so that the result in Eq.\,(\ref{eq:results}) provides its first direct determination (updating the value quoted in \cite{Blum:2011ng}). 

A detailed discussion of the determination of the systematic errors will be presented in the following sections. As explained in section \ref{subsec:parameters}, the statistical error was obtained by analysing configurations each separated by 8
molecular dynamics time units.  With the aim of reducing the correlations between successive measurements,
the gauge fields were shifted by 16 lattice spacings in the time direction relative to the previous 
configuration prior to measuring the quark propagators. In order to check that shifting the gauge fields is sufficient to overcome potential autocorrelations, we have repeated the entire analysis, including the determination of the physical quark masses and lattice spacings, by binning all quantities over four successive measurements (32 molecular dynamics time units). This is a natural choice as it
matches the periodicity of the quark propagator measurements. The effects of the binning are completely negligible. For illustration we show in Tab.\,\ref{tab:binA2} a comparison of the results for $A_2$ obtained with and without the binning.

\begin{table}
\centering
\begin{tabular}{|c|c|c||c|c|}
\hline
&\multicolumn{2}{|c||}{Re\,$A_2$}&\multicolumn{2}{c|}{Im\,$A_2$}\\ 
\hline
$\delta t$& 146 bins & 36 bins & 146 bins & 36 bins \\
\hline
20 &1.411(56)& 1.418(52)& -6.59(19) & -6.55(16)\\
24 &1.345(64)& 1.344(57)&-6.67(22) & -6.60(20)\\
28 &1.427(73)& 1.411(83)&-6.28(25) & -6.23(29)\\
32 &1.295(94)& 1.28(10) & -6.56(33) & -6.58(31)\\
\hline
EWA(a) &1.381(38)& 1.386(34)&-6.54(15) & -6.52(14)\\
\hline
\end{tabular}
\caption{\label{tab:binA2} Final results for Re\,$A_2$ in units of $10^{-8}$~GeV and Im\,$A_2$ in units of $10^{-13}$~GeV. The table shows a comparison between the results obtained as in Tab.\,\ref{tab:A2} (146 bins each with a single configurations) and those with bin-size 4 (36 bins each with 4 configurations). The errors are statistical ones only.} 
\end{table}

In the remainder of the section we present the results for each of the three matrix elements which contribute to $A_2$ (Sec.\,\ref{subsec:me}) and also deduce the value of the unknown quantity Im\,$A_0$ by combining our result for Im\,$A_2$ with the experimental values of $\epsilon^\prime/\epsilon$ and other quantities (Sec.\,\ref{subsec:a0}). In order to explain fully our conventions, we also present the explicit expressions for $A_0$, $A_2$ and the partial widths for the $K\to\pi\pi$ decays in terms of the matrix elements. 

\subsection{Results for the matrix elements}\label{subsec:me}
Eq.\,(\ref{eq:results}) contains our final results for $A_2$ within the Standard Model. In order to facilitate detailed comparisons with results from future computations and to enable our results to be used in extensions of the Standard Model for which the Wilson coefficient functions are different, we now present the results for the matrix elements themselves. The results are presented for operators renormalized in the $\msbar$-NDR scheme at a renormalization scale of 3\,GeV. Our convention is that 
$\sqrt{2}\,A_2 = \left \langle (\pi\pi)^{I=2}_{I_z =0} \vert H_W \vert K^0\right \rangle$. With this 
definition $|A_2| = \sqrt{\frac{2}{3}} |A_{+0}|$, where $A_{+0} =\langle \pi^+ \pi^0\vert H_W \vert K^+ \rangle$ and the corresponding partial width is given by
\begin{equation}\label{eq:kpippi0width}
\Gamma(K^+\to\pi^+\pi^0)=\frac{1}{8\pi}\,\left|A_{+0}\right|^2\,\frac{p_{+0}}{m_{K^+}^2}\,,
\end{equation}
where 
\begin{equation}
p_{+0}=\sqrt{\frac{m_{K^+}^2}{4}-\frac{m_{\pi^+}^2+m_{\pi^0}^2}{2}+\frac{(m_{\pi^0}^2-m_{\pi^+}^2)^2}{4m_{K^+}^2}
}\,.
\end{equation}

\subsubsection{$K^+ \to \pi^+ \pi^+$ matrix elements}

We start with the results for the $K^+\to\pi^+\pi^+$ matrix elements of the operators defined in Eq.\,(\ref{eq:Qdef}) in terms of which $A_2$ is given by
\begin{equation}
 A_2 = \frac{G_F}{\sqrt{2}} V_{ud} V^*_{us} \frac{\sqrt{3}}{2} \sum_i C_i(3\text{ GeV}) \mathcal{A}^{\msbar\text{-NDR}}_{i}(3\text{ GeV})  
\label{eq:A2_pp}\,,
\end{equation}
where $\mathcal{A}^{\msbar\text{-NDR}}_{i}=\langle\,\pi^+\pi^+\,|\,Q_i\,|K^+\rangle$ and the label $i$ runs over (27,1), (8,8) and (8,8)$_\mathrm{mix}$\,. The $\mathcal{A}_i$ take the values
\begin{subequations}
 \begin{eqnarray}
   \mathcal{A}^{\msbar\text{-NDR}}_{(27,1)}(3\text{ GeV}) & = & 0.03071(97)~\rm{GeV}^3 \\
   \mathcal{A}^{\msbar\text{-NDR}}_{(8,8)}(3\text{ GeV}) & = & 0.583(33)~\rm{GeV}^3 \\  
 \mathcal{A}^{\msbar\text{-NDR}}_{(8,8)_{\rm{mix}}}(3\text{ GeV}) & = & 2.64(15)~\rm{GeV}^3\,.
 \end{eqnarray}
\end{subequations}

\subsubsection{$K^+ \to \pi^+ \pi^0$ matrix elements}
Alternatively we may express $A_2$ in terms of the matrix elements for the physical $K^+ \to \pi^+ \pi^0$ decay. In this case 
\begin{equation}
A_2 = \frac{G_F}{\sqrt{2}} V_{ud} V^*_{us} \frac{1}{\sqrt{3}} \sum_i C_i(3\text{ GeV}) \mathcal{A^{\prime}}^{\msbar\text{-NDR}}_{i}(3\text{ GeV}).  
\label{eq:A2_p0s}
\end{equation}
where the two-pion final state is symmetrised ($\frac{1}{\sqrt{2}}\,(\langle \pi^+(\vec{p})\pi^0(-\vec{p})|+
\langle \pi^+(-\vec{p})\pi^0(\vec{p})|)$.
We find the matrix elements to be 
\begin{subequations}
 \begin{eqnarray}
   \mathcal{A^{\prime}}^{\msbar\text{-NDR}}_{(27,1)}(3\text{ GeV}) & = & 0.0461(14)~\rm{GeV}^3 \\
   \mathcal{A^{\prime}}^{\msbar\text{-NDR}}_{(8,8)}(3\text{ GeV}) & = & 0.874(49)~\rm{GeV}^3 \\  
 \mathcal{A^{\prime}}^{\msbar\text{-NDR}}_{(8,8)_{\rm{mix}}}(3\text{ GeV}) & = & 3.96(23)~\rm{GeV}^3\,.
\end{eqnarray}
\label{eq_Mp0}
\end{subequations}
\subsubsection{Contributions to $A_2$ from the Matrix Elements}
Finally we present the separate contributions to $A_2$ in Eq.\,(\ref{eq:results}) from the matrix elements of the three different operators:
\begin{equation}
 \begin{array}{lcclcc}
\rm{Re}(A_2)_{(27,1)}  & =  &(1.398 \pm 0.044)\,10^{-8}\,\rm{GeV};&  \rm{Im}(A_2)_{(27,1)} &=& (1.55 \pm 0.36)\,10^{-13}\,\rm{GeV} \\  
\rm{Re}(A_2)_{(8,8)}  &= & (4.29 \pm 0.24)\,10^{-11}\,\rm{GeV};&   \rm{Im}(A_2)_{(8,8)} &=& (4.47 \pm 0.25)\,10^{-14}\,\rm{GeV} \\  
 \rm{Re}(A_2)_{(8,8)_{\rm{mix}}} & = & (-2.14 \pm 0.12)\,10^{-10}\,\rm{GeV};&\rm{Im}(A_2)_{(8,8)_{\rm{mix}}}& =& (-8.14 \pm 0.47)\,10^{-13}\,\rm{GeV}\,.
 \end{array}\label{eq:a2components}
\end{equation}

\subsection{Prediction for Im\,$A_0$}\label{subsec:a0}

Before describing our indirect determination of the unknown quantity Im\,$A_0$, we summarise our conventions. $A_0$ is defined by $\sqrt{2}\,A_0 = \left \langle (\pi\pi)^{I=0}_{I_z =0} \vert H_W \vert K^0\right \rangle$. Defining the amplitudes $A_{+-}$ and $A_{00}$ by 
\begin{equation}
A_{+-}= \langle \pi^+ \pi^-\vert H_W \vert K_S \rangle\quad\mathrm{and}\quad A_{00}= \langle \pi^0 \pi^0\vert H_W \vert K_S \rangle\,,
\end{equation}
the corresponding partial widths are given by 
\begin{eqnarray}\label{eq:kpippimwidth}
\Gamma(K_S\to\pi^+\pi^-)&=&\frac{1}{8\pi}\,\left|A_{+-}\right|^2\,\frac{p_{+-}}{m_{K_S}^2}\,,\\ 
\Gamma(K_S\to\pi^0\pi^0)&=&\frac{1}{16\pi}\,\left|A_{00}\right|^2\,\frac{p_{00}}{m_{K_S}^2}\,,
\label{eq:kpi0pi0width}
\end{eqnarray}
where the relative momenta are given by
\begin{equation}
p_{+-}=\frac12\sqrt{m_{K_S}^2-4m_{\pi^+}^2}\quad\mathrm{and}\quad p_{00}=\frac12\sqrt{m_{K_S}^2-4m_{\pi^0}^2}\,.
\end{equation} $A_{+-}$ and $A_{00}$ are given in terms of $A_0$ and $A_2$ by
\begin{eqnarray}
A_{+-}&=&\sqrt{\frac23}\,A_2\,e^{i\delta_2}+\frac2{\sqrt{3}}\,A_0\,e^{i\delta_0}\\ 
A_{00}&=&2\sqrt{\frac23}\,A_2\,e^{i\delta_2}-\frac2{\sqrt{3}}\,A_0\,e^{i\delta_0}\,, 
\end{eqnarray}
where $\delta_I$ is the s-wave $\pi\pi$ phase shift for isospin $I$. With these definitions we now evaluate Im\,$A_0$.

Having obtained $A_2$, and recalling that Re\,$A_0$ is known from experiment, the remaining unknown quantity is Im\,$A_0$. We now determine this by combining our result for $\text{Im}\,A_2/\text{Re}\,A_2$ from Tab.\,\ref{tab:results}, with the experimental values of 
\begin{equation}
\label{eq:cp}
 \text{Re}\left ( \frac{\epsilon'}{\epsilon} \right) = \frac{\omega}{\sqrt{2}\left\vert \epsilon \right \vert} \left[ \frac{\text{Im}\,A_2}{\text{Re}\,A_2} - \frac{\text{Im}\,A_0}{\text{Re}\,A_0} \right]\,,
\end{equation}
$\omega$, $\left \vert \epsilon \right \vert$ and Re\,$A_0$, where $\omega = \text{Re}\,A_2/\text{Re}\,A_0$ and 
\begin{equation}\label{eq:epsilondef}
\epsilon=\frac{2\eta_{+-}+\eta_{00}}{3}\quad\textrm{where}\quad\eta_{ij}=\frac{A(K_L\to\pi^{i}\pi^{j})}{A(K_S\to\pi^{i}\pi^{j})}\,.
\end{equation}
The numerical values which we use for these quantities are given in Tab.\,\ref{tab:results}.
The systematic error on $\text{Im}\,A_2/\text{Re}\,A_2$ is
found by combining in quadrature the systematic error on $\text{Re}\,A_2$ and $\text{Im}\,A_2$ with the error due to lattice
artefacts excluded. We then add in quadrature a single estimate of 5\% systematic error on $\text{Im}\,A_2/\text{Re}\,A_2$ due to lattice artefacts. 
This estimate is based on the Symanzik theory of improvement which implies that the artefacts are proportional to $a^2$ and in the absence of any knowledge of the constant of proportionality we use the spread of the derived values of the lattice spacing in Tab\,\ref{tab:different_as} below as a guide. Our result and error for Im\,$A_0/$Re\,$A_0 $ are very insensitive to the estimate of the artefacts in  $\text{Im}\,A_2/\text{Re}\,A_2$.

Rearranging Eq.\,\eqref{eq:cp} we determine the unknown quantity Im\,$A_0$ within the Standard Model, finding 
\begin{equation}
\text{Im}\,A_0 = -5.34(62)_{\text{stat}}(68)_{\text{syst}}\times 10^{-11}\text{ GeV}. \end{equation}
The error on Im$\,A_0$ is obtained by combining the errors on the quantities in Tab.\,\ref{tab:results} in quadrature. 
The relative contribution to $\text{Im}\,A_0/\text{Re}\,A_0$ from $\text{Im}\,A_2/\text{Re}\,A_2$ and the term containing the experimentally known contributions is given by:
\begin{equation}
\label{eq:rearrange}
 \begin{array}{ccccc}
  \dfrac{\text{Im}\,A_0}{\text{Re}\,A_0} & = &\dfrac{\text{Im}\,A_2}{\text{Re}\,A_2} &- &\dfrac{\sqrt{2} \left\vert \epsilon \right \vert}{\omega} \dfrac{\epsilon'}{\epsilon}\\
&&\\
-1.61(19)_{\mathrm{stat}}(20)_{\mathrm{syst}}\times10^{-4}  & = &-4.42(31)_{\mathrm{stat}}(89)_{\mathrm{syst}}\times 10^{-5} & -&1.16(18) \times 10^{-4}~.
 \end{array}
\end{equation}
Thus we see that while the error on the determination of Im\,$A_0$ is dominated by the uncertainty in the experimental value of $ \epsilon^\prime/ \epsilon$, the contribution of Im\,$A_2$/Re\,$A_2$ is significant (about 25\%). Of course our ultimate aim is to calculate $A_0$ directly and we hope to be able to report on this soon; an important step towards this goal was presented in \cite{Blum:2011pu}.

The ratio Im${A_0}/$Re$A_0$ allows us to determine the effect of direct CP-violation in $K_L\to\pi\pi$ on $\epsilon$, customarily denoted by $\kappa_{\epsilon}$~\cite{Buras:2008nn}, $(\kappa_{\epsilon})_{\mathrm{abs}} = 0.924 \pm 0.006$. where the subscript ``abs" denotes that at present only the absorptive long-distance contribution (Im $\Gamma_{12}$) is included~\cite{Buras:2010pza} (the error is now dominated by the experimental uncertainty in $\epsilon^\prime/\epsilon$). The analogous contribution from the dispersive part (Im $M_{12}$)~\cite{Buras:2010pza} is yet to be determined in lattice QCD, but we describe progress towards being able to do this in~\cite{Christ:2010zz}.

Using our value of Im$\,A_2$ in Eq.\,(\ref{eq:results}) and taking the experimental value given above for Re\,$A_2$ from $K^+$ decays 
we obtain the electroweak penguin (EWP) contribution to $\epsilon^\prime/\epsilon$,  
Re$(\epsilon^\prime/\epsilon)_{\mathrm{EWP}} = -(6.25 \pm 0.44_{\textrm{stat}} \pm 1.19_{\textrm{syst}}) \times 10^{-4}$ (the experimental value for the complete Re$(\epsilon^\prime/\epsilon)$ is $1.65(26)\times 10^{-3}$~\cite{Nakamura:2010zzi}). Even though we have labelled this contribution EWP, and indeed it is dominated by the matrix element of the EWP operator $Q_{(8,8)_{\mathrm{mix}}}$, the result contains contributions from all three components to Im\,$A_2$ in Eq.\,(\ref{eq:a2components}). The (renormalization-group invariant) sum of the contributions from the two EWP operators $Q_{(8,8)}$ and $Q_{(8,8)_{\mathrm{mix}}}$ is $-(7.34\pm 0.52_{\textrm{stat}} \pm 1.39_{\textrm{syst}}) \times 10^{-4}$. 

We end this section with a brief comparison of an earlier result obtained using finite-energy sum rules\,\cite{Cirigliano:2002jy}, where the contribution to $\epsilon^\prime/\epsilon$ from the operator $Q_{(8,8)_{\mathrm{mix}}}$ (renormalized at 2\,GeV) was found to be $-(11.0\pm 3.6)\times 10^{-4}$. Our result for this particular contribution is $(-7.88\pm 0.43)\times 10^{-4}$. (Note that the contribution from $Q_{(8,8)_{\mathrm{mix}}}$ by itself is not renormalization group invariant.)
We also note that our result is consistent with expectations based on the vacuum saturation approximation at scales around 2\,GeV~\cite{Cirigliano:2002jy,Buras:2003zz}.
For a comprehensive general recent review on kaon decays we refer the reader to Ref.\,\cite{Cirigliano:2011ny}.

\begin{center}
\begin{table}[t]
\begin{tabular}{|c|c|}
 \hline
$\epsilon'/\epsilon$ & $(1.65\pm0.26)\times10^{-3}$\\
$\omega$ & 0.04454(12)\\
$\left \vert \epsilon \right \vert$ & $(2.228\pm 0.011) \times 10^{-3}$ \\
Re$\,A_0$ & $3.3201(18)\times 10^{-7}$~GeV\\
$\text{Im} \,A_2/\text{Re}\,A_2$ (lattice) & $-4.42(31)_{\mathrm{stat}}(89)_{\mathrm{syst}} \times 10^{-5}$\\
\hline
\end{tabular}
\caption{\label{tab:results} Experimental values of the quantities in Eq.\,\eqref{eq:cp} which is used in the determination of Im\,$A_0$, together with the result for $\text{Im} \,A_2/\text{Re}\,A_2$ from this paper.}
\end{table}
\end{center}

\section{Renormalization of the Lattice Operators}\label{sec:npr}

We have seen in Sec.\,\ref{sec:analysis} above, that in order to determine the physical amplitudes we need to combine the $K\to\pi\pi$ matrix elements with Wilson coefficient functions. The coefficient functions are calculated in perturbation theory and most often correspond to renormalization schemes 
based on dimensional regularization, such as the $\msbar$-NDR scheme. We therefore need to determine the matrix elements of the weak operators also renormalized in the $\msbar$-NDR scheme and schematically this is done as follows:
\begin{equation*}\left\{\begin{array}{c}
\mathrm{bare}\\ \mathrm{lattice}\\ \mathrm{operators}\end{array}\right\}
\begin{array}{c}\\ \stackrel{\mathrm{NPR}}{\longrightarrow}\\ \\ \end{array}
\left\{\begin{array}{c}
\mathrm{operators~renormalized~in}\\ \mathrm{intermediate~scheme(s)}\\ \mathrm{(RI-MOM,~RI-SMOM)}\end{array}
\right\}\begin{array}{c}\\ \stackrel{\mathrm{Pert.Th.}}{\longrightarrow}\\ \\ \end{array}
\left\{\begin{array}{c}
\mathrm{operators}\\ \mathrm{renormalized~in}\\\msbar-\mathrm{NDR~scheme.}\end{array}\right\}
\end{equation*}

\noindent In the first step we perform non-perturbative renormalization (NPR) on the bare lattice operators to obtain operators defined in a renormalization scheme which can be simulated numerically, such as the RI-MOM or RI-SMOM schemes discussed below~\cite{Martinelli:1994ty,Aoki:2007xm,Sturm:2009kb}. Since we cannot perform
simulations in a non-integer number of space-time dimensions, the introduction of intermediate schemes is necessary. In the second step continuum perturbation theory is used to relate the operators in these intermediate schemes to the $\msbar$-NDR scheme. In this way we avoid the use of lattice perturbation theory, which frequently converges more slowly than its 
continuum counterpart and for which it is more difficult to calculate the higher-order corrections.

Of course the relations between the bare lattice operators and those renormalized in the $\msbar$-NDR scheme are, in principle, independent of the choice of the intermediate scheme. In practice, in addition to the 
remaining lattice systematic uncertainties, the fact that the matching between the operators in the intermediate schemes and  $\msbar$-NDR is performed only at a relatively low order of perturbation theory means that there
is a small dependence on the choice of intermediate scheme. As explained in the following subsections, we find it useful to use a number of intermediate schemes and to use the spread of results as an indication of the uncertainties due to the truncation of perturbation theory.

\subsection{The Intermediate Renormalization Schemes}\label{subsec:intermediate}

The intermediate renormalization schemes we use are natural extensions of those we introduced in our recent study of the $B_K$ parameter of neutral kaon mixing~\cite{Aoki:2010pe}. These in turn were based on the schemes we had introduced for quark bilinear operators in which there are no \emph{exceptional} channels, i.e.~no channels with small or zero momenta~\cite{Aoki:2007xm,Sturm:2009kb}. By imposing the renormalization conditions on quark and gluon Green functions with no exceptional channels we suppress the systematic errors due to the breaking of chiral symmetry by infrared effects. We now explicitly explain the schemes which we use. For all the operators we introduce two ways of treating the vertex renormalization and two ways of defining the wave function renormalization, leading to four renormalization schemes for the operators themselves.

\begin{figure}[t]
\begin{center}
\includegraphics[width =0.25\textwidth]{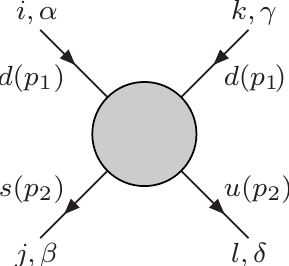}
\caption{\label{fig:nprkinematics} Schematic diagram illustrating the process in Eq.\,(\ref{eq:nprkinematics}). In the diagram the arrows refer both to the flow of the indicated flavor quantum number and also to the indicated momentum. The spinor ($ \alpha,\,\beta,\,\gamma,\,\delta$) and color ($i,\, j,\, k,\, l$) labels are also indicated.}\end{center}
\end{figure}

The three operators which we need to renormalize are defined in Eq.\,(\ref{eq:Qdef}). $Q_{(27,1)}$ renormalizes multiplicatively, whereas the two electroweak penguin operators $Q_{(8,8)}$ and $Q_{(8,8){\mathrm{mix}}}$ mix under renormalization so that there is a corresponding $2\!\times\! 2$ matrix of renormalization constants. We start with a discussion of the renormalization of $Q_{(27,1)}$ for which we compute the Green function for the process
\begin{equation}\label{eq:nprkinematics}
d(p_1)\bar{s}(-p_2)\to\bar{d}(-p_1)u(p_2)
\end{equation}with $p_1^2=p_2^2=(p_1-p_2)^2=\mu^2$ for a variety of momenta satisfying this condition. The process is illustrated in the diagram of Fig\,\ref{fig:nprkinematics} and $\mu$ is taken to be the renormalization scale.

Let $ \Lambda^{(27,1)\,ij,kl}_{ \alpha\beta, \gamma \delta}(p_1,p_2)$ be the amputated Landau-gauge Green function of the bare lattice $Q_{(27,1)}$, where $ \alpha, \beta, \gamma$ and $ \delta$ are the spinor labels corresponding to the incoming $d$ and $\bar{s}$ quarks and outgoing $\bar{d}$ and $u$ quarks respectively and $i,j,k,l$ are the corresponding color labels. Since $Q_{(27,1)}$ is multiplicatively renormalizable, the relation between the bare lattice and renormalized operator is of the form:
\begin{equation}
Q_{(27,1)}^{\mathrm{(I_v,I_q)}}=Z_{(27,1)}^{\mathrm{(I_v,I_q)}}\,Q_{(27,1)}^{\mathrm{(latt})}\,,
\end{equation}
where $\mathrm{I_v}$ labels the choice of the intermediate (one-particle irreducible) vertex renormalization scheme and $\mathrm{I_q}$ the intermediate scheme for the wave function renormalization. The index ``$\mathrm{latt}$" reminds us that the operator on the right-hand side is the bare lattice operator. The overall renormalization constant is obtained by evaluating a trace of $ \Lambda$ with a projection operator $P^{(\mathrm{I_v})}$
\begin{equation}
Z_{(27,1)}^{\mathrm{(I_v,\mathrm{I}_q})}=Z_q^{\mathrm{(I_q)}\,2}\,\frac{1}{P^{(\mathrm{I_v)}\,ij,kl}_{ \alpha\beta, \gamma\delta}\,
\Lambda^{(27,1)\,ij,kl} _{ \alpha\beta, \gamma\delta}}\,,
\end{equation}
where $Z_q^{\mathrm{(I_q)}}$ is the wave function renormalization constant which will be discussed below. The two choices we make for the projection operators are labelled by $\mathrm{I_v}=\gamma_\mu$ or $\mathrm{I_v}=\qslash$~\cite{Aoki:2007xm}:
\begin{eqnarray}
P^{(\gamma^{\mu})\,ij, kl}_{\alpha \beta, \gamma \delta} &= &\frac{1}{128N(N+1)}\,
\left[ (\gamma^{\mu})_{\beta\alpha} (\gamma^{\mu})_{\delta\gamma} + (\gamma^\mu \gamma^5)_{\beta\alpha}
 (\gamma^\mu\gamma^5)_{\delta\gamma} \right]
 \delta^{ij} \delta^{kl}\label{eq:p271}\\
P^{(\qslashs)\,ij, kl}_{\alpha \beta, \gamma \delta} &=& \frac{1}{32q^{2}N(N+1)}\left[ (\qslash)_{\beta\alpha} (\qslash)_{\delta\gamma} + (\qslash \gamma^{5})_{\beta\alpha} (\qslash\gamma^{5})_{\delta\gamma} \right]
 \delta^{ij} \delta^{kl} \,, \label{eq:p272}
\end{eqnarray}
where $N=3$ is the number of colors. These projectors are constructed to give 1 when contracted with the tree-level results for $\Lambda^{(27,1)\,ij,kl} _{ \alpha\beta, \gamma\delta}$.

For the wave function renormalization we use the schemes defined as RI-SMOM and RI-SMOM$_{\gamma_\mu}$ in ref.\,\cite{Sturm:2009kb}, which for compactness of notation, we label as $\mathrm{I_q}=\not{\!\hspace{-0.3pt}q}$ and $\mathrm{I_q}=\gamma_\mu$ respectively. The corresponding renormalization constants are defined as 
\begin{equation}
Z_q^{(\qslashs)}=\frac{q^\mu}{12q^2}\,\mathrm{Tr}[\Lambda_V^\mu\qslash]\qquad\mathrm{and}\qquad
Z_q^{(\gamma_\mu)}=\frac{1}{48}\,\mathrm{Tr}[\Lambda_V^\mu\gamma^\mu]\,,
\end{equation}
where $\Lambda_V^\mu$ is the amputated Green function of the conserved vector current. This completes the description of the determination of the renormalization constant for $Q_{(27,1)}$ in the four schemes in which each of $\mathrm{I_q}$ and  $\mathrm{I_v}$ are either $\qslash$ or $\gamma_\mu$.

 We now turn to the determination of the renormalization constants of the electroweak penguin operators $Q_7=Q_{(8,8)}$ and $Q_8=Q_{(8,8){\mathrm{mix}}}$, where the notation $Q_{7}$ and $Q_8$ is another standard one and will prove convenient in the following discussion. In this case we define two projection operators $P_{7}^{(\mathrm{I_v})}$ and $P_{8}^{(\mathrm{I_v})}$ for each scheme ($\mathrm{I_v}=\gamma_\mu$ or $\qslash$):
 \begin{eqnarray}
\left[P^{(\gamma^{\mu})}_{7}\right]^{ij, kl}_{\alpha \beta, \gamma \delta} &=& 
\left[ (\gamma^{\mu})_{\beta\alpha} (\gamma^{\mu})_{\delta\gamma} - (\gamma^\mu\gamma^5)_{\beta\alpha} (\gamma^\mu\gamma^5)_{\delta\gamma} \right]
 \delta^{ij} \delta^{kl}\\
\left[P^{(\gamma^{\mu})}_{8}\right]^{ij, kl}_{\alpha \beta, \gamma \delta} &= &
\left[ (\gamma^{\mu})_{\beta\alpha} (\gamma^{\mu})_{\delta\gamma} - (\gamma^\mu\gamma^5)_{\beta\alpha} (\gamma^\mu\gamma^5)_{\delta\gamma} \right]
\delta^{il} \delta^{jk}\\
\left[P^{(\qslashs)}_{7}\right]^{ij, kl}_{\alpha \beta, \gamma \delta} &=& \frac{1}{q^{2}}\left[ (\qslash)_{\beta\alpha} (\qslash)_{\delta\gamma} - (\qslash \gamma^{5})_{\beta\alpha} (\qslash \gamma^{5})_{\delta\gamma} \right]\delta^{ij} \delta^{kl} \\
\left[P^{(\qslashs)}_{8}\right]^{ij, kl}_{\alpha \beta, \gamma \delta} &=& \frac{1}{q^{2}}\left[ (\qslash)_{\beta\alpha} (\qslash)_{\delta\gamma} - (\qslash \gamma^{5})_{\beta\alpha} (\qslash \gamma^{5})_{\delta\gamma} \right]
 \delta^{il} \delta^{jk} \, .
\end{eqnarray}
For each scheme, let $M_{ab}$ ($a,b=7,8$) be the matrix obtained by tracing $P_b$ with the amputated Green function $\Lambda^a$ over spinor and color indices:
\begin{equation}
M_{ab}\equiv \left[P_{b}\right]^{ij, kl}_{\alpha \beta, \gamma \delta} \left[\Lambda^a\right]^{ij, kl}_{\alpha \beta, \gamma \delta}
\end{equation}
with an implicit sum over all repeated indices and we have suppressed the index $I_v=\gamma^\mu$ or $\qslash$ defining the renormalization scheme. The matrix of renormalization constants $Z_{ab}$ ($a,b=7,8$) is defined by 
\begin{equation}
\frac{1}{Z_q^2}\,ZM=M^0\,,
\end{equation}where the matrix $M^0$ is the free-field expression for $M$.

With a single choice of boundary conditions, the components of momenta are quantized in steps of $2\pi/L$, where $L$ is the spatial extent of the lattice. In order to study the momentum dependence of the Green functions from which the renormalization constants are calculated we need to take a range of values for each component of momentum. The presence of lattice artefacts which are not invariant under the $O(4)$ group (but which are invariant under the lattice discrete symmetry group) leads to irregularities in the computed momentum dependence. Examples of such contributions are terms proportional to $a^2 (\sum_\mu p_\mu^4)/(\sum_\mu p_\mu^2)$. Such terms are not proportional to $a^2p^2$ (where $p^2\equiv\sum_\mu p_\mu^2$) and introduce a scatter in Green functions when plotted as functions of $p^2$, making it difficult to extrapolate the results to the continuum limit. The use of partially twisted boundary conditions~\cite{Sachrajda:2004mi,Bedaque:2004ax} allows us to scale the components of the momenta (almost) continuously, so that $(\sum_\mu p_\mu^4)/(\sum_\mu p_\mu^2)$ and $p^2$ scale in the same way and the scatter is eliminated. This technique was used in our recent calculation of the $B_K$ parameter~\cite{Aoki:2010pe} where it is described in detail and it is used throughout the present calculation of the renormalization constants.

 \subsection{Step Scaling}\label{subsec:stepscaling}
In the preceding subsection we described how we obtain the renormalized operators $Q^{(\mathrm{I_v,I_q})}_{(27,1)}(\mu)$, $Q^{(\mathrm{I_v,I_q})}_{(8,8)}(\mu)$ and $Q^{(\mathrm{I_v,I_q})}_{(8,8)_{\mathrm{mix}}}(\mu)$ on the coarse IDSDR lattice, where the renormalization scale $\mu^2=p_1^2=p^2_2=(p_1-p_2)^2$ (see the discussion around Eq.\,(\ref{eq:nprkinematics})). In order to limit the lattice artefacts on this coarse lattice
($a\simeq$ 0.14\,fm) $\mu$ should not be very large. On the other hand if we choose $\mu$ to be too small then perturbation theory cannot be used reliably to relate the operators in the intermediate schemes to those in the $\msbar$-NDR scheme. The use of \textit{step scaling}\,\cite{Luscher:1993gh,Luscher:1991wu}, and in particular its recent generalization to the RI-SMOM schemes being used in this work~\cite{Aoki:2010pe,Arthur:2010ht,Arthur:2010hy},
overcomes this last limitation as explained below. This step scaling
approach can also be generalised to operators which mix under renormalization 
\cite{Boyle:2011kn,Arthur:2011cn} and this is applied in our calculation.

Imagine that we use the procedure of Subsec.\,\ref{subsec:intermediate} to obtain the renormalization constants 
$Z^{\mathrm{(I_v,I_q)}}_{(27,1)}(\mu_0)$ and $Z^{\mathrm{(I_v,I_q)}}_{ab}(\mu_0),\,(a,b=7,8)$, on the IDSDR lattice 
for a renormalization scale $\mu_0$ which is sufficiently small that lattice artefacts can be neglected and which is therefore likely to be outside of the perturbative regime. We then repeat the same renormalization procedure to obtain the corresponding renormalization constants, and hence the corresponding operators, on the finer Iwasaki lattices mentioned in Sec.\,\ref{sec:simulation}. (Renormalization constants on the Iwasaki ensembles were presented
in~\cite{Boyle:2011kn}.) The benefit of doing this is that on the finer lattices we can run the renormalization constants non-perturbatively from $\mu_0$ to a larger scale $\mu$ at which perturbation theory can be applied. Taking $Q_{(27,1)}$ as an example, we define a step scaling function on the finer lattices:
\begin{equation}
\Sigma_{(27,1)}^{\mathrm{(I_v,I_q)}}(\mu,\mu_0,a)=\lim_{m\to0}\,\left[Z^{\mathrm{(I_v,I_q)}}_{(27,1)}(\mu,a,m)
\left(Z^{\mathrm{(I_v,I_q)}}_{(27,1)}(\mu_0,a,m)\right)^{-1}\right]\,,
\end{equation}where $m$ is the quark mass. Since we have results at two different lattice spacings on the finer Iwasaki lattices we can perform the continuum extrapolation and define the continuum step scaling functions as
\begin{equation}
\sigma_{(27,1)}^{\mathrm{(I_v,I_q)}}(\mu,\mu_0)=\lim_{a\to 0}\Sigma_{(27,1)}^{\mathrm{(I_v,I_q)}}(\mu,\mu_0,a)\,.
\end{equation}
The step scaling function $\sigma_{(27,1)}(\mu,\mu_0)$ describes the continuum
non-perturbative running of the 4 quark operator $Q_{(27,1)}$ in a given scheme.
Because it does not depend on the lattice action, we can use it to run
the Z factor obtained from the IDSDR lattice at a low scale $\mu_0$ to a
higher energy $\mu$ where perturbation theory is more convergent.
Finally, the operator $Q_{(27,1)}$ renormalized in the intermediate scheme $(\mathrm{I_v,I_q})$ at a perturbative scale $\mu$ is related to the IDSDR lattice operator by: 
\begin{equation}\label{eq:aux1}
Q_{(27,1)}^{\mathrm{(I_v,I_q)}}(\mu)=\sigma_{(27,1)}^{\mathrm{(I_v,I_q)}}(\mu,\mu_0)\,
Z_{(27,1)}^{\mathrm{(I_v,I_q)}}(\mu_0)\,Q_{(27,1)}^{\mathrm{(latt})}.
\end{equation}

Having obtained the operator renormalized in the intermediate schemes at a perturbative renormalization scale, we convert it to the $\msbar$-NDR scheme using one-loop perturbation theory
\begin{equation}
Q^{\overline{\text{MS}}}_{(27,1)}(\mu) = S^{(\mathrm{I_v, I_q}) \rightarrow \overline{\text{MS}}}_{(27,1)}(\mu) \,\, Q_{(27,1)}^{\mathrm{(I_v,I_q)}}(\mu)\,.
\end{equation}
The expressions for the conversion factors $S^{(\mathrm{I_v, I_q}) \rightarrow \overline{\text{MS}}}_{(27,1)}(\mu)$ can be found in ref.\cite{Lehner:2011fz}. 
Since these are known to $O(\alpha_{s})$ the determinations of
$Q^{\overline{\text{MS}}}_{(27,1)}(\mu)$ via different intermediate schemes $(\mathrm{I_v, I_q})$
will differ from one another at $O(\alpha_{s}^{2})$.  
The difference of results calculated via different intermediate schemes provides an estimate for the size of this effect.

For the electroweak operators the above equations become $2 \! \times \! 2$ matrix equations with the constants $Z_{(27,1)}^{\mathrm{(I_v,I_q)}}$ replaced by the matrices $Z_{ab}^{\mathrm{(I_v,I_q)}}$ and similarly for the step scaling factors.

\subsection{Numerical Evaluation of the Renormalization Constants}\label{subsec:numerical}

\newcommand{\ph}[1]{\phantom{#1}}

We now present the numerical results for the  conversion matrices that relate our bare lattice operators, $Q_{i}^{(\text{latt})}$  ($i = (27, 1), \,(8,8), \,(8,8)_{\mathrm{mix}}$), to those renormalized in the $\overline{\text{MS}}$-NDR scheme at the renormalization scale ${\mu}$, $Q_{i}^{\overline{\text{MS}}}\,(\mu)$,
\begin{align}\label{eq:convfactors}
Q^{\overline{\text{MS}}}(\mu) &= S^{(\mathrm{I_v, I_q}) \rightarrow \overline{\text{MS}}}(\mu) \,\,
\sigma^{(\mathrm{I_v, I_q})}(\mu, \mu_0) \,\, Z^{(\mathrm{I_v, I_q})}(\mu_0) \,\, Q^{(\text{latt})} \\
&\equiv Z^{\overline{\text{MS}},(\text{latt})}_{(\mathrm{I_v, I_q})}(\mu) \,\,  Q^{(\text{latt})} \,.
\end{align}
As explained in Sec.\,\ref{subsec:stepscaling}, the conversion matrix
$[ Z^{\overline{\text{MS}},(\text{latt})}_{(\mathrm{I_v, I_q})}(\mu)]_{ab}$ 
is a product of the three factors explicitly exhibited in Eq.\,(\ref{eq:convfactors}).
We have studied the four different intermediate schemes $(\mathrm{I_v, I_q})$ introduced in Sec.\,\ref{subsec:intermediate} in order to 
estimate the uncertainty
from perturbative truncation errors in the $(\mathrm{I_v, I_q})$ to $\overline{\text{MS}}$ matching factors $S^{(\mathrm{I_v, I_q}) \rightarrow \overline{\text{MS}}}(\mu)$,  which are known at one loop~\cite{Lehner:2011fz}, and also the uncertainty from discretisation effects.

In the evaluation of $B_K$~\cite{Aoki:2010pe}, our study of the renormalization of the $(27,1)$ operator concluded that of the four choices of 
$\mathrm{(I_v,I_q)}$ it was the non-perturbative running functions 
 $\sigma^{(\qslashs, \qslashs)}_{(27,1)}(\mu, \mu_{0})$ and 
 $\sigma^{(\gamma^{\mu}, \gamma^{\mu})}_{(27,1)}(\mu, \mu_{0})$ which 
were best approximated by perturbation theory for 
$\mu \approx3 \text{ GeV}$.
These two intermediate schemes were then chosen for the determination of  the matrix elements of $Q^{\overline{\text{MS}}}_{(27,1)}$ and in estimating the truncation uncertainty.
In the current work, we again find that the running functions in the 
$(\slashed{q}, \slashed{q})$ and $(\gamma^{\mu}, \gamma^{\mu})$  schemes,
now $3\!\times\!3$ matrices, are generally well described by perturbation theory.  We therefore choose to 
adopt the same procedure as in~\cite{Aoki:2010pe}:
we take the results from the $(\slashed{q}, \slashed{q})$ 
intermediate scheme as our central values for 
$\text{Re}\,A_{2}$ and $\text{Im}\,A_{2}$,
and use the difference between
these and the results obtained in the $(\gamma^{\mu}, \gamma^{\mu})$  scheme as an estimate of the uncertainty.

In order to minimise discretisation effects in the calculation of the $Z$-factors on the IDSDR
lattices, where $a$ is large and only one lattice spacing is available, we take as a matching point 
the low scale $\mu_{0} = 1.136 \text{ GeV}$.  We obtain:
\begin{eqnarray} \label{eq:zggmu0}
Z^{(\gamma^{\mu}, \gamma^{\mu})} (\mu_0) &=
\begin{pmatrix}
0.443\,(1)	&	0	&	0	\\
0	&	\ph{-}0.505\,(1)	&	-0.114\,(1)	\\
0	&	-0.022\,(3)	&	\ph{-}0.231\,(2)	\\
\end{pmatrix} \\[3mm]
Z^{(\qslashs, \qslashs)} (\mu_0) &=
\begin{pmatrix}
0.489\,(1)	&	0	&	0	\\
0	&	\ph{-}0.510\,(2)	&	-0.116\,(1) 	\\
0	&	-0.077\,(6)	&	\ph{-}0.305\,(4) 	\\
\end{pmatrix} \,,\label{eq:zqqmu0}
\end{eqnarray}
where the quoted errors are statistical only.  Here and in the remainder of this section we estimate and propagate the statical errors by using 100 bootstrap samples.

The block structure of the matrices given in Eqs.\,(\ref{eq:zggmu0})
and (\ref{eq:zqqmu0}) is justified by the short-distance chiral
symmetry of the DWF formulation which implies that
changes in action and lattice spacing can be compensated
by multiplicative renormalization of $Q_{(27,1)}$ and
mixing between $Q_7$ and $Q_8$.  The method described
above determines the five elements of the $3\times 3$
matrix $Z$ which are expected to be non-zero and the
remaining four are set to zero because the chiral symmetry of the theory implies that operators with different chirality do not mix under renormalization.

The renormalization constants at $\mu_0$ are converted to the higher scale $\mu = 3 \text{ GeV}$ using 
step-scaling functions calculated on the Iwasaki lattices,
extrapolated to the continuum limit. When performing the continuum extrapolation, we match scales on the different lattices
by interpolating the simulated data, which are very smooth on account of our use of twisted boundary conditions.  
Twisted boundary conditions also ensure that the data lie along a continuum trajectory, and with two Iwasaki ensembles we can attempt to remove $O(a^{2})$ artefacts using a straight-line fit.
Since we have only two lattice spacings, we choose to quote a conservative systematic error: the difference between 
the results on our finest Iwasaki lattice and those obtained by extrapolating to the continuum. In this way we obtain
\begin{eqnarray} \label{eq: step_scale_results}
\sigma^{(\gamma^{\mu}, \gamma^{\mu})} (3\,\mathrm{GeV}, \mu_0) &=
\begin{pmatrix}
0.942\,(4)(1)	&	0	&	0	\\
0	&	0.964\,(9)(13)	&	0.386\,(20)(79) 	\\
0	&	0.038\,(23)(17)	&	2.210\,(76)(103) 	\\
\end{pmatrix} \\[3mm]
\sigma^{(\qslashs, \qslashs)} (3\,\mathrm{GeV}, \mu_0) &=
\begin{pmatrix}
0.876\,(7)(9)	&	0	&	0	\\
0	&	0.973\,(11)(6)	&	0.309\,(16)(67) 	\\
0	&	0.166\,(38)(50)	&	1.884\,(84)(45) 	\\
\end{pmatrix} \,.
\end{eqnarray}
The first quoted errors are statistical, while the second are the systematic ones from the continuum extrapolation.

The matching to the $\overline{\text{MS}}$-NDR scheme is performed at the scale $\mu = 3 \text{ GeV}$ where perturbation theory is more convergent than at the conventional scale of 
$\mu = 2 \text{ GeV}$.
Using $\alpha^{\overline{\text{MS}}}_{s}(3 \text{ GeV}) = 0.24544$,
we obtain for the matching factors:
\begin{eqnarray}
S^{(\gamma^{\mu}, \gamma^{\mu}) \rightarrow \overline{\text{MS}}}(3 \text{ GeV}) &=
\begin{pmatrix}
1.00414	&	0	&	0	\\
0	&	\ph{-}1.00084	&	-0.00253 	\\
0	&	-0.03152&	\ph{-}1.08781 	\\
\end{pmatrix} \\[3mm]
S^{(\qslashs, \qslashs) \rightarrow \overline{\text{MS}}}(3 \text{ GeV}) &=
\begin{pmatrix}
0.99112	&	0	&	0	\\
0	&	\ph{-}1.00084	&	-0.00253	\\
0	&	-0.01199	&	\ph{-}1.02921	\\
\end{pmatrix} \,.
\end{eqnarray}
Multiplying these results together and propagating the systematic errors in quadrature gives our final result:
\begin{eqnarray} \label{eq:Z_MSbar_results}
 Z^{\overline{\text{MS}},(\text{latt})}_{(\gamma^{\mu}, \gamma^{\mu})}(3\,\mathrm{GeV})  &=
\begin{pmatrix}
0.419\,(2)(1)	&	0	&	0	\\
0	&	\ph{-}0.479\,(5)(8)	&	-0.022\,(5)(20) 	\\
0	&	-0.047\,(13)(11)	&	\ph{-}0.552\,(19)(28) 	\\
\end{pmatrix} \\ [3mm] \label{eq:Z_MSbar_results2}
 Z^{\overline{\text{MS}},(\text{latt})}_{(\qslashs,\qslashs)}(3\,\mathrm{GeV})  &=
\begin{pmatrix}
0.424\,(4)(4)	&	0	&	0	\\
0	&	\ph{-}0.472\,(6)(8)	&	-0.020\,(5)(21) 	\\
0	&	-0.067\,(23)(30)	&	\ph{-}0.572\,(28)(20) 	\\
\end{pmatrix} \,.
\end{eqnarray}
For each result, the first quoted error is statistical errors, while the second is the systematic uncertainty due to
the continuum extrapolation in~(\ref{eq: step_scale_results}). 

\subsection{Is 3\,GeV a sufficiently large momentum for perturbative matching?}\label{subsec:checks}

In the previous subsections we described how we calculate the renormalization constants relating the bare lattice operators on the IDSDR lattices to those renormalized in the RI-SMOM schemes at a renormalization scale of 3\,GeV. This calculation is entirely non-perturbative. In order to obtain the physical amplitude $A_2$
the matrix elements of these renormalized operators have to be combined with the Wilson coefficient functions which are calculated in perturbation theory, most often in schemes based on dimensional regularisation. We therefore convert our results to the $\msbar$-NDR scheme at $\mu=3$\,GeV and, since we cannot perform simulations in a non-integer number of dimensions,  this conversion has necessarily to be performed using (continuum) perturbation theory. At present we know the conversion factor to one-loop order and the difference of the results in Eqs.\,(\ref{eq:Z_MSbar_results}) and (\ref{eq:Z_MSbar_results2}) provides an estimate of the systematic error due to the truncation of the perturbative matching to one-loop order in going from the RI-SMOM to the $\msbar$-NDR schemes. In this subsection we investigate further whether 3\,GeV is a sufficiently large scale at which to use perturbation theory. We do this in two ways. Firstly we study how well the non-perturbative running tracks perturbation theory in the vicinity of $\mu=3\,$GeV.  We then check whether the infra-red chiral symmetry breaking effects are small at 3\,GeV.

\subsubsection{Comparing perturbative and non-pertubative running.}

\begin{figure}
\includegraphics[width =0.4\textwidth]{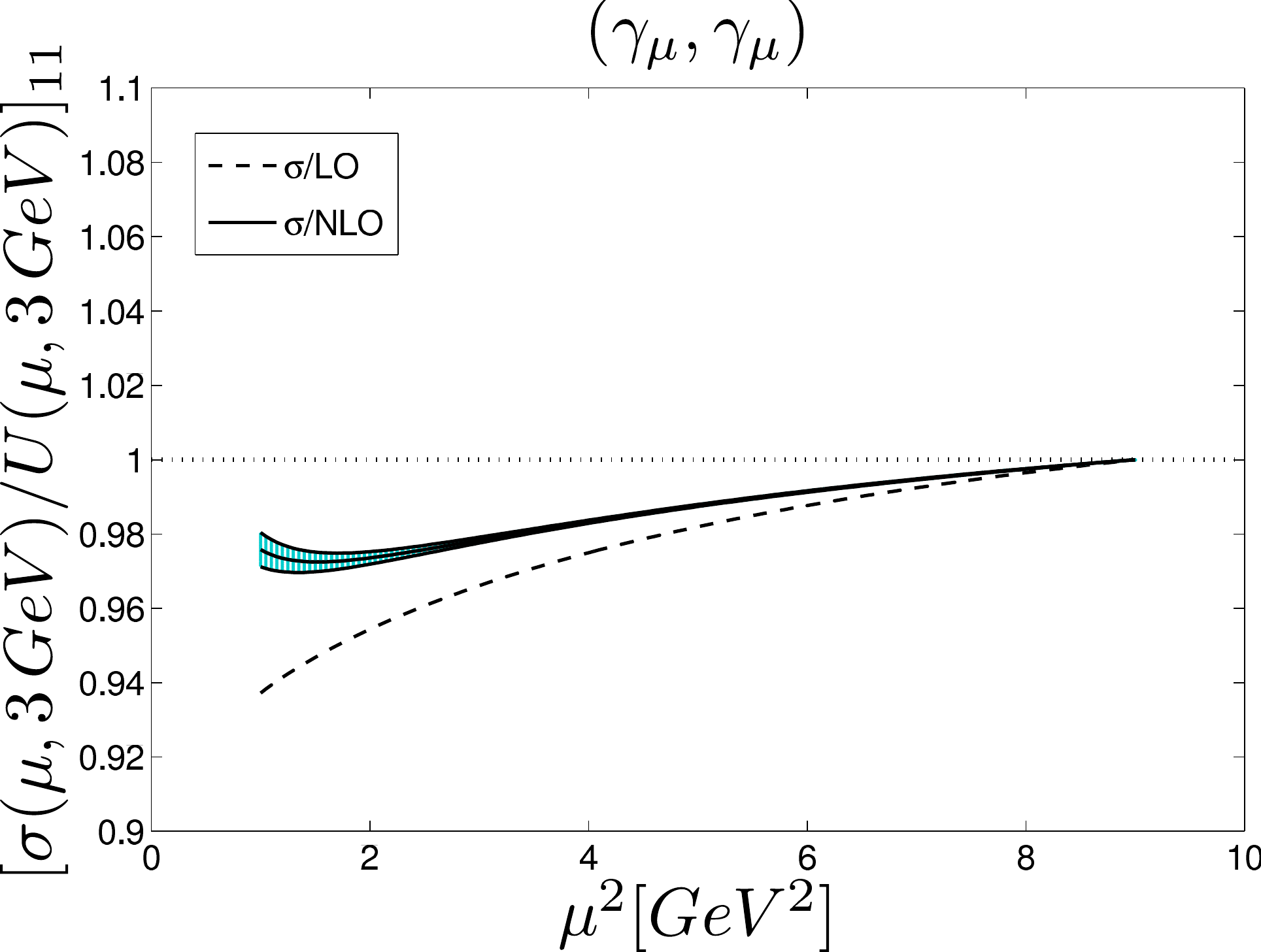}\qquad
\includegraphics[width =0.4\textwidth]{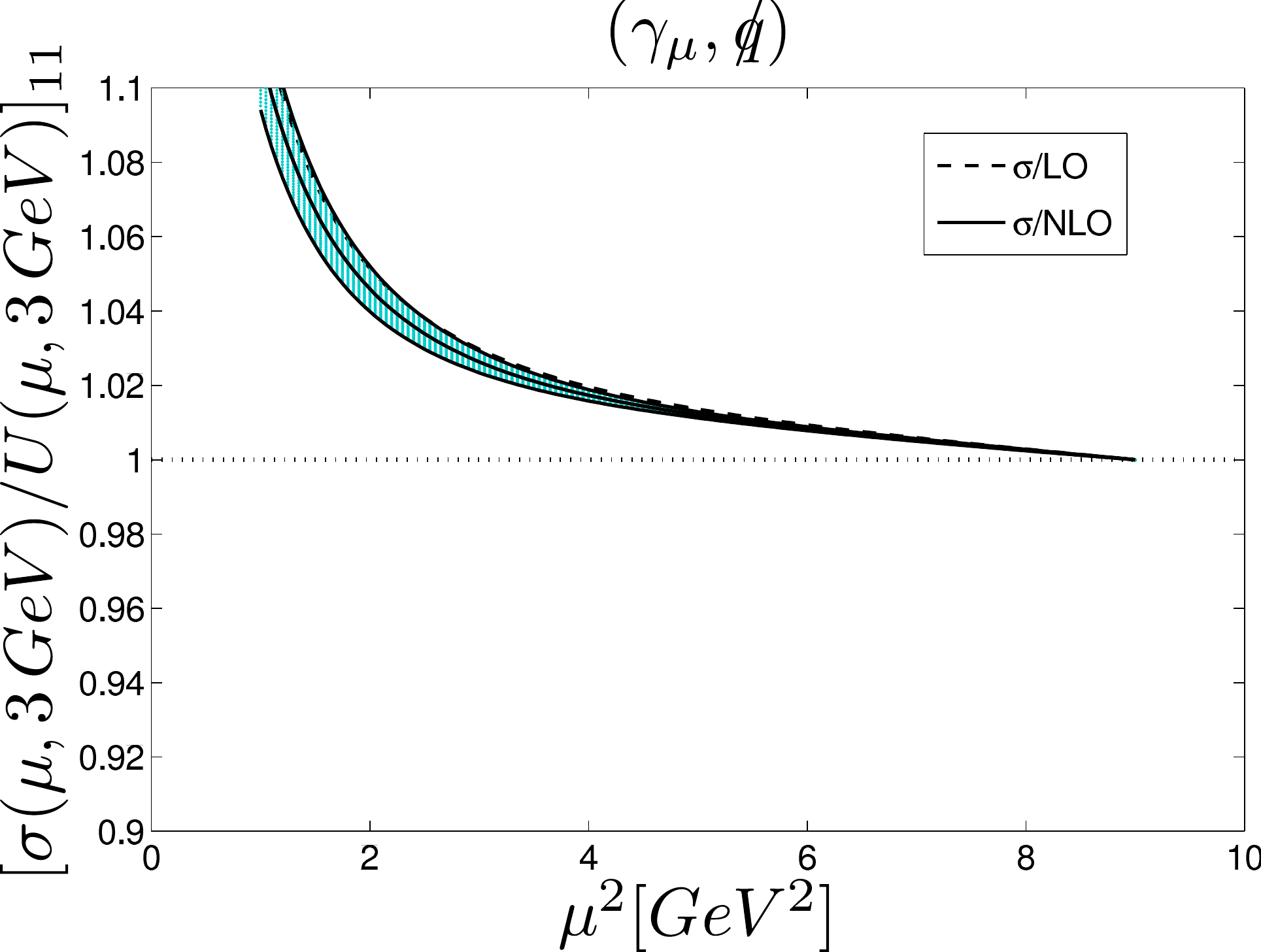}\\ 
\includegraphics[width =0.4\textwidth]{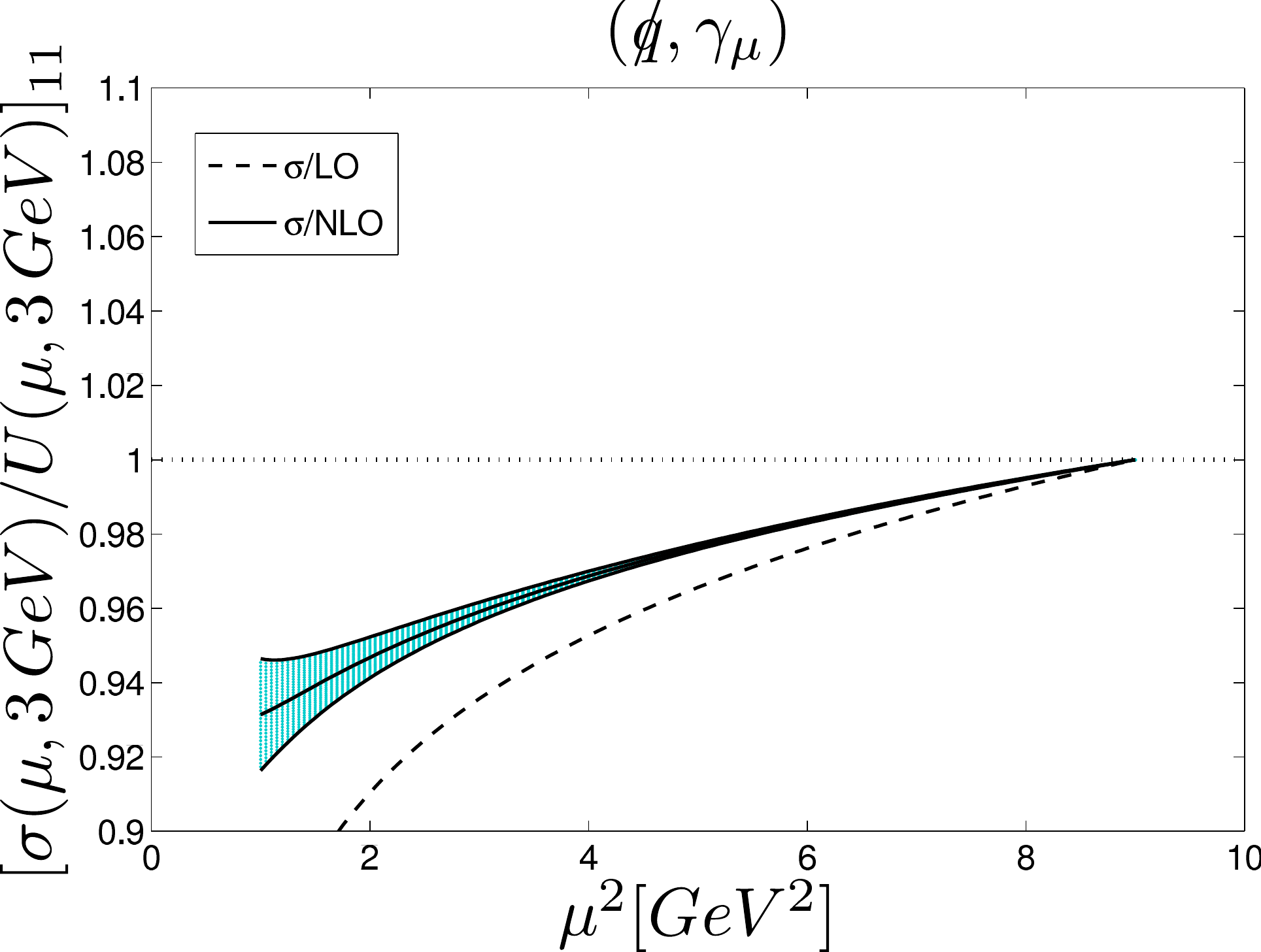}\qquad
\includegraphics[width =0.4\textwidth]{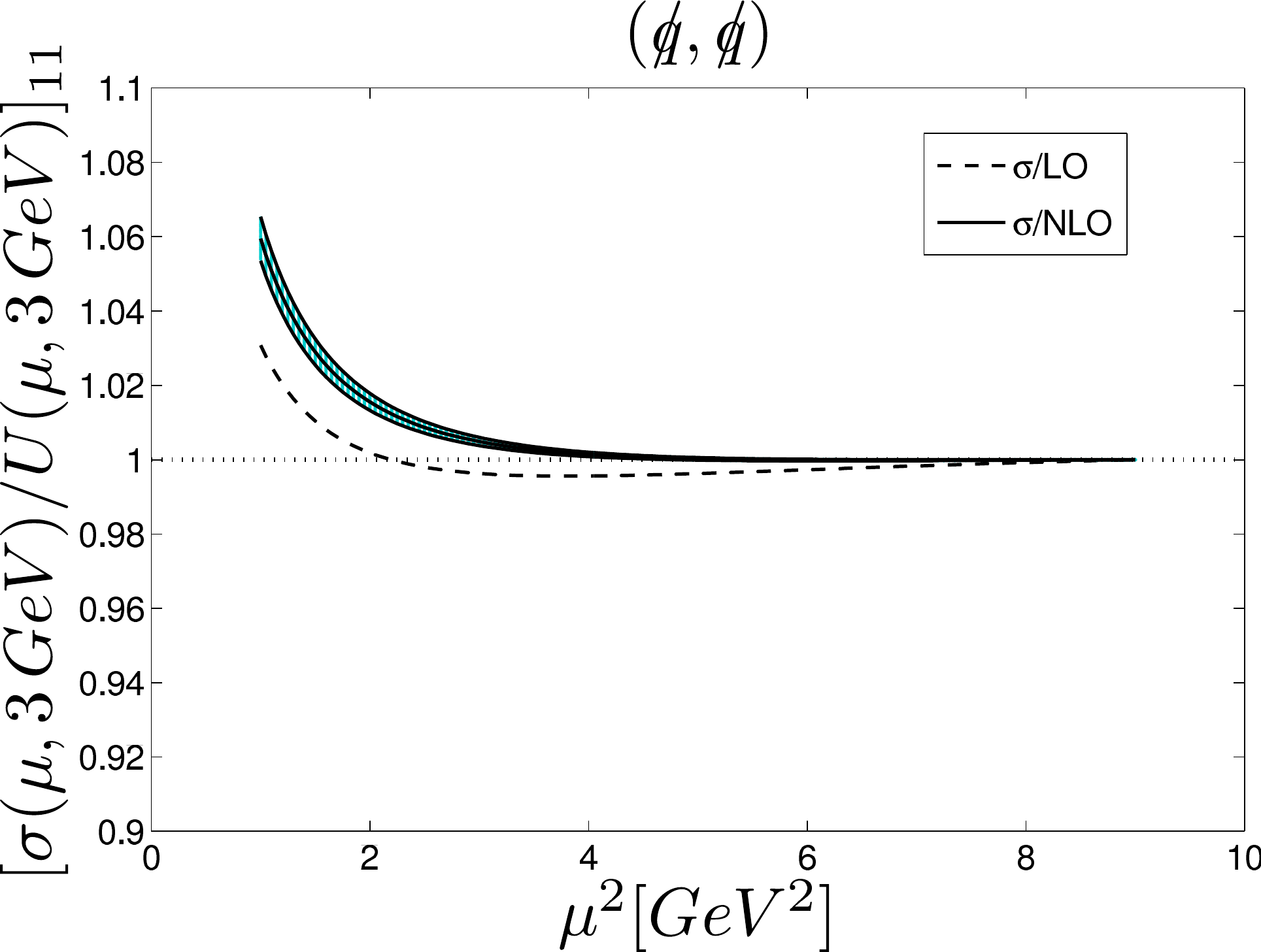}
\caption{\label{fig:sigmapert27} The running of the step-scaling function divided by the LO or NLO perturbative expression for the operator $Q_{(27,1)}$ for the four RI-SMOM schemes considered in this paper. The ratio is set to 1 at $\mu=3$\,GeV. The non-perturbative results have been extrapolated to both the chiral and continuum limits.}
\end{figure}

It is instructive to start with the four plots in Fig.\,\ref{fig:sigmapert27}. These represent the running of the step scaling functions for $Q_{(27,1)}$, determined non-perturbatively, normalized by the LO or NLO perturbative expressions for the four RI-SMOM schemes considered in this paper. The ratios are fixed to be 1 at $\mu=3$\,GeV where we match perturbatively to the $\msbar$-NDR scheme. We see that for the $(\qslash,\qslash)$ scheme the running is very much as expected from NLO perturbation theory (and indeed LO perturbation theory) in the vicinity of 3\,GeV and this was the primary reason why our central values for $B_K$ (which is also obtained from the matrix element of an operator which transforms as an SU(3)$_L\times$SU(3)$_R$ (27,1) and is related to the operator studied here by a chiral rotation)
were quoted using $(\qslash,\qslash)$ as the intermediate scheme~\cite{Aoki:2010pe}. The $(\gamma,\gamma)$ scheme shows a reasonable agreement between the perturbative and non-perturbative running in the vicinity of 3\,GeV and we used the results with this intermediate scheme to estimate the truncation error of the matching to the $\msbar$-NDR scheme. 

\begin{figure}[t]
\includegraphics[width =0.4\textwidth]{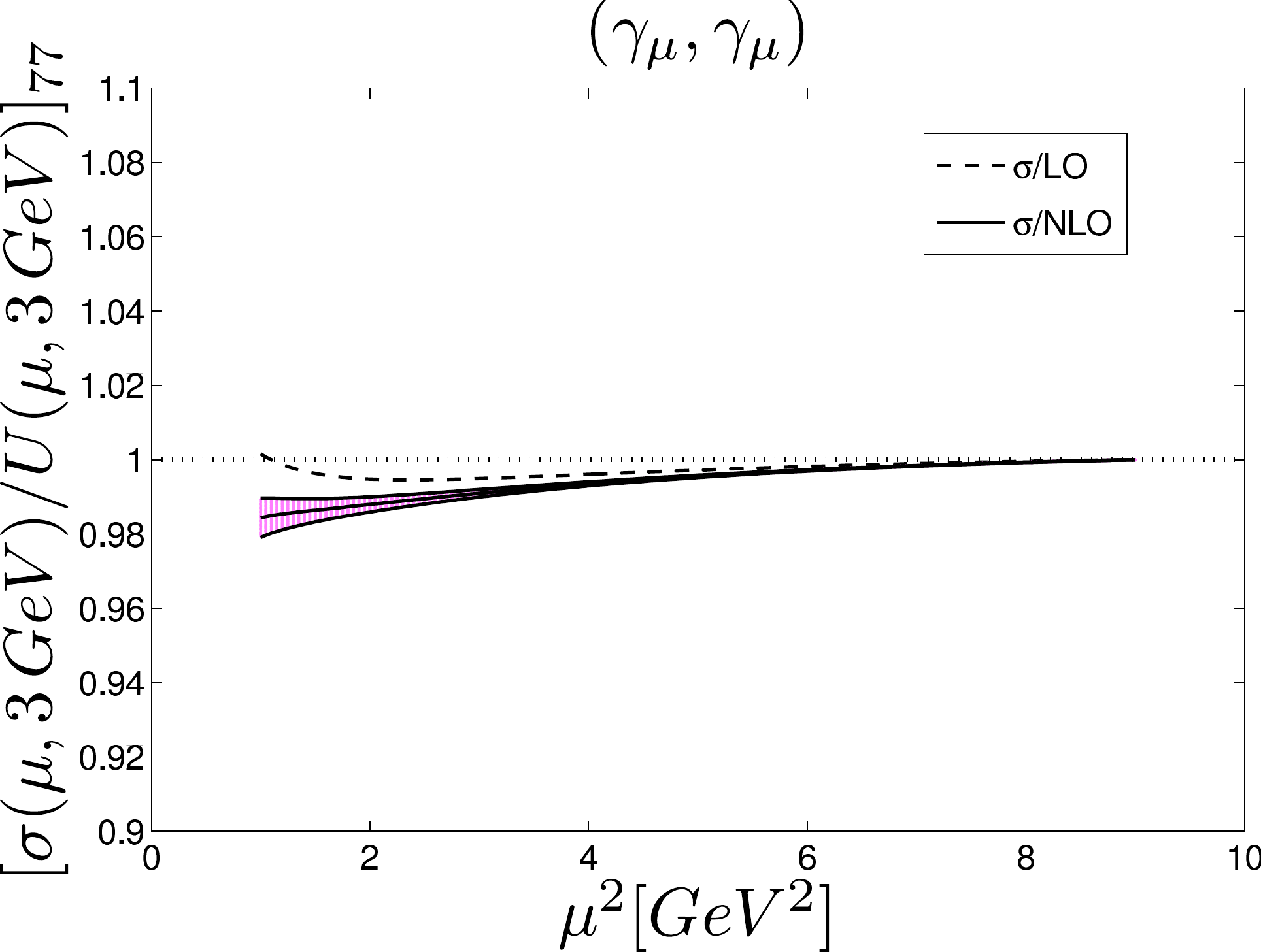}\qquad
\includegraphics[width =0.4\textwidth]{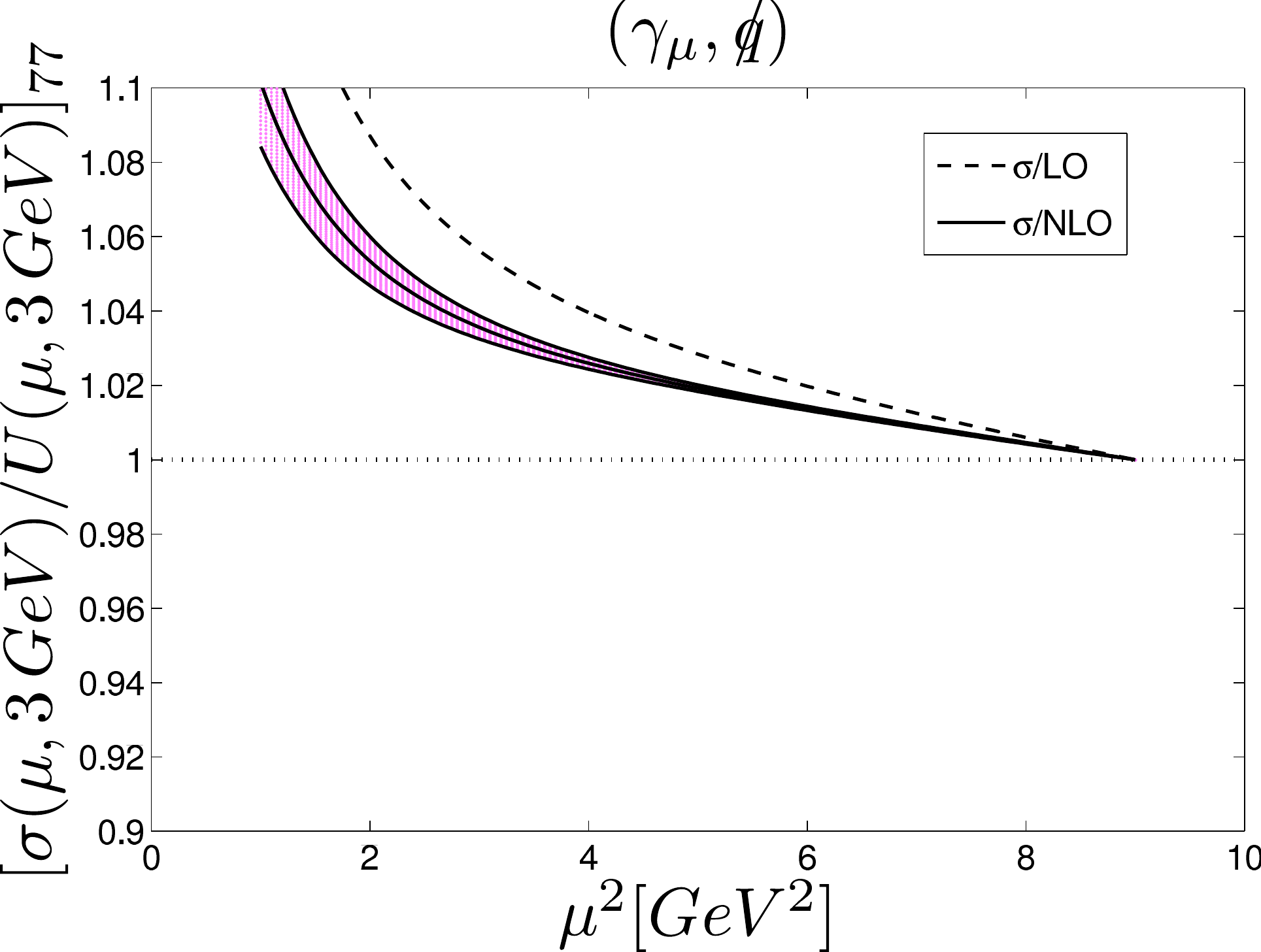}\\ 
\includegraphics[width =0.4\textwidth]{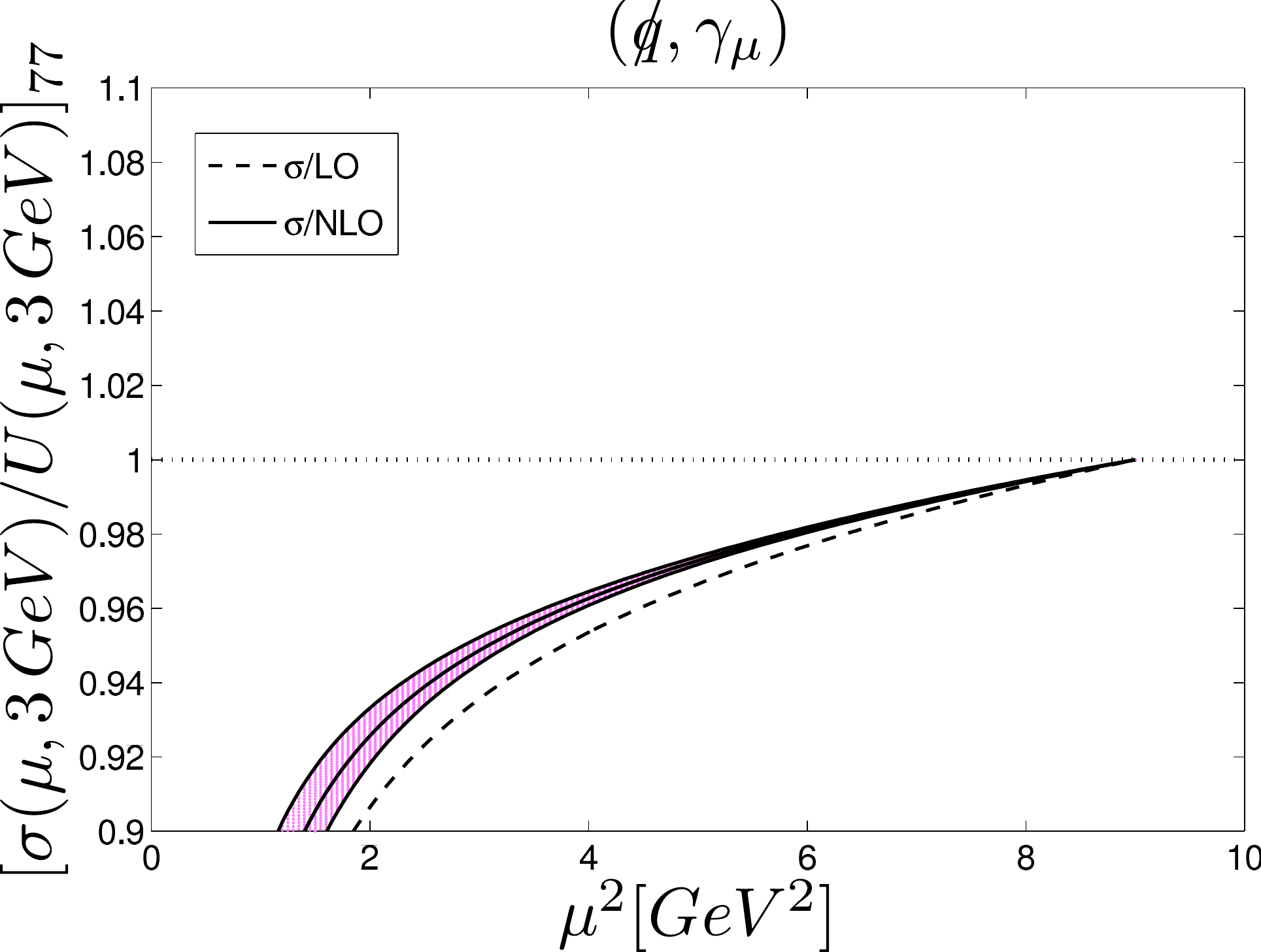}\qquad
\includegraphics[width =0.4\textwidth]{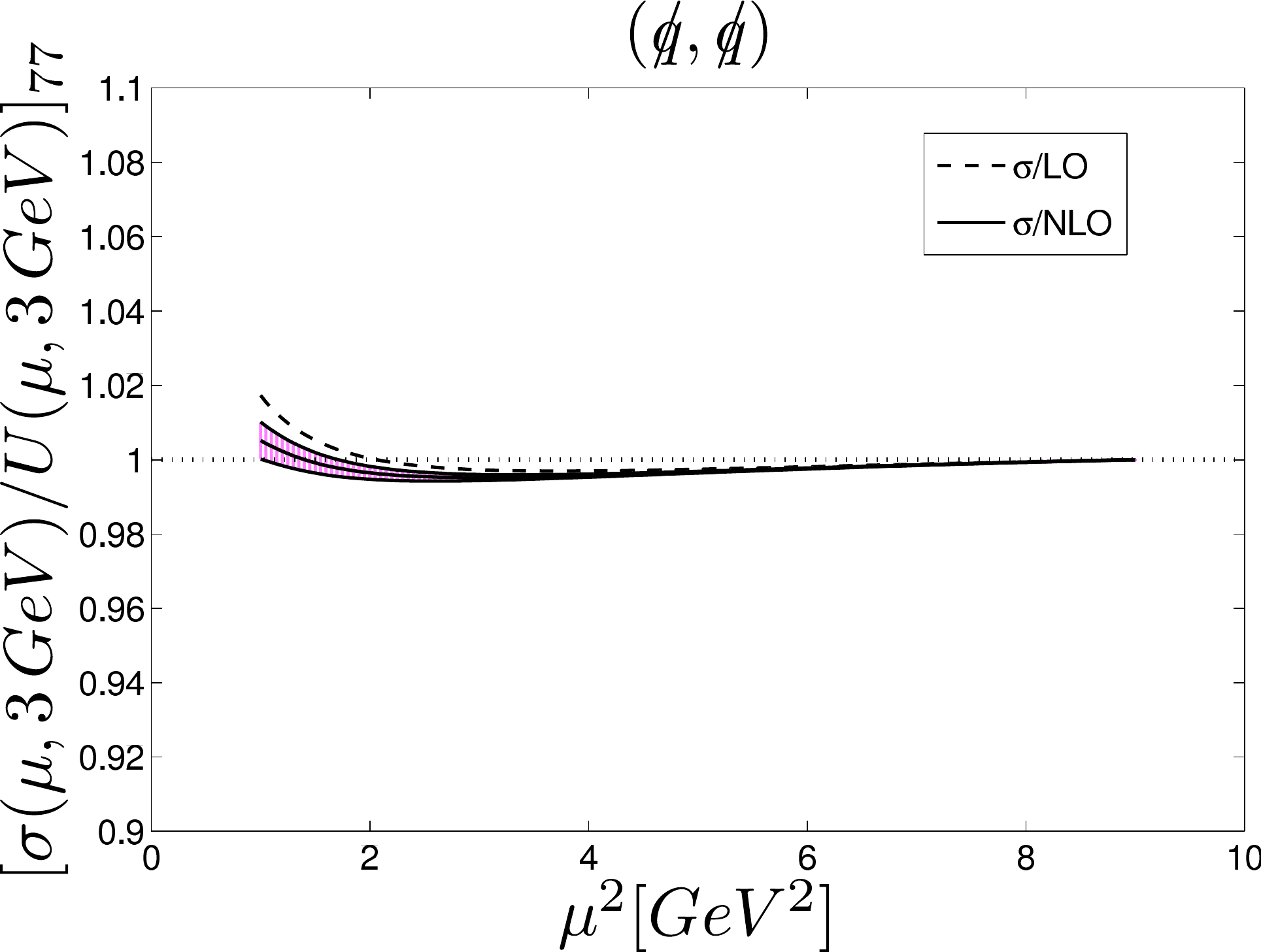}
\caption{\label{fig:sigmapert77} The running of the step-scaling function divided by the LO or NLO perturbative expression for the 77 element of the step scaling function for the four RI-SMOM schemes considered in this paper. The ratio is set to 1 at $\mu=3$\,GeV. The non-perturbative results have been extrapolated to both the chiral and continuum limits.}
\end{figure}

\begin{figure}[t]
\includegraphics[width =0.4\textwidth]{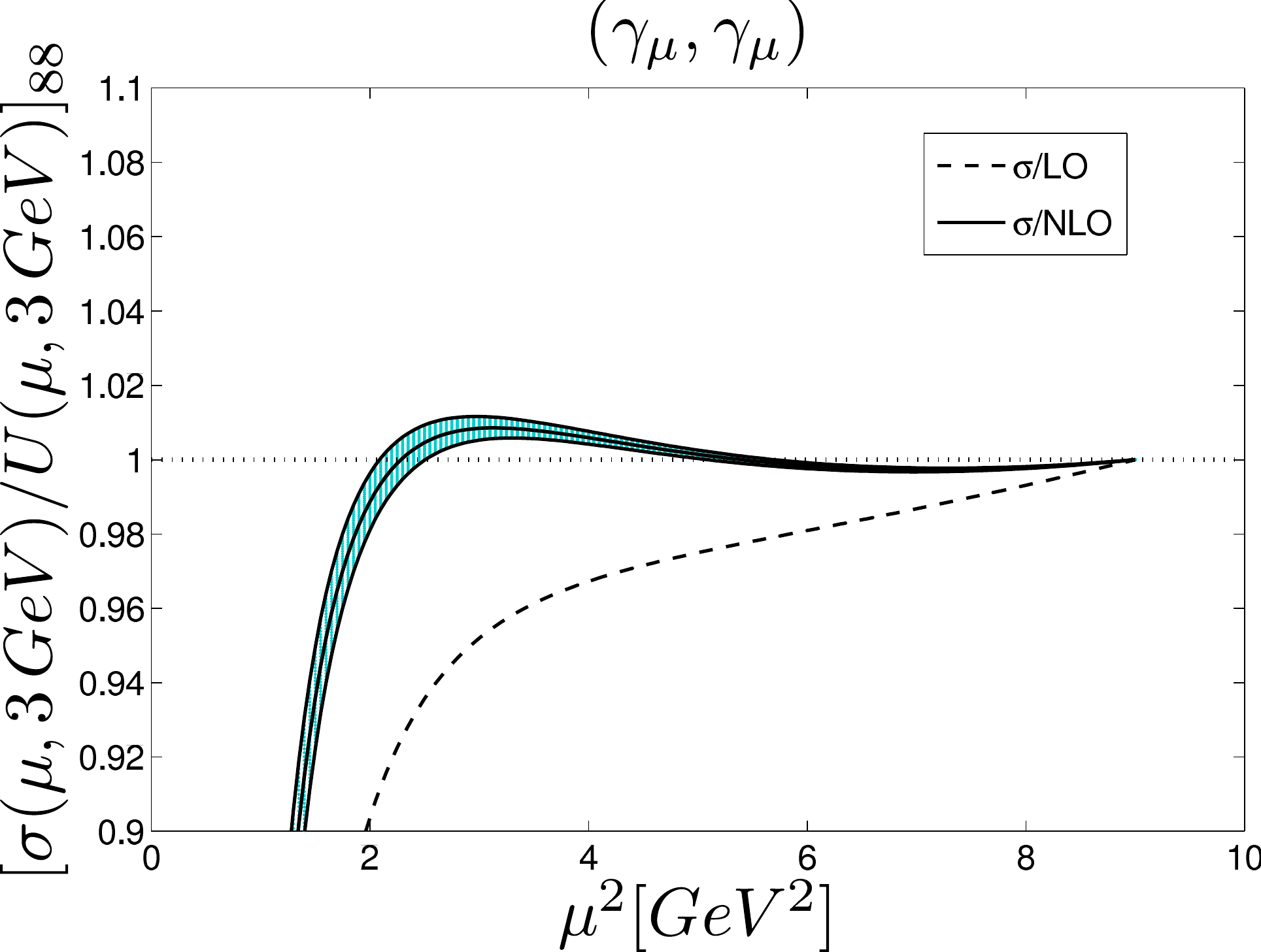}\qquad
\includegraphics[width =0.4\textwidth]{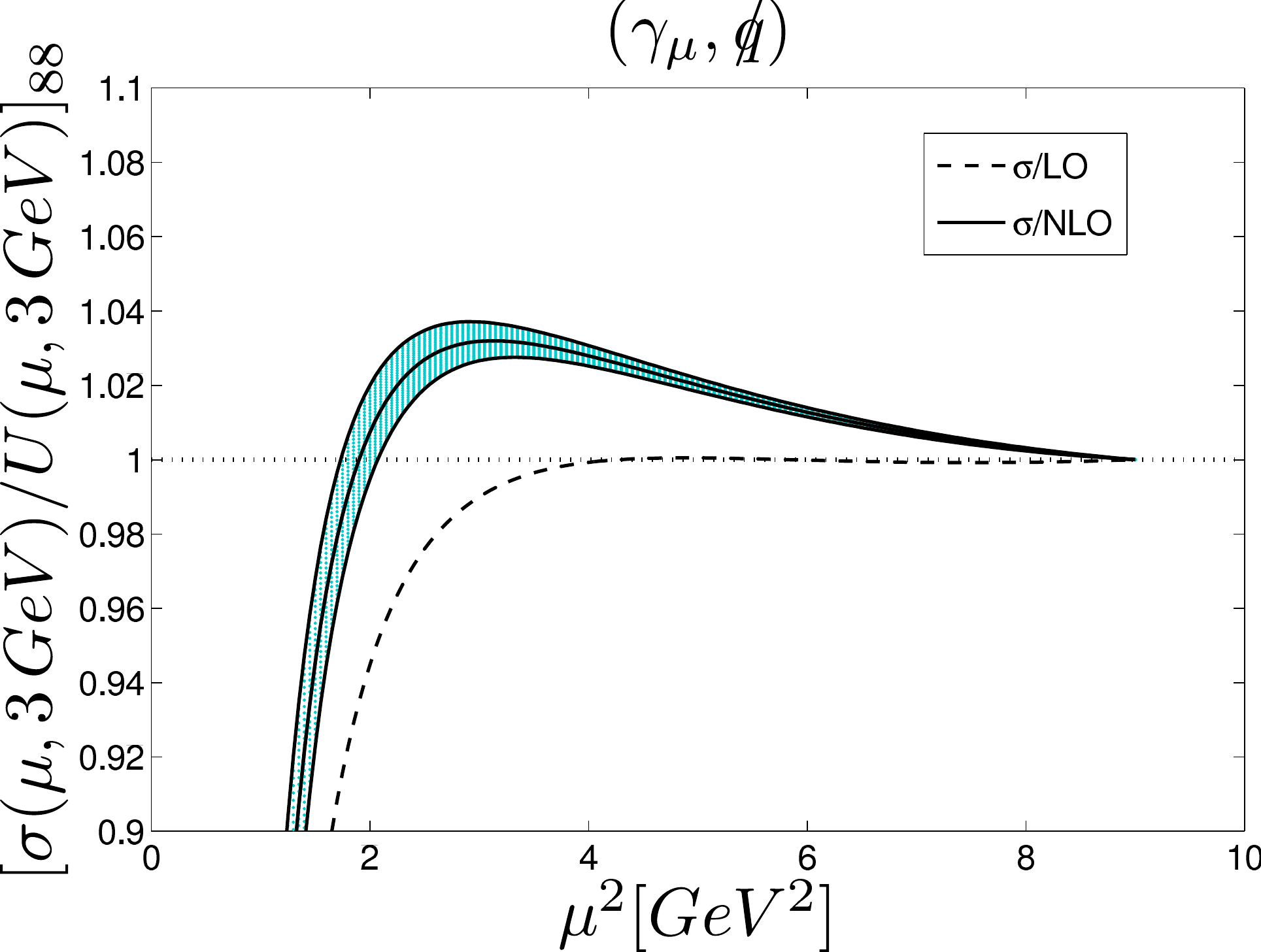}\\ 
\includegraphics[width =0.4\textwidth]{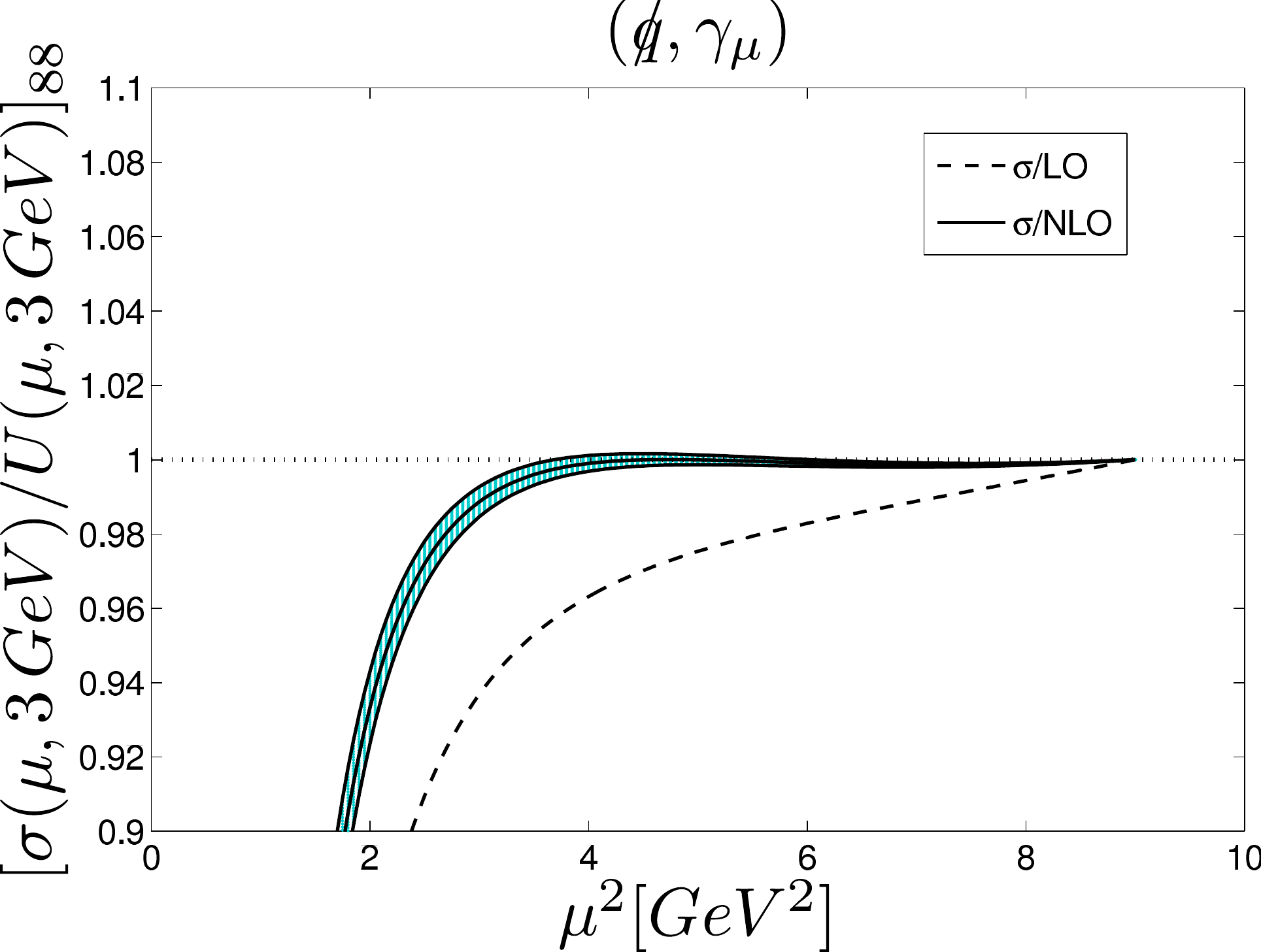}\qquad
\includegraphics[width =0.4\textwidth]{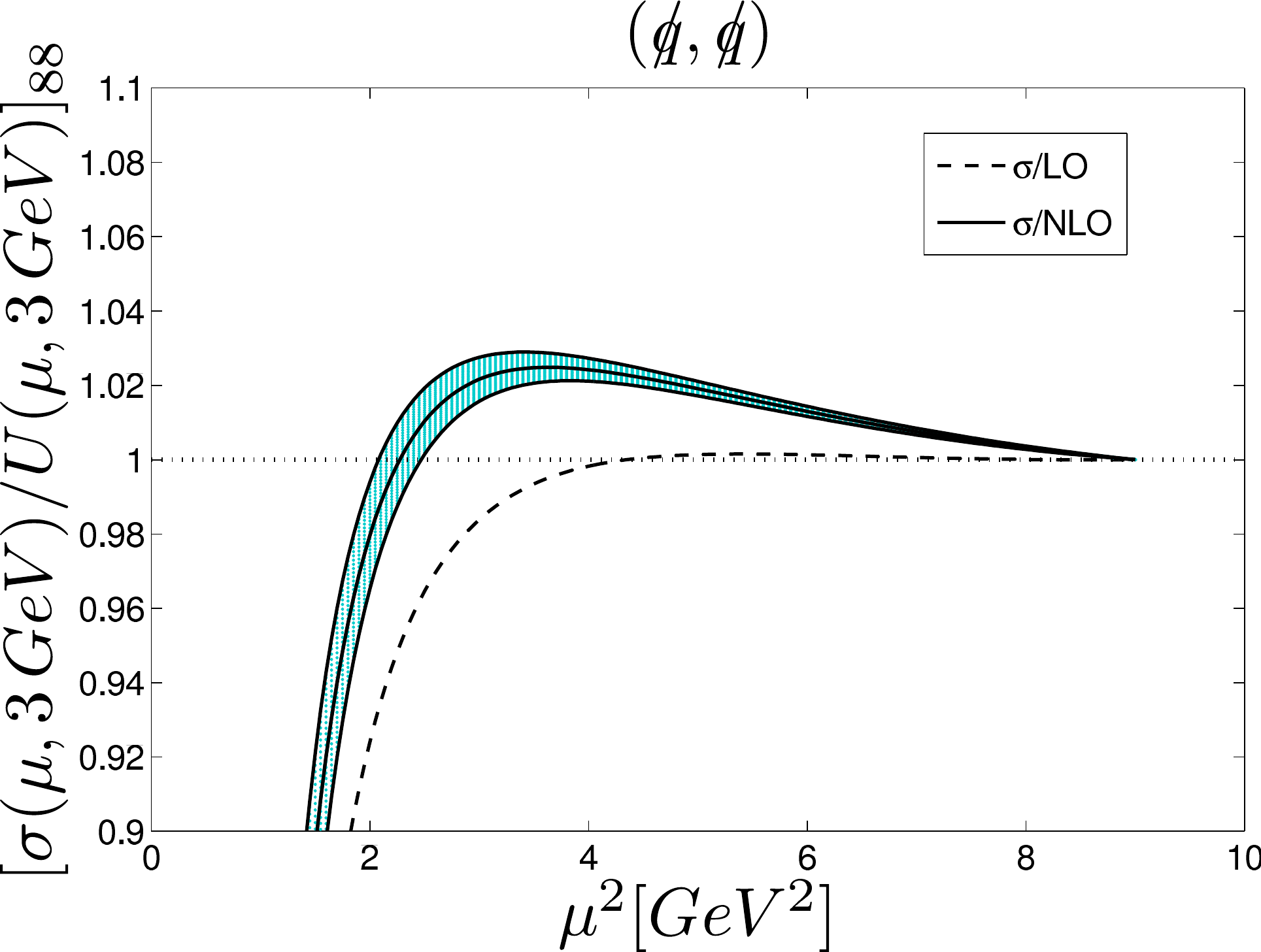}
\caption{\label{fig:sigmapert88} The running of the step-scaling function divided by the LO or NLO perturbative expression for the 88 element of the step scaling function for the four RI-SMOM schemes considered in this paper. The ratio is set to 1 at $\mu=3$\,GeV. The non-perturbative results have been extrapolated to both the chiral and continuum limits.}
\end{figure}

For the electroweak penguin operators, while the numerical details are different, the same general features are also present. For illustration we present the results for the diagonal terms, which are the most important ones, in Figs.\,\ref{fig:sigmapert77} and \ref{fig:sigmapert88} and we follow the same procedure in quoting our central values and systematic errors. More details can be found in ref.\,\cite{Arthur:2011cn}.

\subsubsection{Infrared chiral symmetry breaking effects}\label{subsubsec:ireffects}

We initially impose the RI-SMOM renormalization conditions at the relatively low scale of $\mu_0=1.136$\,GeV where we might expect that infrared effects due to the spontaneous breaking of chiral symmetry may not be negligible, even after the quark masses are set to zero. This does not matter however, since we do not need to introduce perturbation theory until we have run all our results to 3\,GeV. Recall that we have determined the renormalization constants needed to relate the bare lattice operators defined on the coarse IDSDR lattices to the RI-SMOM renormalized operators completely non-perturbatively, including the infrared effects. At 3\,GeV we would expect that these effects are very small, indeed we require this to be the case in order later to apply perturbation theory at this scale. To illustrate that this is indeed the case, we study the size of the ``wrong chirality traces" as we now describe. 

For the purposes of this discussion it is convenient to modify the projection operators defined in Sec.\,\ref{subsec:intermediate} so that the tree-level projections give the identity. The 5 renormalization conditions imposed in Sec.\,\ref{subsec:intermediate} can be written in the schematic form:
\begin{equation}\label{eq:npraux1}
P_{(27,1)}\Lambda^{\mathrm{R}}_{(27,1)}(\mu)=F\quad\mathrm{and}\quad P_j\Lambda_i^{\mathrm{R}}(\mu)=G_{ij}\quad\mathrm{where\ }i,j=7,8\,.
\end{equation}
The $\Lambda^\mathrm{R}$ are the Green functions of the operators renormalized in one of the RI-SMOM schemes at the renormalization scale $\mu$ (the superscript $R$ stands for \emph{Renormalized}), the $P$ are the projectors as defined in Sec.\,\ref{subsec:intermediate} and the constant $F$ and constant $2\times 2$ matrix $G$ correspond to the tree-level values of the traces (with the normalization factor in Eqs.\,(\ref{eq:p271}) and (\ref{eq:p272}) the constant $F=1$, but we leave the value unspecified for this general discussion). We now modify the projectors to 
\begin{equation}\label{eq:pprimedef}
P^\prime_{(27,1)}=\frac{1}{F}\,P_{(27,1)}\quad\mathrm{and}\quad P^\prime_j=(P\,G^{-1})_j\quad\mathrm{for}~j=7,8\,,
\end{equation}
in terms of which the conditions (\ref{eq:npraux1}) on the Green functions read
\begin{equation}\label{eq:npraux2}
P^\prime_{(27,1)}\Lambda^{\mathrm{R}}_{(27,1)}(\mu)=1\quad\mathrm{and}\quad P^\prime_j\Lambda_i^{\mathrm{R}}(\mu)=\delta_{ij}\quad\mathrm{where\ }i,j=7,8\,.
\end{equation}
We now introduce the $3\times 3$ matrix $M_{ij}(\mu)$, where the labels $i,j=1,2,3$ correspond to the three operators (27,1), 7 and 8 respectively;
\begin{equation}
M_{ij}(\mu)\equiv P^\prime_j\,\Lambda^\mathrm{R}_i(\mu)=\begin{pmatrix}
1&a(\mu)&b(\mu)\\ c(\mu)&1&0\\ d(\mu)&0&1
\end{pmatrix}\,.
\end{equation}
At 3\,GeV, as explained above, we require the wrong chirality constants $a(3\,\mathrm{GeV})$, $b(3\,\mathrm{GeV})$, $c(3\,\mathrm{GeV})$ and $d(3\,\mathrm{GeV})$ to be small and indeed this is what we find. For example, in our preferred $(\qslash,\qslash)$ scheme on the $32^3$ Iwasaki lattice in the chiral limit we obtain
\begin{equation}\label{eq:mprime3}
M_{32^3}(3\,\mathrm{GeV})=
\begin{pmatrix}
1&-2\,(2)\!\times\!10^{-5}&2\,(2)\!\times\!10^{-5}\\ 
0\,(2)\!\times\!10^{-5}&1&0\\ 
-4\,(2)\!\times\!10^{-5}&0&1
\end{pmatrix}\,.
\end{equation}
Had the wrong-chirality traces not been small, we would have expected similar infrared effects in the renormalization conditions themselves and not been able to apply perturbation theory at this scale.

At the lower scale of $\mu_0=1.136\,$GeV we expect the wrong chirality traces to be larger and this is indeed the case, although we find that they are actually still small. The key point here is that they are physical and therefore should be the same for all lattices. We find
\begin{equation}\label{eq:mprimemu0idsdr}
M_{\mathrm{IDSDR}}(1.136\,\mathrm{GeV})=
\begin{pmatrix}
1&-0.002\,(2)&-0.004\,(2)\\ 
\ph{-}0.002\,(2)&1&0\\ 
-0.002\,(4)&0&1
\end{pmatrix}\,
\end{equation}
for the IDSDR lattices and
\begin{eqnarray}
M_{24^3}(1.136\,\mathrm{GeV})&=&
\begin{pmatrix}
1&-0.001\,(1)&-0.007\,(1)\\\ 
\ph{-}0.001\,(1)&1&0\\ 
-0.004\,(2)&0&1
\end{pmatrix}
\quad\mathrm{and}\\ [3mm]
M_{32^3}(1.136\,\mathrm{GeV})&=&
\begin{pmatrix}
1&0.000\,(1)&-0.006\,(2)\\\ 
\ph{-}0.003\,(1)&1&0\\ 
-0.008\,(3)&0&1
\end{pmatrix}\label{eq:mprimemu0iw}
\end{eqnarray}
for the two Iwasaki lattices. Within the errors, the results are indeed consistent with our expectations.

\section{Estimating the Error due to Lattice Artefacts}\label{sec:artefacts}

We now begin a detailed examination of the systematic uncertainties leading to the estimates in Tab.\,\ref{tab:errors}. In this section we study the largest single contribution to the systematic uncertainty, that due to the artefacts.

Our calculations of the $K\to\pi\pi$ amplitudes were performed at a single, rather large, value of the lattice spacing, $a^{-1}=1.364(9)$\,GeV. As described earlier, this value of the lattice spacing was obtained in our standard way using the mass of the $\Omega$-baryon to set the scale and the masses of the pion and kaon to determine the physical quark masses. With the action which we are using, all other computed physical quantities have errors of $O(a^2)$, but  without a simulation at a second lattice spacing we cannot determine these lattice artefacts directly. In this section we describe our indirect estimates of the $O(a^2)$ effects.

\begin{table}[t]
\centering
\begin{tabular}{|c|c|c|}\hline
Quantity&  ChPTFV&  Analytic\\ \hline 
$m_{\Omega}$&  1.364(8)\,GeV&  1.362(11)\,GeV\\ 
$f_\pi$&  1.410(27)\,GeV&  1.386(19)\,GeV\\ 
$f_K$&  1.413(29)\,GeV&  1.392(28)\,GeV\\ 
$r_0$&  1.357(4)\,GeV&  1.362(7)\,GeV\\ \hline 
\end{tabular}
\caption{Values of the inverse lattice spacing obtained using different physical quantities to set the scale. For the Sommer scale $r_0$ we use the value $r_0=2.433(50)(18)(13)\,\mathrm{GeV}^{-1}=0.4795(99)(35)(26)$\,fm from our detailed analysis in \cite{dsdrpaper}. The two columns of results correspond to the use of finite-volume SU(2) chiral perturbation theory and the analytic ansatz for the light-quark mass dependence.~\label{tab:different_as}}
\end{table}

We use two (related) methods to estimate the artefacts. In the first of these we imagine using quantities other than $m_\Omega$ to set the scale and observe the corresponding variation which we ascribe to artefacts. The results are presented in Tab.\,\ref{tab:different_as}. The difference between the largest and smallest entry in the table is about 4\%. Recalling that the $K\to\pi\pi$ matrix elements are of dimension 3, we would estimate the corresponding uncertainty in the amplitudes to be 10-15\%.
On the other hand, it could be argued that we don't know the physical value of $r_0$ very well and that we should simply impose that we obtain the same value of $r_0$ on the Iwasaki and IDSDR lattices. This then fixes the ratio of lattice spacings on the two ensembles. Combining this ratio with the well determined lattice spacing on the Iwasaki  ensembles from $m_{\Omega}$ leads to the IDSDR value $a^{-1}=1.363(22)$\,GeV, closer to those obtained from $m_\Omega$ and the decay constants. Although this may suggest that the 10-15\% estimate is conservative, because of the indirect nature of these estimates, we prefer to be conservative when quoting the uncertainties.

As a second approach we set the scale from $m_\Omega$ as usual and study the matrix element 
$M^{\Delta S=2}=\langle\bar{K}^0|(\bar{s}\gamma^\mu(1-\gamma^5)d)\,(\bar{s}\gamma^\mu(1-\gamma^5)d)|K^0\rangle$ on the Iwasaki and IDSDR lattices.  This matrix element gives the dominant contribution to the indirect CP-violation parameter $\epsilon$ and is in the same representation of the chiral symmetry as $Q_{(27,1)}$. We perform global chiral and continuum fits using the form
\begin{equation}
M^{\Delta S=2}=c_0(1+c^{\mathrm{I,IDSDR}}_a\,a^2)+c_l\tilde{m}_l+c_h(\tilde{m}_h-\tilde{m}_{h_0})+c_x\tilde{m}_x+c_{y}(\tilde{m}_y-\tilde{m}_{h_0})\,,
\end{equation}
where $\tilde{m}_l$ and $\tilde{m}_x$ are the sea and valence light-quark masses, $\tilde{m}_h$ and $\tilde{m}_y$ the corresponding strange-quark masses and $\tilde{m}_{h_0}$ is the physical bare strange quark mass. The coefficients $c_a$ depend on the action as indicated. By performing the global fits, $c^{\mathrm{IDSDR}}_a$ can be determined and the size of the lattice artefacts can be determined. Using all our data we find that the artefacts are 12\% in the SU(2) chiral limit and 18\% at the physical quark masses. If we restrict the data to pions with masses less than 350\,MeV, we find artefacts of 10\% in the chiral limit and 14\% for physical quark masses.

Based on these calculations we estimate the uncertainty due to the lattice artefacts as being 15\%, which we combine with the remaining uncertainties in quadrature. This estimate of the discretization error includes possible artefacts in the conversion of the renormalization constants from the IDSDR to the Iwasaki lattices. We stress that while lattice artefacts are the dominant source of systematic uncertainty in the present work, they will be reliably reduced when the calculations are repeated at a second lattice spacing.

\section{Estimating the Error due to Partial Quenching}\label{sec:pq}

The calculations described in this paper were designed to have almost physical kinematics, i.e. the kaon and pions have masses which are close to their physical values. This is achieved however, by the sea and valence quark masses being different; the sea-quark masses are $m_l^{\mathrm{sea}}=0.001$ and $m_h^{\mathrm{sea}}=0.045$ and the valence masses are $m_l^{\mathrm{valence}}=0.0001$ and $m_h^{\mathrm{valence}}=0.049$. Although we do not expect the dependence on the sea-quark to be very significant, in this section we report on some studies to check this. We start by describing an investigation of the sea-quark mass dependence performed with the $32^3$ Iwasaki lattices and in Subsec.\,\ref{subsec:pqreweighting} we report on the results obtained by reweighting $m_l^{\mathrm{sea}}$ from 0.001 to the valence value of 0.0001.  Note that as the bare mass decreases from 0.001 to 0.0001, $m_l+m_{\mathrm{res}}$ decreases by a relatively smaller ratio, from 0.0028 to 0.0019.

\subsection{Sea-quark mass dependence on the $32^3$ Iwasaki lattices}\label{subsec:pqiwasaki}
\begin{table}[t]
\begin{center}\begin{tabular}{|c|c|c|c|} \hline
& $m_l=0.004$ & $m_l=0.006$ & $m_l=0.008$ \\
\hline
Re(A$_2$)$\times 10^8$ GeV & 0.697(44)& 0.748(41) & 0.719(38)\\
$\text{Im(A}_2)\times 10^{13}$ GeV &-14.73(37) & -14.99(35) & -15.23(34) \\
\hline
\end{tabular}\end{center}
\caption{\label{tab:Iwasaki_results} The amplitude $A_2$, computed on the Iwasaki ensembles, after extrapolation to physical kaon and pion masses. The two pions in the final state are at rest (up to finite-volume effects) and energy is not conserved in these amplitudes (see text).}
\end{table}

$K \rightarrow \pi\pi$ correlation functions were also computed on the $32^3\times 64$, $L_s=16$ Iwasaki lattices ($a^{-1} = 2.285(29)$~GeV) with three different light sea-quark masses $m_l^{\text{sea}} = 0.004, 0.006, 0.008$~\cite{Lightman:2009ka, Lightman:thesis}. For each of the sea-quark masses, the correlation functions were calculated using several valence masses: $m^{\text{valence}} = 0.002, 0.004, 0.006, 0.008, 0.025, 0.03$. Periodic boundary conditions were used, so the pions have zero momentum,
resulting in a decay which does not conserve energy. For each of the three sea-quark masses, a chiral extrapolation was performed over the valence masses to determine the $K \rightarrow \pi\pi$ amplitudes corresponding to physical kaon and pion masses (for the strange quark in the kaon this was an interpolation). The results are summarised in Tab.\,\ref{tab:Iwasaki_results}.

From the table we see that any dependence on the light sea-quark mass is small, and generally within the statistical uncertainties. As an estimate of the uncertainty we take the 
standard deviation of the results obtained with the different sea light-quark masses; $3.5\%$ for Re(A$_2$) and
$1.7\%$ for Im(A$_2$). 
Although the kinematics are different from those for the physical decay on the IDSDR lattice, we take this to be an estimate of the error due to partial quenching. The range of sea-quark masses on the Iwasaki lattice and the long length of the extrapolation suggest that this may be a conservative estimate. 
We do not attempt to estimate the error due to the partial quenching of the strange quark, but note that the deviation from unitarity in the strange-quark mass is relatively small ($m_h^{\mathrm{sea}}=0.045$ compared to $m_s^{\mathrm{valence}}=0.049$) .

\subsection{Reweighting the light sea quarks}\label{subsec:pqreweighting}

The technique of reweighting allows us to change the sea-quark masses \emph{a posteriori}, i.e. after the generation of the configurations~\cite{Hasenfratz:2008fg}, albeit at a loss of statistical precision. It is commonly used to correct for any difference between the simulated and physical strange-quark masses, see for example 
\cite{Aoki:2010dy}. Here we reweight the light-quark mass in order to investigate the effects of its partial quenching.
\begin{figure}[t]
\centering
\subfigure[~Reweighting Re\,A$_2$\label{fig:rwReA2}]{\includegraphics[width=0.45\textwidth]{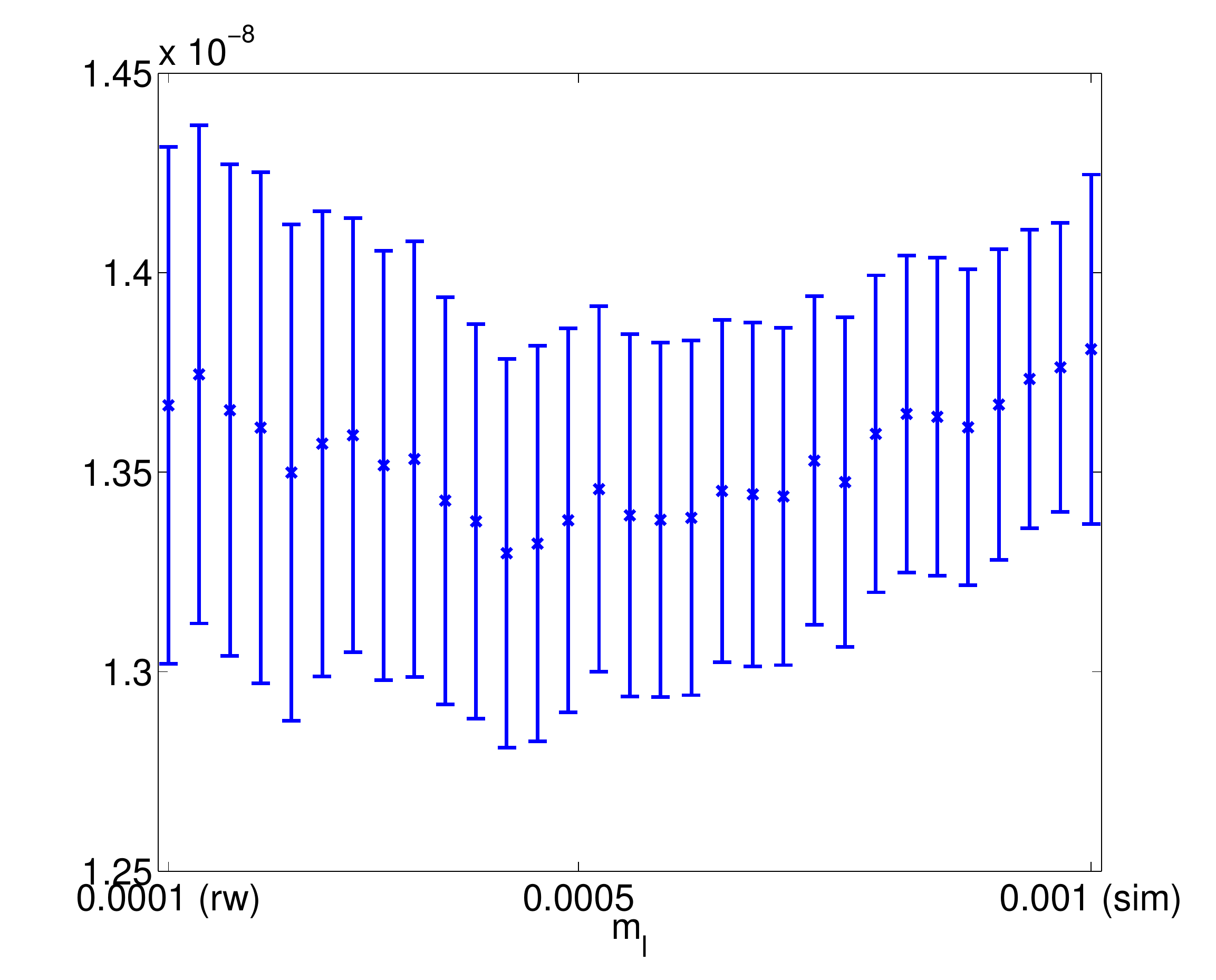}}
\subfigure[~Reweighting Im\,A$_2$\label{fig:rwImA2}]{\includegraphics[width=0.45\textwidth]{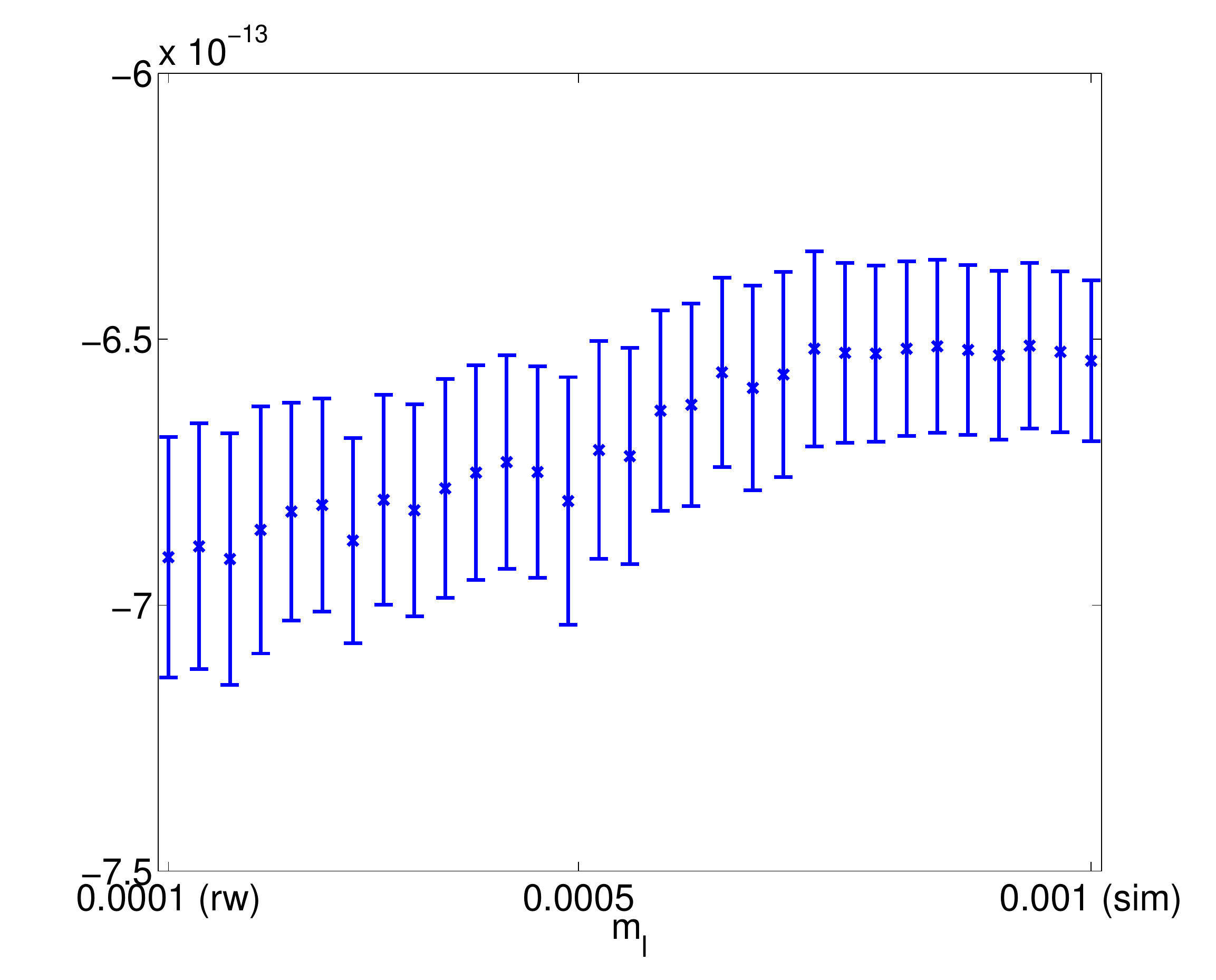}}
\caption{ Reweighting $\text{A}_2$ from $m_l^{\mathrm{sea}}=0.001$ to $m_l^{\mathrm{sea}}=0.0001$. 
\label{fig:rw}}
\end{figure}

The reweighting is performed in 30 increments from the simulated mass $m_l^{\text{sea}}=0.001$ down to a value of $m_l^{\text{sea}}=0.0001$ which corresponds to the valence light-quark mass and the results 
are shown in Fig.\,\ref{fig:rw}.
The rightmost point in Fig.\,\ref{fig:rw}(a) shows the result for Re\,$A_2$ before reweighting, while the remaining points show the results after reweighting to the mass indicated on the $x$-axis, ending with $m_l^{\mathrm{sea}}=0.0001$ for the leftmost point. 
Similarly Fig.\,\ref{fig:rw}(b) shows the effects of reweighting on Im\,$A_2$.   The final results after reweighting are shown in Tab.\,\ref{tab:rw_A2} where they are compared with the results before reweighting. In this table, for illustration of the effects of reweighting, we only quote the statistical error from the correlation functions themselves; we do not include the statistical errors from the determination of the lattice spacing or renormalization or any of the systematic errors.

\begin{table}[t]
\centering
\begin{tabular}{|c|c|c|}
\hline
 & $m_l = 0.001$ & $m_l = 0.0001$ (reweighted) \\ \hline
Re\,$A_2$& $1.381(38)\times 10^{-8}$ GeV& $1.367(65)\times 10^{-8}$ GeV \\
Im\,$A_2$ & $-6.54(15)\times 10^{-13}$ GeV& $-6.91(23)\times 10^{-13}$ GeV\\
\hline
\end{tabular}
\caption{\label{tab:rw_A2} $A_2$ before and after reweighting. The quoted errors correspond to the statistical fluctuations in the correlation functions only. The statistical uncertainties in the determination of the lattice spacing and non-perturbative renormalization have been omitted here.}
\end{table}

Examining the figures, it can be seen that, as expected, the statistical errors on Re\,$A_2$ and Im\,$A_2$ grow.
Table \ref{tab:rw_A2} shows that the real part of $A_2$ remains unchanged whereas the central value of the imaginary part decreases by 5.7\% which is more than the 1.7\% estimated in Sec\,\ref{subsec:pqiwasaki} which we take to be our main estimate of the error due to partial quenching. 
An alternative approach would be to eliminate the systematic error due to partial quenching by using
the reweighted values for our final results. In doing this the systematic errors on Re\,$A_2$ and Im\,$A_2$ are unchanged at 18\% and 19\% respectively.
Using the reweighted values, we would obtain the following results for the complex amplitude $A_2$:
\begin{equation}\label{eq:rwresults}
\textrm{Re}\,A_2=1.367(70)_{\textrm{stat}}(246)_{\textrm{syst}}\,10^{-8}\,\textrm{GeV},
\quad\textrm{Im}\,A_2=-6.91(51)_{\textrm{stat}}(131)\,_{\textrm{syst}}10^{-13}\,{\rm GeV}\,.
\end{equation}
The results of Eq.\,(\ref{eq:rwresults}) should be compared with Eq.\,(\ref{eq:results}), and it is clear that the differences due to reweighting
are well within the total error.

\section{Error Budget}\label{sec:errors}

\begin{table}[t]
\begin{center}
\begin{tabular}{|c|c|c|}
\hline
& $\textrm{Re}A_2$ & Im$A_2$ \\
\hline
lattice artefacts& 15\% & 15\% \\
finite-volume corrections& 6.0\%& 6.5\%\\
partial quenching & 3.5\%& 1.7\%\\
renormalization & 1.8\%& 5.6\%\\
unphysical kinematics & 0.4\%& 0.8\%\\
derivative of the phase shift&0.97\%& 0.97\%\\
Wilson coefficients &6.6\%& 6.6\% \\
\hline
Total & 18\%& 19\%  \\
\hline
\end{tabular}
\caption{Systematic error budget for Re\,$A_2$ and Im\,$A_2$. \label{tab:errors}}
\end{center}\end{table}

The sources of systematic error in the calculation of Re\,$A_2$ and Im\,$A_2$ include those from
lattice artefacts, finite-volume effects, partial quenching, the uncertainty in the non-perturbative renormalization, 
the unphysical kinematics used in the calculation, the determination of the derivative of the phase shift and the Wilson coefficients. Although some of these uncertainties have been estimated in previous sections (NPR in Sec.\ref{sec:npr}, lattice artefacts in Sec.\ref{sec:artefacts} and partial quenching in Sec.\ref{sec:pq}), here we summarise the conclusions of sections\,\ref{sec:npr}-\ref{sec:pq} and briefly
discuss the remaining sources of uncertainty before finally combining them all into a total systematic error. The results can be found in Tab.\,\ref{tab:errors}.

\subsection{Lattice Artefacts}\label{subsec:artefacts}
The estimate of the systematic error due to lattice artefacts is described in Sec.\ref{sec:artefacts} and was estimated to be 15\%. Comparing this with the other errors in Tab.\,\ref{tab:errors}, we see that lattice artefacts are the dominant source of systematic error. They would be very significantly reduced by repeating the calculation at a second value of the lattice spacing.

\subsection{Finite-Volume Corrections}\label{subsec:fv}
In order to estimate the systematic error due to the finite volume of the lattice, we use 
SU(3) finite-volume chiral perturbation theory, in which the loop-integrals in Feynman diagrams are replaced by discrete sums over the allowed momenta. Expressions for the $\Delta I=3/2$ $K\to \pi\pi$ matrix elements, 
${\cal M}_{(27,1)} = \left \langle \pi^+ \pi^- | Q_{(27,1)} |K^0 \right \rangle$
and
$\mathcal{M}_{(8,8)} = \left \langle \pi^+ \pi^- | Q_{(8,8)} |K^0 \right \rangle$ 
are known to next-to-leading order
in $SU(3)$ chiral perturbation theory. Since in chiral perturbation theory to leading order there is a single $\Delta I=3/2$ operator constructed from the Goldstone boson fields which transforms as the (8,8) representation, the estimates derived below are the same for $Q_{(8,8)}$ and $Q_{(8,8)_\mathrm{mix}}$. There is also a single operator at lowest order which transforms as the (27,1) representation.
We will be considering the leading order terms (labelled by ``LO'')  and leading (one-loop) logarithmic terms (labelled by ``log''). 
The LO expressions are well known and can be found in \cite{Laiho:2002jq} and \cite{Aubin:2008vh}.  For $\mathcal{M}_{(27,1)}^{\log}$ we use Eq.\,(C5) in \cite{Laiho:2002jq}, (where we have added logarithmic terms from $(m_K^2-m_{\pi}^2)_{\text{1-loop}}$ by hand as necessitated by Eq.\,(\ref{eq:results}) and 
corrected a factor of $1/f^2$ in equation (A2)), and for
$\mathcal{M}_{(8,8)}^{\log}$ we use Eq.\,(E3) in \cite{Aubin:2008vh}.

We denote the finite-volume corrections to the logarithmic terms in $\mathcal{M}_{(27,1)}$ and $\mathcal{M}_{(8,8)}$ by $\Delta \mathcal{M}^{\log}_{(27,1)}$ and  $\Delta \mathcal{M}^{\log}_{(8,8)}$ respectively.  
We estimate the relative size of these  corrections, by using the pion and kaon masses in our 
lattice calculation finding,
\begin{equation}\label{eq:M_fracLO}
 \frac{\Delta \mathcal{M}^{\log}_{(27,1)}}{\mathcal{M}^{\text{LO}}_{(27,1)}} = 0.0597
\qquad\mathrm{and}\qquad \frac{\Delta \mathcal{M}^{\log}_{(8,8)}}{\mathcal{M}^{\text{LO}}_{(8,8)}} = 0.0649
\end{equation}
if we normalize to the leading order expressions of the matrix elements, and
\begin{equation}
\label{eq:M_fracLog}
 \frac{\Delta \mathcal{M}^{\log}_{(27,1)}}{\left \vert \mathcal{M}^{\text{LO}}_{(27,1)}+\mathcal{M}^{\log}_{(27,1)}\right \vert} = 0.0352
\qquad\mathrm{and}\qquad 
 \frac{\Delta \mathcal{M}^{\log}_{(8,8)}}{\left \vert \mathcal{M}^{\text{LO}}_{(8,8)}+\mathcal{M}^{\log}_{(8,8)}\right \vert} = 0.0438
\end{equation}
if we normalize to the leading order plus leading logarithmic expressions.  More details can be found in \cite{Lightman:thesis}.

Evidently the leading logarithmic terms make significant corrections to the leading order terms.
To have confidence that the chiral perturbation theory is converging we should check the size of
the next-to-leading-order terms, but as these have unknown coefficients we are unable to make  
a numerical estimate. We therefore make a 
conservative estimate by taking the larger relative finite-volume correction of Eq.\,(\ref{eq:M_fracLO})  and conclude that the (27,1) operator carries a 6.0\% finite-volume correction
and that the (8,8) operator carries a 6.5\% finite-volume correction. Since Re\,$A_2$ is
dominated by the (27,1) operator and  
Im\,$A_2$ is dominated by the $(8,8)_\mathrm{mix}$ operator, these are the percentage errors due to finite-volume effects we assign to Re\,$A_2$ and Im\,$A_2$ respectively.

\subsection{Partial Quenching}
The effects of partial quenching have been discussed in detail in section \ref{sec:pq}. 
Here we simply remind the reader that we neglect any systematic error due to 
partial quenching of the heavy-quark and attribute a $3.5\%$ error to Re\,$A_2$ and 
a $1.7\%$ error to Im\,$A_2$ due to the partial quenching in the light-quark sector
of this calculation.

\subsection{Uncertainties due to the Renormalization}

We consider two sources of systematic error from the calculation of the renormalization  constants. The
first is designed to take into account lattice artefacts of higher order
than $\mathcal{O}(a^2)$ in the
continuum extrapolation of the step-scaling function using the Iwasaki lattices, as described in section \ref{subsec:stepscaling}, 
and corresponds to the second error in equation\,(\ref{eq:Z_MSbar_results}).
This systematic error is estimated in the same way that the
statistical NPR error on $A_2$ is calculated, i.e. Eq.\,(\ref{eq:DeltaZdef}) is used, 
but in this case $\delta Z$ denotes the systematic errors on the Z-factors. 
The resulting error is displayed in Tab.\,\ref{tab:A2_schemes} and is labelled NPR-sys. We find this to be a 
1.1\% effect for Re\,$A_2$ and a 5.0\% effect for Im\,$A_2$ (see the second row of the table).

The second source of systematic error in the renormalization constants is due to the 
truncation error in the perturbative matching 
to the $\overline{\text{MS}}$ scheme and to $\mathcal{O}(a^2)$ scaling errors since we only have one lattice spacing and the 
Z-factors in the different schemes need not approach the continuum limit along the same scaling
trajectory. Following conversion to the $\overline{\text{MS}}$ scheme, the four intermediate NPR schemes
described in Sec.\ref{subsec:intermediate} 
should give equivalent answers. We estimate the resulting systematic error by considering the spread in results when $A_2$ is calculated in the 
$\text{RI-SMOM}(\gamma_{\mu}, \gamma_{\mu})$ scheme and in the $\text{RI-SMOM}(\qslash, \qslash)$ scheme.

The results for $A_2$ in the $\text{RI-SMOM}(\gamma_{\mu}, \gamma_{\mu})$ and $\text{RI-SMOM}(\slashed{q}, \slashed{q})$
schemes are presented in Tab.\,\ref{tab:A2_schemes}. 
We observe a 1.4\% spread for Re\,$A_2$ and a 2.5\% spread for Im\,$A_2$.
Combining the two sources of error in quadrature, we find a 1.8\% error for Re\,$A_2$ and
a 5.6\% error for Im\,$A_2$.

\begin{table}[t]
\centering
\begin{tabular}{|l|c|c|}
\hline
& Re\,$A_2$ $\times 10^{8}$ GeV& Im\,$A_2$ $\times 10^{13}$ GeV\\
\hline
RI-SMOM($\qslash,\qslash$) &$1.381(46)_{\text{stat}}(15)_{\text{(NPR-sys)}}$&  $-6.54(46)_{\text{stat}}(33)_{\text{(NPR-sys)}}$\\
 RI-SMOM($\gamma_{\mu}, \gamma_{\mu}$) & 
$1.362(44)_{\text{stat}}(03)_{\text{(NPR-sys)}}$& $-6.35(34)_{\text{stat}}(42)_{\text{(NPR-sys)}}$ \\ 
\hline
\end{tabular}
\caption{\label{tab:A2_schemes} Re\,$A_2$ and Im\,$A_2$ calculated in the two different schemes.}
\end{table}

\subsection{Uncertainties due to the Unphysical Kinematics}
When choosing the parameters of the simulation, including the quark masses, the coupling constant and even the volume, we aim to obtain physical kaon and pion masses and $E_{\pi\pi}=m_K$. Once the simulation has been performed, we naturally find that this is not quite the case (see Tab.\,\ref{tab:masses}) and we now attempt to estimate the systematic error that
these non-physical kinematics contribute to our calculation.

In addition to the results from the current simulation, we have a large collection of $K\to\pi\pi$ amplitudes calculated on quenched lattices
with a variety of light and strange-quark masses and pion momenta. We use the observed dependence of the amplitudes with the parameters to estimate our uncertainty due to the unphysical kinematics. On the quenched lattices we
have a total of 60 values for the  $K\to\pi\pi$ amplitudes, obtained with all combinations of
$am_l = 0.0023, \, 0.0047, \, 0.0071$, $am_s = 0.046, \, 0.062, \, 0.078, \, 0.094, \, 0.110$
and with
$n = 0, \, 1, \, 2 \text{ and }3$, where $n$ is the number of spatial directions in which antiperiodic boundary conditions are imposed. $n$ parametrizes the pion momenta as briefly explained in Sec.\ref{subsec:sources}. 

The procedure for estimating the systematic error due to non-physical kinematics uses
these quenched amplitudes, extrapolating the results in $am_l$ and interpolating them in  
$a m_s$ and $n$, first to physical
kinematics, and then to the kinematics simulated on the IDSDR lattices. This procedure is described in
detail in \cite{Lightman:thesis}, and is very similar to the extrapolation procedure described
in section \ref{subsec:pqiwasaki} when computing the error due to partial quenching. The difference here is that we can now interpolate to the 
correct pion-momenta. This is achieved by fitting the 
two-pion energy as a function of $n$, and interpolating to find $n^{\text{phys}}$, the value of $n$ which corresponds to the desired two-pion energy.
This in turn allows the decay amplitude to be interpolated and evaluated at $n^{\text{phys}}$. 

For the extrapolation to physical kinematics we find
from the quenched lattices:
\begin{equation}
\label{eq:quenched_phys}
\text{Re}\,A_2 = 2.25\times10^{-8}\,\text{GeV}, \quad \text{Im}\,A_2 = -13.45 \times10^{-13}\,\text{GeV} \,,
\end{equation}
 while the extrapolation to $m_{\pi}$, $m_K$ and $E_{\pi\pi}$ simulated in this article gives
\begin{equation}
\label{eq:quenched_sim}
\text{Re}\,A_2 =  2.26\times10^{-8}\,\text{GeV},\quad \text{Im}\,A_2 = -13.56 \times10^{-13}\,\text{GeV} \,.
\end{equation}
We take the percentage differences between the two extrapolations
as a measure of the systematic error due to simulating at non-physical kinematics, and find
0.4\% for $\text{Re}\,A_2$ and 0.8\% for  $\text{Im}\,A_2$.

\subsection{Uncertainty in the Derivative of the Phase Shift}
The derivative of the s-wave phase shift $\partial \delta/\delta k$ appearing in the Lellouch-L\"uscher
factor was found by evaluating the derivative of the phenomenological curve at the momentum simulated
in our lattice calculation. This was discussed in section \ref{sec:analysis} and illustrated in Fig.\,\ref{fig:phase_shift}. Alternatively we could have made a crude estimate of the derivative by taking
the slope of the straight line between the phase shift at 17.63~MeV and 196.8~MeV. 
(c.f. the results of Tab\,\ref{tab:p_delta}). 
We estimate the systematic error to be 0.97\%, which we find by calculating the percentage difference between the final 
results as obtained by the two different approaches. Since the derivative of the phase-shift only contributes a small fraction 
to the Lellouch-L\"uscher factor (see Tab.\,\ref{tab:derivs}) it is not surprising that the corresponding error is negligible. 
We note also that the derivative of the phase-shift can be calculated directly using the method proposed in \cite{Kim:2010sd}.

\subsection{Uncertainties in the evaluation of the Wilson coefficients}
The Wilson coefficients, which are calculated in perturbation theory and hence are not part of our lattice computations, are a necessary ingredient in the determination of the amplitude $A_2$. The values presented in 
Tab.\,\ref{tab:wilson} were calculated at next-to-leading order (NLO) following the procedure outlined in
\cite{Buchalla:1995vs}. In this section we estimate the systematic error due to the truncation of perturbation theory. To this
end we calculate the Wilson coefficients to leading order (LO), following the procedure in \cite{Buchalla:1995vs} and measure the effect this has on the final results for 
$\rm{Re}\,A_2$ and $\rm{Im}\,A_2$. The LO contribution to the Wilson coefficients is defined according to the following procedure:
\begin{enumerate}
 \item\label{it:alphaLO} A value is chosen for the $\Lambda$ parameter of four-flavor QCD. In ref.\,\cite{Buchalla:1995vs} a range of values from 215\,MeV to 435\,MeV was used. In this paper we use the value of 328\,MeV, which is close to the value corresponding to $\alpha_s(M_Z)=0.1184$\cite{Nakamura:2010zzi}.
 \item In setting the initial conditions for the Wilson coefficients at the scale of the W mass, 
corrections of $O(\alpha)$ and $O(\alpha_s)$ are only included when they depend on the top-quark mass.
This also applies when calculating the coefficients $z_i$ at the scale of the charm mass (Eq.(VII.17) in \cite{Buchalla:1995vs}).
\item In the QCD running to lower energies the one-loop expressions for the anomalous dimension matrix and $\beta$-function
are used. In the presence of electromagnetic interactions, the LO anomalous dimension matrix also includes the term 
$\dfrac{\alpha}{4\pi}\gamma_e^{(0)}$.
\item At leading order the Wilson coefficients are continuous when crossing quark-mass thresholds.
\end{enumerate}

Tab.\,\ref{table:amps_with_LO_wilson} shows how the decay amplitude varies when the LO Wilson coefficients are used instead of
the NLO Wilson coefficients. The error in $A_2$ due to the truncation in the perturbative calculation of the 
Wilson coefficients is very conservatively estimated by taking the difference
between the NLO result and the LO result, and calculating this as a percentage of the LO result. 
We find the error to be $7.1\%$ for $\rm{Re}\,A_2$ and $8.1\%$ for $\rm{Im}\,A_2$.

\begin{table}[t]
 \centering
\begin{tabular}{|c|c|c|}
\hline
 & LO & NLO \\
\hline
$\rm{Re}\,A_2$ & 1.289(42)$\times 10^{-8}$ GeV & 1.381(46)$\times 10^{-8}$ GeV\\
$\rm{Im}\,A_2$ & -6.11(36)$\times 10^{-13}$ GeV & -6.54(46) $\times 10^{-13}$ GeV\\
\hline
\end{tabular}
\caption{\label{table:amps_with_LO_wilson}$\rm{Re}(A_2)$ and $\rm{Im}(A_2)$ as calculated with LO Wilson coefficients and NLO Wilson coefficients. The errors quoted here represent the total statistical uncertainty.}
\end{table}

\section{Summary and Conclusions}\label{sec:concs}

In ref.\,\cite{Blum:2011ng} and the present paper we have presented the results of the first \emph{ab initio} calculation of the complex $K\to(\pi\pi)_{I=2}$ decay amplitude $A_2$ and our results can be found in Eq.\,(\ref{eq:results0}). It is very encouraging that our result for Re\,$A_2$ agrees with the known experimental value and we are also able to determine Im\,$A_2$ for the first time. 
The calculation was made possible by the major theoretical advances and technical progress which has been achieved over many years as described in the text above. Much of the important particle physics phenomenology,  including the description of the weak interactions of the quarks in terms of matrix elements of  
specified four-quark operators multiplied by Wilson coefficients, has been understood since the 1970's,
even before the methods of lattice QCD
had been invented.
However, it was only after major advances in lattice techniques that this calculation has become possible.  The good control
of chiral symmetry provided by the 5-dimensional domain wall formulation,
the ability to translate from lattice to continuum normalization of operators
using non-perturbative methods and the finite-volume techniques capable
of creating the proper, interacting $\pi\pi$ final state with the correct
energy are all essential ingredients in the calculation presented here. In addition, improvements in computational algorithms, and teraflops-scale computing resources, enable us to perform the simulations with nearly physical $u$ and $d$ quark masses. 

The error on our result is dominated by lattice artefacts due to the fact that the calculation was performed at a single, rather course, lattice spacing $a^{-1}\simeq 1.4\,$GeV. The most important extension of the calculation of $A_2$ is therefore to repeat it at different values of $\beta$, or at least at a second lattice spacing,
so that the discretization errors can be
essentially eliminated by extrapolating to zero lattice spacing. In addition, since the methods to compute $A_2$ are now well
in hand, more refined calculations using a larger lattice volume and 
physical light-quark masses (for the sea quarks as well as the valence ones) should be possible.
These enhancements to the calculation reported here are well within reach of
the next generation of high performance computers and should reduce the
errors on the result for $A_2$ by nearly an order of magnitude.

Much more challenging but of even greater interest is the application of
these methods to the calculation of the complex $I=0$ amplitude $A_0$.
The calculation of both  $A_0$ and  $A_2$ from first principles will allow a 
direct comparison of $\epsilon^\prime/\epsilon$ with the experimental result,
giving new sensitivity to the search for physics beyond the Standard Model.
The computational framework presented here will also
support the calculation of $A_0$.  However, serious obstacles must be
overcome.   Much larger Monte Carlo samples will be required to remove
the large statistical fluctuations remaining after the contribution
of the vacuum state has been removed.  The device of applying anti-periodic
boundary conditions to a single quark field used in this paper cannot be
used in the case of the $I=0$ $\pi\pi$~state.  More sophisticated
boundary conditions mixing quarks and anti-quarks and an isospin rotation,
the so called {\it G-parity} boundary conditions, must be used instead for both the
valence and the sea quarks.  Exploratory studies \cite{Blum:2011pu} suggest that
obtaining adequate Monte Carlo statistics will be practical with the next
generation of high performance computers and efforts are presently
underway to develop the necessary boundary conditions.  We anticipate
that a complete calculation of CP violation in $K\to\pi\pi$ decay within
the Standard Model will be achieved before the fiftieth anniversary of its
original discovery.

\section*{Acknowledgements}We thank R.\,Arthur for help with generating the non-perturbative renormalization data and A.\,Buras for helpful discussions and support. Critical to this calculation were the BG/P facilities of the Argonne Leadership Computing Facility (supported by DOE contract DE-AC02-06CH11357).  
Also important were the DOE USQCD and RIKEN BNL Research Center
QCDOC computers at the Brookhaven National Lab., the DiRAC facility (supported by STFC grant ST/H008845/1) and the Univ. of Southampton's Iridis cluster (supported by STFC grant ST/H008888/1).
T.B. was supported by U.S. DOE grant DE- FG02-92ER40716,  P.B. and N.G. by STFC grant ST/G000522/1, N.C., C.K., M.L., Q.L. and R.M. by US DOE grant DE-FG02-92ER40699, E.G., A.L. and C.T.S. by STFC Grant ST/G000557/1, C. J., T. I. and A. S. by U.S. DOE contract DE-AC02-98CH10886, T.I by JSPS Grants  22540301 and 23105715 and  C.L. by the RIKEN FPR program.


\begin{thebibliography}{99}
\bibitem{Blum:2011ng}
  T.~Blum, P.~A.~Boyle, N.~H.~Christ, N.~Garron, E.~Goode, T.~Izubuchi, C.~Jung and C.~Kelly {\it et al.},
  Phys.\ Rev.\ Lett.\  {\bf 108} (2012) 141601
  [arXiv:1111.1699 [hep-lat]].
\bibitem{Christenson:1964fg}
  J.~H.~Christenson, J.~W.~Cronin, V.~L.~Fitch and R.~Turlay,
  Phys.\ Rev.\ Lett.\  {\bf 13} (1964) 138.
 \bibitem{Burkhardt:1988yh}
  H.~Burkhardt {\it et al.} [ NA31 Collaboration ],
  Phys.\ Lett.\  {\bf B206 } (1988)  169.
\bibitem{Gibbons:1993zq}
  L.~K.~Gibbons, A.~Barker, R.~A.~Briere, G.~Makoff, V.~Papadimitriou, J.~R.~Patterson, B.~Schwingenheuer, S.~V.~Somalwar {\it et al.},
  Phys.\ Rev.\ Lett.\  {\bf 70 } (1993)  1203-1206.
\bibitem{Fanti:1999nm}
  V.~Fanti {\it et al.} [ NA48 Collaboration ],
  Phys.\ Lett.\  {\bf B465 } (1999)  335-348.
  [hep-ex/9909022].
\bibitem{AlaviHarati:2002ye}
  A.~Alavi-Harati {\it et al.}  [KTeV Collaboration],
  Phys.\ Rev.\  D {\bf 67}, 012005 (2003)
  [Erratum-ibid.\  D {\bf 70}, 079904 (2004)]
  [arXiv:hep-ex/0208007].
\bibitem{Blum:2011pu}
  T.~Blum, P.~A.~Boyle, N.~H.~Christ, N.~Garron, E.~Goode, T.~Izubuchi, C.~Lehner, Q.~Liu {\it et al.},
 [arXiv:1106.2714 [hep-lat]].
 %
\bibitem{Kim:2009fe}
  C.~Kim and N.~H.~Christ,
  PoS LAT {\bf 2009} (2009) 255
  [arXiv:0912.2936 [hep-lat]].
%
\bibitem{Buras:2008nn}
A.~J.~Buras, D.~Guadagnoli,
Phys.\ Rev.\ {\bf D78}, 033005 (2008).
[arXiv:0805.3887 [hep-ph]].
\bibitem{Buras:2010pza}
A.~J.~Buras, D.~Guadagnoli, G.~Isidori,
Phys.\ Lett.\ {\bf B688}, 309-313 (2010).
[arXiv:1002.3612 [hep-ph]].
%
\bibitem{Colangelo:2010et}
  G.~Colangelo, S.~Durr, A.~Juttner, L.~Lellouch, H.~Leutwyler, V.~Lubicz, S.~Necco, C.~T.~Sachrajda {\it et al.},
[arXiv:1011.4408 [hep-lat]].
%
\bibitem{Kaplan:1992bt}
  D.~B.~Kaplan,
  Phys.\ Lett.\  {\bf B288 } (1992)  342-347.
  [hep-lat/9206013].
\bibitem{Shamir:1993zy}
  Y.~Shamir,
  Nucl.\ Phys.\  {\bf B406 } (1993)  90-106.
  [hep-lat/9303005].
\bibitem{Furman:1994ky}
  V.~Furman, Y.~Shamir,
  Nucl.\ Phys.\  {\bf B439 } (1995)  54-78.
  [hep-lat/9405004].
\bibitem{Antonio:2008zz}
  D.~J.~Antonio {\it et al.}  [RBC Collaboration and UKQCD Collaboration],
  Phys.\ Rev.\  D {\bf 77} (2008) 014509
  [arXiv:0705.2340 [hep-lat]].
  \bibitem{Allton:2008pn}
  C.~Allton {\it et al.} [ RBC-UKQCD Collaboration ],
  Phys.\ Rev.\  {\bf D78 } (2008)  114509.
  [arXiv:0804.0473 [hep-lat]].
\bibitem{Aoki:2010dy}
  Y.~Aoki {\it et al.}  [RBC Collaboration and UKQCD Collaboration],
  Phys.\ Rev.\  D {\bf 83} (2011) 074508
  [arXiv:1011.0892 [hep-lat]].
\bibitem{Vranas:1999rz}
  P.~M.~Vranas,
  arXiv:hep-lat/0001006.
\bibitem{Vranas:2006zk}
  P.~M.~Vranas,
  Phys.\ Rev.\  D {\bf 74} (2006) 034512
  [arXiv:hep-lat/0606014].
\bibitem{Fukaya:2006vs}
  H.~Fukaya {\it et al.} [ JLQCD Collaboration ],
  Phys.\ Rev.\  {\bf D74 } (2006)  094505.
  [hep-lat/0607020].
\bibitem{Renfrew:2009wu}
  D.~Renfrew, T.~Blum, N.~Christ, R.~Mawhinney and P.~Vranas,
  PoS {\bf LATTICE2008} (2008) 048
  [arXiv:0902.2587 [hep-lat]].
\bibitem{dsdrpaper} \emph{Domain wall QCD with near-physical pions}, R.Arthur et al.,
\{RBC and UKQCD Collaborations\}, (in preparation).
\bibitem{Bratt:2010jn}
  J.~D.~Bratt {\it et al.}  [LHPC Collaboration],
  Phys.\ Rev.\ D {\bf 82} (2010) 094502
  [arXiv:1001.3620 [hep-lat]].
\bibitem{Sachrajda:2004mi}
  C.~T.~Sachrajda, G.~Villadoro,
  Phys.\ Lett.\  {\bf B609}, 73-85 (2005).
  [hep-lat/0411033].
\bibitem{Kim:2003xt}
  C.~Kim,
  Nucl.\ Phys.\ Proc.\ Suppl.\  {\bf 129 } (2004)  197-199.
  [hep-lat/0311003].
\bibitem{Kim:2005gka}
  C.~H.~Kim,
  Nucl.\ Phys.\ Proc.\ Suppl.\  {\bf 140 } (2005)  381-383.
\bibitem{Lellouch:2000pv}
  L.~Lellouch, M.~Luscher,
  Commun.\ Math.\ Phys.\  {\bf 219 } (2001)  31-44.
  [hep-lat/0003023].
\bibitem{Lin:2001ek}
  C.~J.~D.~Lin, G.~Martinelli, C.~T.~Sachrajda, M.~Testa,
  Nucl.\ Phys.\  {\bf B619 } (2001)  467-498.
  [hep-lat/0104006].
%
\bibitem{Luscher:1990ux}
  M.~Luscher,
  Nucl.\ Phys.\  {\bf B354 } (1991)  531-578.
%
\bibitem{Hoogland:1977kt}
  W.~Hoogland, S.~Peters, G.~Grayer, B.~Hyams, P.~Weilhammer, W.~Blum, H.~Dietl, G.~Hentschel {\it et al.},
  Nucl.\ Phys.\  {\bf B126 } (1977)  109.
%
\bibitem{Losty:1973et}
  M.~J.~Losty, V.~Chaloupka, A.~Ferrando, L.~Montanet, E.~Paul, D.~Yaffe, A.~Zieminski, J.~Alitti {\it et al.},
  Nucl.\ Phys.\  {\bf B69 } (1974)  185-204.
 %
\bibitem{schenk_curve}
A. Schenk, \emph{Nucl. Phys. B} {\bf{363}} (1991) 97.
%
\bibitem{Colangelo:2001df}
  G.~Colangelo, J.~Gasser, H.~Leutwyler,
  Nucl.\ Phys.\  {\bf B603 } (2001)  125-179.
  [hep-ph/0103088].
%
\bibitem{Buchalla:1995vs}
  G.~Buchalla, A.~J.~Buras, M.~E.~Lautenbacher,
  Rev.\ Mod.\ Phys.\  {\bf 68 } (1996)  1125-1144.
  [hep-ph/9512380].
\bibitem{Buras:1993dy}
  A.~J.~Buras, M.~Jamin, M.~E.~Lautenbacher,
  Nucl.\ Phys.\  {\bf B408 } (1993)  209-285.
  [hep-ph/9303284].
\bibitem{Ciuchini:1992tj}
  M.~Ciuchini, E.~Franco, G.~Martinelli, L.~Reina,
  Phys.\ Lett.\  {\bf B301 } (1993)  263-271.
  [hep-ph/9212203].
\bibitem{Ciuchini:1995cd}
  M.~Ciuchini, E.~Franco, G.~Martinelli, L.~Reina, L.~Silvestrini,
  Z.\ Phys.\  {\bf C68 } (1995)  239-256.
  [hep-ph/9501265].
\bibitem{vanRitbergen:1997va}
  T.~van Ritbergen, J.~A.~M.~Vermaseren, S.~A.~Larin,
  Phys.\ Lett.\  {\bf B400 } (1997)  379-384.
  [hep-ph/9701390].
\bibitem{Nakamura:2010zzi}
  K.~Nakamura {\it et al.} [ Particle Data Group Collaboration ],
  J.\ Phys.\ G {\bf G37 } (2010)  075021.
 %
\bibitem{Christ:2010zz}
  N.~H.~Christ [ RBC and UKQCD Collaboration ],
  PoS {\bf LATTICE2010 } (2010)  300.
%
\bibitem{Cirigliano:2002jy}
  V.~Cirigliano, J.~F.~Donoghue, E.~Golowich and K.~Maltman,
  Phys.\ Lett.\ B {\bf 555} (2003) 71
  [hep-ph/0211420].
\bibitem{Buras:2003zz}
  A.~J.~Buras and M.~Jamin,
  JHEP {\bf 0401} (2004) 048
  [hep-ph/0306217].
\bibitem{Cirigliano:2011ny}
  V.~Cirigliano, G.~Ecker, H.~Neufeld, A.~Pich and J.~Portoles,
  Rev.\ Mod.\ Phys.\  {\bf 84} (2012) 399
  [arXiv:1107.6001 [hep-ph]].
\bibitem{Martinelli:1994ty}
  G.~Martinelli, C.~Pittori, C.~T.~Sachrajda, M.~Testa, A.~Vladikas,
  Nucl.\ Phys.\  {\bf B445 } (1995)  81-108.
  [hep-lat/9411010].
\bibitem{Aoki:2007xm}
  Y.~Aoki, P.~A.~Boyle, N.~H.~Christ, C.~Dawson, M.~A.~Donnellan, T.~Izubuchi, A.~Juttner, S.~Li {\it et al.},
  Phys.\ Rev.\  {\bf D78 } (2008)  054510.
  [arXiv:0712.1061 [hep-lat]].
%
\bibitem{Sturm:2009kb}
  C.~Sturm, Y.~Aoki, N.~H.~Christ, T.~Izubuchi, C.~T.~C.~Sachrajda, A.~Soni,
  Phys.\ Rev.\  {\bf D80 } (2009)  014501.
  [arXiv:0901.2599 [hep-ph]]. 
\bibitem{Aoki:2010pe}
  Y.~Aoki, R.~Arthur, T.~Blum, P.~A.~Boyle, D.~Brommel, N.~H.~Christ, C.~Dawson, T.~Izubuchi {\it et al.},
   [arXiv:1012.4178 [hep-lat]].
%
\bibitem{Bedaque:2004ax} 
  P.~F.~Bedaque and J.~-W.~Chen,
  Phys.\ Lett.\ B {\bf 616}, 208 (2005)
  [hep-lat/0412023].
%
\bibitem{Luscher:1993gh}
  M.~Luscher, R.~Sommer, P.~Weisz, U.~Wolff,
  Nucl.\ Phys.\  {\bf B413 } (1994)  481-502.
  [hep-lat/9309005].
\bibitem{Luscher:1991wu}
  M.~Luscher, P.~Weisz, U.~Wolff,
  Nucl.\ Phys.\  {\bf B359 } (1991)  221-243.
 \bibitem{Arthur:2010ht}
   R.~Arthur and P.~A.~Boyle  [RBC Collaboration and UKQCD Collaboration],
   Phys.\ Rev.\  D {\bf 83} (2011) 114511.
   [arXiv:1006.0422 [hep-lat]].
\bibitem{Arthur:2010hy}
   R.~Arthur and P.~A.~Boyle,
   PoS {\bf LATTICE2010} (2010) 244.
   [arXiv:1010.6140 [hep-lat]].
\bibitem{Boyle:2011kn}
   P.~Boyle and N.~Garron,
   PoS {\bf LATTICE2010} (2010) 307.
   [arXiv:1101.5579 [hep-lat]].
\bibitem{Arthur:2011cn}
  R.~Arthur, P.~A.~Boyle, N.~Garron, C.~Kelly and A.~T.~Lytle,
 [arXiv:1109.1223 [hep-lat]].
\bibitem{Lehner:2011fz}
  C.~Lehner, C.~Sturm,
  Phys.\ Rev.\  {\bf D84 } (2011)  014001.
  [arXiv:1104.4948 [hep-ph]].
\bibitem{Lightman:2009ka}
  M.~Lightman [ RBC and UKQCD Collaborations ],
  PoS {\bf LATTICE2008}, 273 (2008).
  [arXiv:0906.1847 [hep-lat]].
  \bibitem{Lightman:thesis}
M.~Lightman, Columbia University Doctoral Thesis (2011).
\bibitem{Hasenfratz:2008fg}
  A.~Hasenfratz, R.~Hoffmann, S.~Schaefer,
  Phys.\ Rev.\  {\bf D78}, 014515 (2008).
  [arXiv:0805.2369 [hep-lat]].
\bibitem{Laiho:2002jq}
  J.~Laiho and A.~Soni,
  Phys.\ Rev.\  D {\bf 65} (2002) 114020.
  [arXiv:hep-ph/0203106].
\bibitem{Aubin:2008vh}
  C.~Aubin, J.~Laiho, S.~Li and M.~F.~Lin,
  Phys.\ Rev.\  D {\bf 78} (2008) 094505.
  [arXiv:0808.3264 [hep-lat]].
\bibitem{Kim:2010sd}
  C.~H.~Kim, C.~T.~Sachrajda,
  Phys.\ Rev.\  {\bf D81 } (2010)  114506.
  [arXiv:1003.3191 [hep-lat]].
\end{thebibliography}
\end{document}